\documentclass[12pt,preprint]{aastex} 
\newcommand\etal{{\it et al.}}
\newcommand\lae{\mathrel{<\kern-1.0em\lower0.9ex\hbox{$\sim$}}}
\newcommand\gae{\mathrel{>\kern-1.0em\lower0.9ex\hbox{$\sim$}}}
\newcommand\kms{km~s$^{-1}$}

\newcommand\mthree{$^{-3}$}

\newcommand{\m}{$\mu$m}
\newcommand{\syone}{\mbox{Sy 1}}
\newcommand{\sytwo}{\mbox{Sy 2}}
\newcommand{\syones}{\mbox{Sy 1's}}
\newcommand{\sytwos}{\mbox{Sy 2's}}
\newcommand{\siii}{\ion{S}{3}}
\newcommand{\nev}{\ion{Ne}{5}}

\newcommand{\oiv}{\ion{O}{4}}
\newcommand{\oiii}{\ion{O}{3}}
\newcommand{\molhyd}{H$_2\,S(1)$}
\newcommand{\fluxunits}{$10^{-14}$ ergs s$^{-1}$ cm$^{-2}$}
\newcommand{\lumunits}{ergs s$^{-1}$}
\newcommand{\nln}{$\nu L_{\nu}$}

\title{Infrared Diagnostics for the Extended 12 \m\ Sample of Seyferts} 
\begin{document}

\author{Stefi A. Baum\altaffilmark{1}, Jack F. Gallimore\altaffilmark{2}, 
Christopher P. O'Dea\altaffilmark{3}, Catherine L. Buchanan\altaffilmark{4},
Jacob Noel-Storr\altaffilmark{1},
David J. Axon\altaffilmark{3}, Andy  Robinson\altaffilmark{3}, 
Moshe Elitzur\altaffilmark{5},
Meghan Dorn\altaffilmark{1,6}, Shawn Staudaher\altaffilmark{1}, 
Martin Elvis\altaffilmark{7}}

\altaffiltext{1}{Chester F. Carlson Center for Imaging Science, Rochester Institute 
of Technology, 
54 Lomb Memorial Drive, Rochester, NY 14623}
\altaffiltext{2}{Department of Physics and Astronomy, Bucknell University, Lewisburg, PA 17837; 
Currently visiting NRAO, 520 Edgemont Rd., Charlottesville, VA 22903}
\altaffiltext{3}{Department of Physics, Rochester Institute of Technology, 
84 Lomb Memorial Drive, 
Rochester, NY 14623}
\altaffiltext{4}{School of Physics, University of Melbourne,
Parkville, Victoria, 3010 Australia}
\altaffiltext{5}{Department of Physics and Astronomy, University of Kentucky,
Lexington, KY 40506}
\altaffiltext{6}{Rush-Henrietta High School,  1799 Lehigh Station Rd., Henrietta, NY 14467}
\altaffiltext{7}{Harvard-Smithsonian Center for Astrophysics, 60 Garden Street, 
Cambridge, MA 02138}

\begin{abstract}
We present an analysis of Spitzer IRS spectroscopy of 83 active galaxies
from the extended 12\m\ sample.
We find rank correlations between several tracers of star formation which 
suggest that (1) the PAH feature is a reliable tracer of star formation,
(2) there is a significant contribution to the heating
of the cool dust by  stars, (3) the H$_2$ emission is also primarily 
excited by star formation.

The 55-90 vs. 20-30 spectral index plot is also a diagnostic
of the relative contribution of Starburst to AGN.
We see there is a large change in spectral index across the sample:
$\Delta \alpha \sim 3$ for both indices. 
Thus, the contribution to the IR spectrum from the AGN and starburst
components can be comparable in magnitude but the relative contribution
also varies widely across the sample.

We find rank correlations between several AGN tracers.
We find correlations of the ratios [OIII]$\lambda$5007/[OIV]26\m\ and
[OIII]$\lambda$5007/[NeV]14\m\ with the Sil strength which we adopt as an
orientation indicator.  This suggests that some of the [OIII]$\lambda$5007 
emission in these Seyferts is subject to orientation dependent obscuration as found
by \citet{haas05} for radio galaxies and quasars.
There is no correlation of [NeV] EW with the Sil
10\m\ strength indicating that the [NeV] emission is not strongly
orientation dependent.	This suggests that the obscuring material
(e.g., torus) is  not very optically thick at 14\m\ consistent with
the results of \citet{buchanan06}.

We search for correlations between AGN and Starburst tracers
and we conclude that the AGN and Starburst tracers are
not correlated. This is consistent with our conclusion that the relative strength
of the AGN and Starburst components varies widely across the sample.
Thus, there is no simple link between AGN fueling and Black Hole Growth
and star formation in these galaxies. 

The density diagnostic [NeV] 14/24 \m\ and [SIII] 18/33 \m\ line
ratios are consistent with the gas being near the low density limit,
i.e., $\sim 10^3$ cm\mthree\ for [NeV] and $n_e \sim $ few hundred
cm\mthree\ for [SIII].

The distribution of Sil 10 \m\ and 18 \m\ strengths is consistent
with the clumpy torus models of \citet{sirocky08}.

We find a rank correlation between the [NeV] 14\m\ line and the 6.7 \m\ continuum 
which may be due to an extended component of hot dust.

The \sytwos\ with a Hidden Broad Line Region (HBLR) have a higher ratio of 
AGN to Starburst contribution to the SED than \sytwos\ without an HBLR.
This may contribute to the detection of the HBLR in polarized light.
The \sytwos\ with an HBLR are more similar to the \syones\ than they are 
to the \sytwos\ without an HBLR. 
\end{abstract}


\section{Introduction}

The basic paradigm for nuclear activity in galaxies, an accretion disc
around a super massive black hole (SMBH), is
well-established. However, we have at best a sketchy understanding of
how SMBH are formed and what role nuclear activity plays in galaxy
evolution.  As we learn more about the formation and evolution of
normal galaxies, and better understand the physics of active galactic
nuclei (AGNs), it is becoming clear that these processes are
intimately connected.  The ubiquity of super massive black holes and
the observed relation between black hole mass and bulge mass
\citep{merritt01, gebhardt00} suggest that nuclear activity may be
closely connected to star formation in all galaxies.  Accretion onto a
black hole, which powers AGNs, requires available fuel in the nuclear
region. Gas infall is also likely to be associated with bursts of star
formation \citep{hernquist95,mihos96}.  Furthermore, AGNs deposit
energy in the surrounding interstellar medium (ISM) via ionizing
radiation, radio jets and perhaps winds, possibly triggering star
formation. Circumnuclear star formation has been observed in AGN and a
link between nuclear activity and star formation is also supported by
a recent finding that the mean stellar age of narrow-line AGNs is much
younger than in similar normal spiral galaxies
\citep{kauffman03}. However, the interaction between active nuclei and
their host galaxies remains poorly understood.

The mid-IR includes important diagnostics such as the PAHs, Sil
features, and fine structure lines
\citep{spinoglio92,voit92b,genzel98,laurent00}.  Previous mid-IR
spectroscopic studies of AGN have shown that these diagnostics allow
separation of AGN and starburst properties, as well as provide
constraints on orientation and obscuration of the central torus
\citep{genzel98,laurent00,clavel00,sturm02,haas05,
  weedman05,ogle06,shi06,hao07,spoon07}.

This paper is part of a series on a Spitzer study of the Extended
12\m\ Sample of Seyfert galaxies.  \citet{buchanan06} presented low
resolution IRS spectra of the first 51 (of 83) sources and
demonstrated the variety of mid-IR spectral shapes due to the relative
mix of comparable contributions from starburst and AGN. They examined
the ratio of IR to nuclear radio flux density and showed that the
ratio is a factor of two larger in \syones\ than \sytwos\ between 15
and 35 \m\ suggesting the torus emission is anisotropic by a factor of
two at these wavelengths. Additional preliminary results have been
presented by \citet{buchanan08}.  Other papers currently planned in
the series will present (1) an atlas of all the data \citep{gallimore09a},
(2) models of the mid- to far- IR SEDs which combine both
starburst and AGN contributions \citep{gallimore09b}, and (3)
UKIRT observations of the near-IR emission and discuss constraints on
the relative mix of starburst and AGN contributions to the spectrum
\citep{buchanan09}.  In this paper we examine diagnostic features
(PAH, Sil, fine structure lines, H$_2$, continuum slope) in the IRS
spectra and discuss the implications for the AGN-Starburst connection
and obscuration of the central engine.  We note that independent
groups have also published results based on our data. \citet{maiolino07}
used observations of some \syones\ from \citet{buchanan06}
as a comparison sample for their study of high luminosity QSOs.  
\citet{deo07} obtained low resolution Spitzer IRS observations 
of 12 Sy 1.8 and 1.9 and combined the results with archival data 
of 57 \syones\ and \sytwos\ from \citet{buchanan06} and \citet{weedman05}. 
High resolution IRS observations of a subsample of 29 
sources are presented by \citet{tommasin08}. Recently \citet{wu09}
published a study of 103 Seyferts which included our IRS data. 
\citet{gallimore09a} show that there are systematic differences between 
our values for the PAH and Sil features and those of \citet{wu09} which are
due to the different analysis techniques used (PAHFIT vs. Spline-fitting). 
Nevertheless, we agree on the overall trends which are common to both papers. 

\section{Spitzer Observations}

\subsection{The Sample of AGN}\label{sec:sample}

The properties of the sample are discussed by \citet{buchanan06} and
\citet{gallimore09a}.  The sample comprises all galaxies identified
as Seyferts or LINERS from the extended 12~\micron\ sample of
\citet{rush93} that have $cz < 10000$~\kms. The potential advantage of
the 12~\micron\ sample is its perceived resistance to
wavelength-dependent selection effects \citep{spinoglio89}. Note that
there may in fact be selection effects in the sample due to
anisotropic emission from the torus at 12
~\micron\ \citep{buchanan06}.  The redshift criterion was included to
ensure that the smallest possible physical size was probed by the
fixed Spitzer apertures in order to better exclude host galaxy
emission.  The data include IRAC imaging, IRS spectral mapping, and
MIPS SED spectra as described by \citet{gallimore09a}.

The spectral classifications were obtained by a thorough search of the 
literature for each galaxy; see Table 1 of \citet{gallimore09a}. For the
purposes of comparing the Seyfert types we consider Seyfert types
1.$n$ (i.e. types 1.0, 1.2, 1.5, 1.8, and 1.9 as well as narrow line
Seyfert 1's) to be \syones, and objects with no broad permitted lines
to be \sytwos. We consider the properties of the Seyferts with hidden
broad line regions (HBLRs) separately and compare them to the
\syones\ and \sytwos. We refer to the sum of the \syones, \sytwos, and
hidden \syones\ as the combined sample.  The resulting sample contains
83 sources of which 7 are now classified as HII galaxies, 6 are LINERS
and the remaining 70 are Seyferts (27 Sy 1-1.5, 17 Sy 2 with HBLR,
20 Sy 2 without HBLR, and 6 Sy 1.8-1.9. The sample is listed in Table 1 of
\citet{gallimore09a}.

\subsection{IRS Observations and Data Reduction}

The sample galaxies were observed by Spitzer IRS \citep{houck04} in
spectral mapping mode, using the Short-Low and Long-Low modules. The
modules cover a wavelength range of $\sim$5 -- 36~\micron, with a
resolving power ($\lambda/\Delta \lambda$) ranging from 64 to 128. The
integration times were 6 seconds per slit pointing. Details of the
observing and reduction are given by \citet{gallimore09a} and are
summarized here.

The spectral maps were constructed to span $> 10''$ in the cross-slit
direction and centered on the target source. The observations were
stepped perpendicular to the slit by half slit-width spacings. For the
SL data, the mapping involved 13 observations stepped by 1.8''
perpendicular to the slit, and for the LL data, 5 observations stepped
by 5.25'' perpendicular to the slit.  The resulting spectral cubes span
roughly 25.3'' x 54.6'' (5.3 - 14.2 \m) for the SL data and 29.1'' x 151''
(14.2 - 36 \m) for the LL data.

The raw data were processed through the Spitzer BCG pipeline, version
S15.3.0. Sky subtraction employed off-source orders. In SL and LL
observations, the source is centered in first or second order
separately, with the ``off-order'' observing the sky at an offset
position along the slit orientation. Sky frames were constructed using
median combinations of the off-source data and subtracted from the
on-source data of matching order. No detectable contamination appeared
in the sky frame images.

Data cubes were constructed and edited for cosmic-rays and spurious
pixels using the CUBISM software package developed by the SINGS IRS
team \citep{smith07b}.  We used synthetic, 20'' diameter circular
apertures to extract the spectra.

To calibrate the flux scale, we employed archival IRS staring and
spectral mapping observations of the calibrator stars HR7341 (SL) and
HR6606 (LL). The staring mode spectra were extracted using SPICE
(Spitzer Science Center 2006), and the spectral mapping observations
were processed as described above. The flux calibration curve was
determined by the ratio of the flux-calibrated staring mode spectra to
the uncalibrated cube-extracted spectra. This flux calibration was
applied to all target sources observed in spectral mapping mode. Based
on inspection of the spectral overlaps between modules and orders the
calibration is self-consistent to within 5\% across the IRS
spectrum. The only exception was NGC 4151, SL order 1, which had to be
scaled down by $\sim 15\%$ to match the neighboring SL order 2 and LL
order 2 spectra, perhaps owing to a residual pipeline error.  After
applying a uniform scaling to SL order 1, the spectrum matched well
the neighboring spectral overlaps.

Archival data of our sample, such as obtained by the SINGS
collaboration, were processed in the same manner. There are however a
few sources for which only staring mode observations are available for
the SL module: NGC 526A, NGC 3227, NGC 4941, IC 5063, NGC 7172, and
NGC 7314. The staring mode spectra were extracted using SPICE and
scaled to match IRAC 5.8 and 8.0 \m\ photometry obtained as part of
this project \citep{gallimore09a}.

\subsection{MIPS-SED Observations and Data Reduction}
The sample galaxies were also observed using the MIPS camera in SED
mode, which provides very low resolution ($R \sim 20$) slit
spectroscopy covering $\sim \lambda 55$--90 \m. The slit-width for SED
mode observations is 20\arcsec. Integration times ranged from 3 to 10
seconds depending on the IRAS 60\m\ flux density. Observations include
six pairs of on-source and off-source positions (1-3\arcmin\ away from source)
for background subtraction.

Basic data processing, including background subtraction and flux and
wavelength calibration, was handled by the {\em Spitzer} pipeline
(version S.14.4). The final, low resolution spectra were extracted
from the long-slit images using a three-pixel ($\sim 30$\arcsec)
synthetic aperture perpendicular to the slit. For each source, the
position of the galaxy nucleus was identified first manually and then
by a Gaussian fit to determine the center of the extraction aperture
to a fraction of a pixel. 

Note that flux calibration is derived from observations of standard
stars and so is most accurate for compact far-infrared
sources. Formally, the accuracy for point sources is better than 10\%,
but degrades to $\sim 15$\% for extended sources with galaxy-like surface
brightness profiles \citep{lu08}.

\subsection{PAHFIT Spectral Decomposition}

We used the PAHFIT spectral decomposition code \citep{smith07a} to fit
and extract line features of the Seyfert IRS spectra. The publicly
available release of PAHFIT permits silicate features only in
absorption. Silicate 10 \m\ and 18 \m\ emission commonly appears among
the sample, however, and we therefore modified PAHFIT to include a
warm, optically thin ($\tau(10\ \mu {\rm m} < 1$) dusty medium. To
accommodate the range of observed Sil emission profiles, the warm dust
model includes two components: (1) an optically thin warm medium,
characterized by a single temperature and 10 \m\ opacity, and (2) a
cold, partially covering dusty medium parameterized solely by covering
fraction (i.e., the cold dust is modeled to produce no significant
emission over the IRS band). Both the cold and warm dust components
obey the extinction law described by \citet{smith07a}. Our goal was to
reproduce the observed silicate emission profile shapes without regard
to determining the physics of the warm dust (other than its presence
and emission strength), and so, for simplicity, the spectrum of the
warm dust, modified by partial foreground absorber, was calculated
assuming a simple slab geometry.

Except for this added warm dust component, we used PAHFIT with its
default settings. PAHFIT robustly decomposes the spectrum into
multiple temperature dust components, silicate features, emission
lines, and PAH dust features. The decomposition proceeds using a
standard non-linear fitting algorithm. PAHFIT was specifically designed to
analyze moderate resolution IRS spectra and carefully limits
parameters, such as line widths and centroids, that might drift to
unphysical values during the fit. 

The final data products include tables of line strengths and
equivalent widths. We further modified PAHFIT to produce
line-subtracted continua. These continua were used to calculate
silicate strengths and spectral indices. In addition, we also modified
the PAHFIT output to produce 3$\sigma$ upper limits for weak or
undetected lines. These data tables are provided in \citet{gallimore09a}.
Note that the line and feature fluxes are not corrected for extinction. 

Table~\ref{tab:median} summarizes the results of the PAHFIT
decomposition, giving the median values of a range of properties for
the continuum emission, Sil feature, star formation tracers, AGN
tracers, and diagnostic line ratios. The median values were calculated
based on the computation of a survival curve
\citep{feigelson1985}. As the line ratios include both upper and lower
limits, we employed the formalism for interval-censoring given by
\citet{finkelstein1986} and \citet{sun1996}, as implemented in the
interval package by M. P. Fay \& P. A. Shaw
(http://cran.r-project.org/web/packages/interval/index.html).

Tables~\ref{tab:ks1}-\ref{tab:ks8} give the results of
statistical tests for the significance of differences in the
distribution of those properties for the different Seyfert
subgroups. These tests comprise Kolmogorov-Smirnov two-sample tests
for uncensored data, or log-rank tests for interval-censored data.

Table~\ref{tab:rank} gives the significance of rank correlations between
the properties for the combined sample. We used Kendall's rank
correlation statistic adapted for interval-censored data
\citep{isobe1986}, reported as a $z$-score and two-sided significance
level. 

In this paper, we search extensively for patterns in the data by
exploring bivariate correlations among measurements: for example, PAH
eqw vs. 20--30~\m\ spectral index. There are well-known parametric and
non-parametric statistical measures of correlation (see, e.g., \citealt{sprent00}).
Parametric methods
specifically apply to cases where the data are believed to be related
through some known and prescribed functional form; for example, the
well-known Pearson correlation coefficient measures the underlying
linear relationship between measurements. One however expects a
fairly complex relationship, if at all, between a spectral index and
an eqw, and so such parametric statistics would not be expected to
apply in the present analyses. For example, data that are tightly but
non-linearly correlated will likely show low linear correlation
measurements \citep{anscombe73}. 

In contrast, rank correlation coefficients provide a robust,
non-parametric measure of association between measurements. They are
better suited for the present study, as they measure simply the degree
to which one measurement increases (or decreases) as another
increases, independent of the degree or functional form of the
underlying association. To our knowledge, only Kendall's rank
correlation statistic (Kendall's $\tau$; \citep{kendall38})
has been generalized to account both for upper limits
(non-detections) and lower limits (e.g., line ratios where the divisor
is an upper limit) in one or both measurements and so is ideal for
this study. We have implemented the generalized Kendall's rank
correlation statistic following the formalism of \citet{isobe1986}.
In the results tables, we report the standard normal
score ($z$) of the statistic as well as the probability ($p$) that
score would have occurred by chance (i.e., the probability of no
correlation). We chose to be conservative in the interpretation of the
correlation statistics and adopt the threshold $p < 1\%$ to indicate
evidence of rank correlation.

\section{RESULTS}

\subsection{The Starburst AGN Connection}

\subsubsection{PAH Emission as a diagnostic of Star Formation}\label{pahsb}

The PAH emission is thought to be produced when small grains
(containing $\sim 50$ Carbon atoms) are heated by single UV photons
from hot stars and radiate though their vibrational modes \citep{tielens99}.
 It is clear that PAH emission is a tracer of
star forming regions in our Galaxy \citep{tielens99}  and
other galaxies \citep{laurent00,peeters04,brandl06,dale06,draine07b,smith07a}
The PAH luminosity is proportional to the star
formation rate in star forming galaxies \citep{roussel01,
brandl06, calzetti07, shi07, farrah07}.  
\citet{voit92a} argues that PAH grains exposed to the
direct hard continuum of an AGN will be destroyed, and the PAHs in AGN
can only survive when they are shielded by high columns of X-ray
absorbing gas, e.g., in the plane of the obscuring torus.  \citet{smith07a} 
notes that the ratio of 7.7/11.3 micron PAH seems to be lower
in some AGN than in starburst galaxies. Since the 11.3 micron PAH are
produced by larger grains than the 7.7 micron PAH \citep{draine07a}
this suggests that the smaller grains are indeed being destroyed by
the AGN.  Thus, PAH emission seems to be a good indicator of star
formation in galaxies with the caveat that it may be significantly
suppressed in AGN.  Siebenmorgen, Kr\"ugel \& Spoon (2004) compare
radiative transfer models which include dust destruction for
starbursts and AGN. In their models, some PAH emission is produced
within clouds illuminated by an AGN, but the strength of the PAH is
lower by a factor of about 5 than that produced in the starburst
models with the same luminosity and A$_v$.  Finally, the equivalent
width of PAH features has been shown to be useful in diagnostic
diagrams as a discriminant between starbursts and AGN as the energy
source for infrared emission \citep{genzel98,laurent00,peeters04,
armus07,spoon07}.

\begin{figure}
\epsscale{0.70}
\plotone{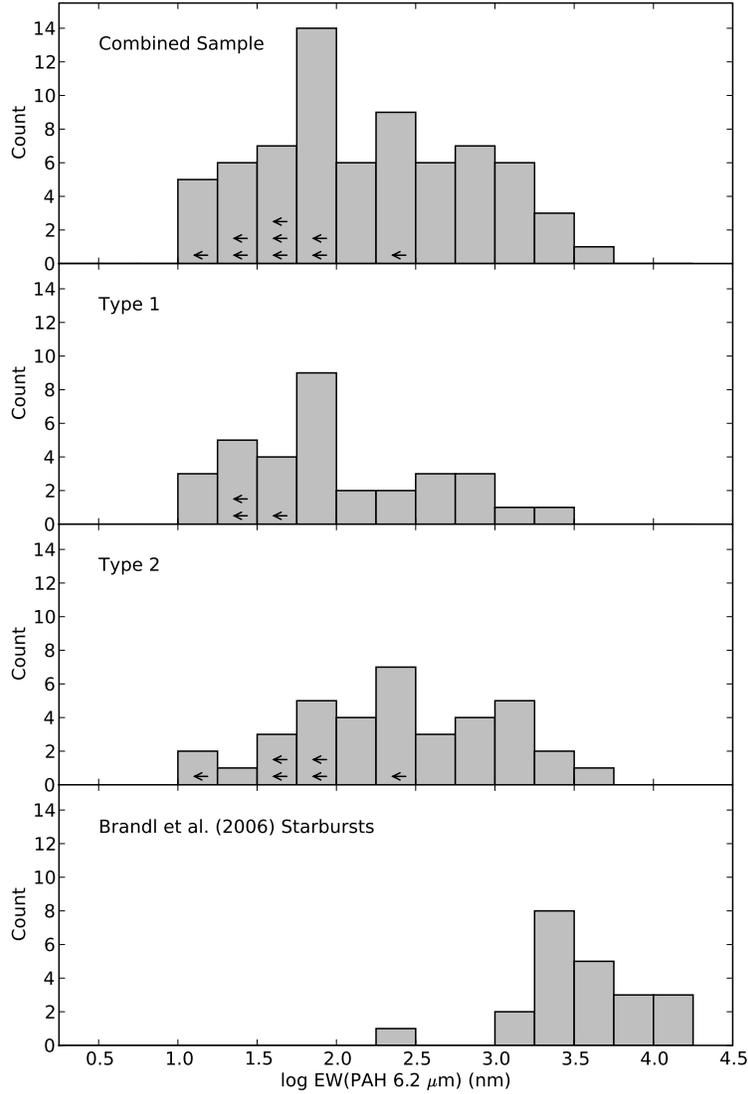}
\caption{The distribution of 6.2 \m\ PAH EW. From the top, the panels
  are (1) \syone\ and \sytwo; (2) \syone\
  (S1.0--1.9, S1n), (3) \sytwo\ (S2, S1h,
  \& S1i), and starbursts from the \citet{brandl06} sample. Note
  that the Brandl et al. sample shares two objects in common with our
  sample, NGC 1097 and NGC1365. The starburst measurements are derived
from a PAHFIT spectral decomposition of the Brandl et al. data. }\label{figpahdist}
\end{figure}

\begin{figure}
\epsscale{0.80}
\plotone{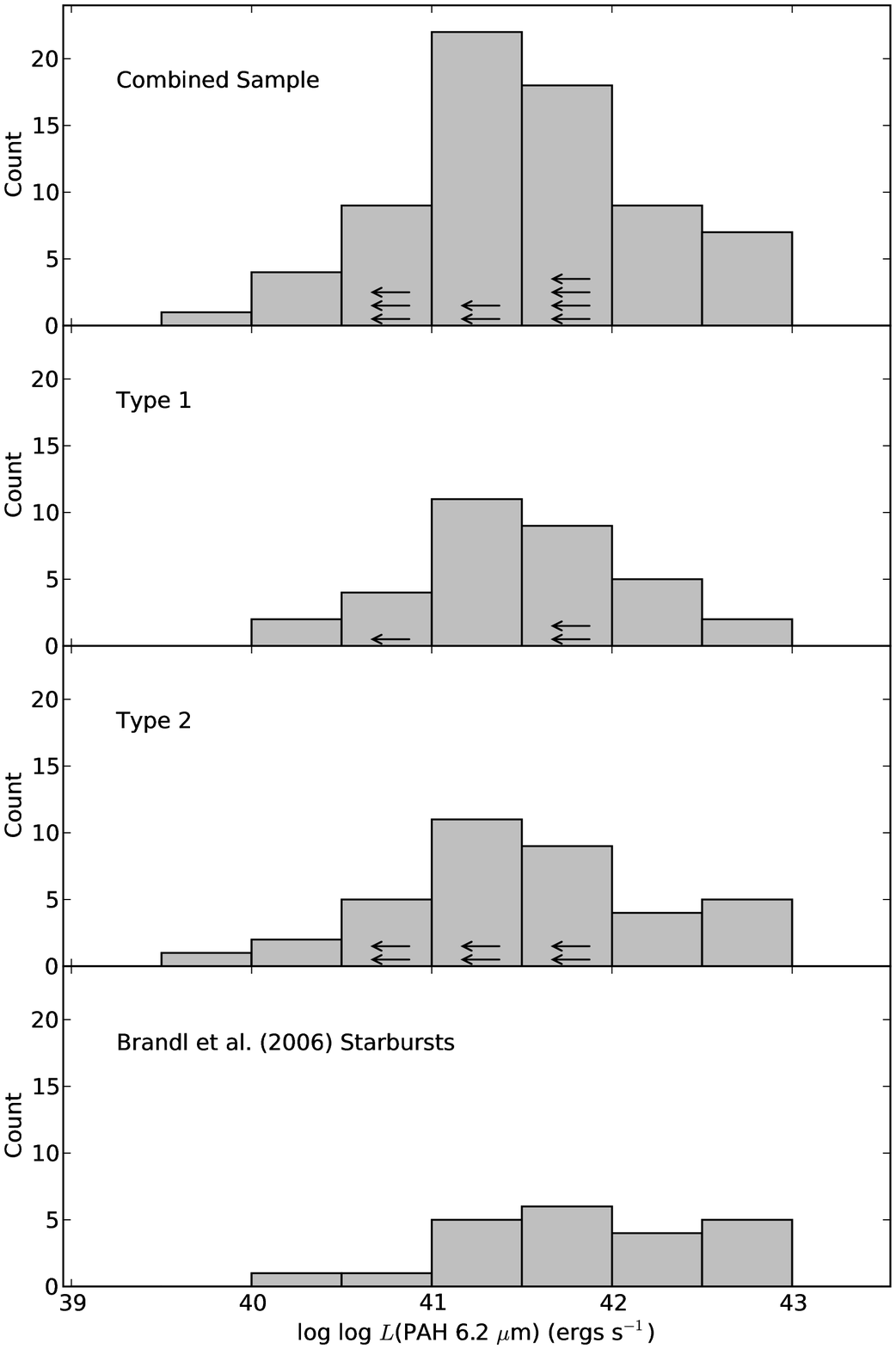}
\caption{The distribution of 6.2 \m\ PAH luminosity. From the top, the panels
  are (1) \syone\ and \sytwo; (2) \syone\
  (S1.0--1.9, S1n), (3) \sytwo\ (S1, S1h,
  \& S1i), and starbursts from the \citet{brandl06} sample. Note
  that the Brandl et al. sample shares two objects in common with our
  sample, NGC 1097 and NGC1365. The starburst measurements are derived
from a PAHFIT spectral decomposition of the Brandl et al. data. }\label{figpahlumdist} 
\end{figure}

The 6.2 micron PAH feature is thought to be relatively ``unconfused''
by other spectral features (e.g., Peeters \etal\ 2004) and so should
be a good measure of the amount of star formation.  We show the
distribution of 6.2 \m\ PAH EW for the \syones, \sytwos\ and hidden
\syones\ in Figure \ref{figpahdist}.  We see that the Sy 2s show a
distribution to higher values of PAH EW. We find that the distribution
of PAH EW in the \syones\ and \sytwos\ are significantly different
(Table \ref{tab:ks1}) in contrast to \citet{wu09} who find no significant
difference in the EWs. This might be due to the systematic differences
in the measured PAH EWs discussed by \citet{gallimore09a}. 
The \sytwos\ which show the high EW are those without an HBLR; the 
\sytwos\ with an HBLR  have a distribution which is similar to that 
of the \syones\ (Tables \ref{tab:median}, \ref{tab:ks1},\ref{tab:ks2}). 
The higher values of PAH EW for the Sy 2s seems mainly due to 
the lower continuum fluxes in the
\sytwos\ relative to the \syones\ since the PAH {\it luminosities}
(Figure~\ref{figpahlumdist}) are similar in the \sytwos\ and the
\syones.  This result is consistent with the findings of Clavel
\etal\ (2000) for a sample of Seyferts studied using ISO.
Figures \ref{figpahdist} and \ref{figpahlumdist} also show that the
Seyferts show a similar distribution in 6.2 \m\ PAH luminosity 
as the \citet{brandl06} starbursts. However, the Seyferts show
lower 6.2 \m\ PAH EW due to the higher continuum contribution from
the AGN. These results are consistent with the results of \citet{buchanan06}
that principle component analysis required a higher star formation
contribution to the SED of \sytwos.

\subsubsection{The PAH 6.2/11.3 ratio}\label{sec:ratio}

The 6.2 \m\ PAH feature is produced by grains with smaller sizes than
the 11.3 \m\ PAH feature \citep{draine07a}.  The smaller grains
are more quickly destroyed by exposure to the AGN spectrum. Thus, the
ratio of 6.2/11.3 PAH should be an indicator of the destruction of
grains by the AGN \citep{smith07a}.  \citet{wu09} show that
their Seyfert sample and the \citet{brandl06} sample of starburst
galaxies have similar average values for the 11.3/6.2 PAH ratio. 

\begin{figure}
\plotone{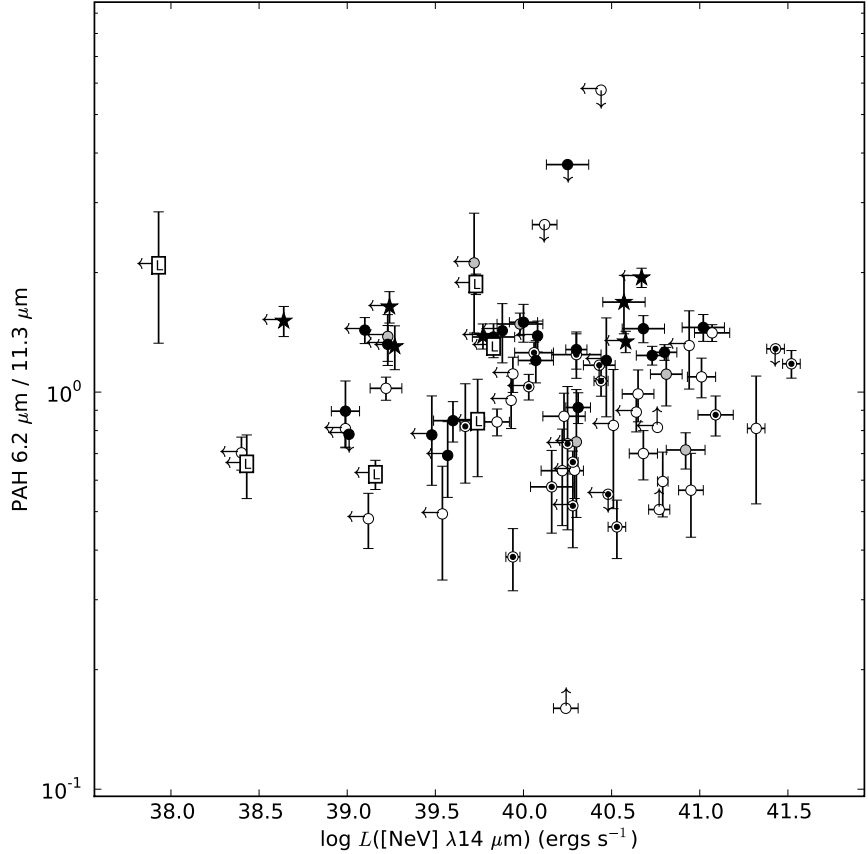}
\caption{The ratio of 6.2/11.3 \m\ PAH strength as a function of [NeV]
  luminosity.  Open circles are S1, S1.5, and S1n, gray-filled
  circles are S1.8 and S1.9, filled circles are Sy 2,
  partially filled circles are ``hidden Seyferts'' (S1h and S1i), squares labeled
  L are LINERs, and filled stars are Starburst/HII galaxies. Limits
  are indicated by appropriately directed arrows. The single lower
  limit in the PAH ratio is MRK 704, for which the 11.3 \m\ feature is
not detected.}\label{figpahratio}
\end{figure}

\begin{figure}
\plotone{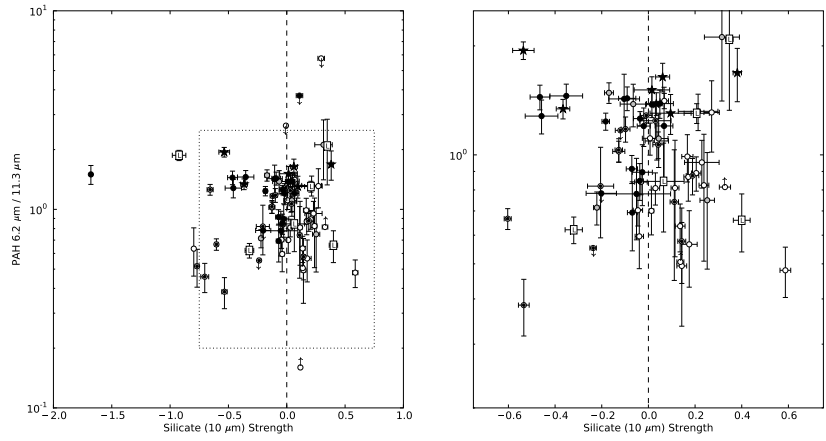}
\caption{Ratio of the 6.2 and 11.3 \m\ PAH lines vs. Sil 10
  \m\ strength.  For the Sil 10 \m\ strength, positive means
  emission.  (Left) the data. (Right) a blow up of the
  region around a Sil strength of zero.  Open circles are S1, S1.5, and S1n, gray-filled
  circles are S1.8 and S1.9, filled circles are Sy 2,
  partially filled circles are ``hidden Seyferts'' (S1h and S1i), squares labeled
  L are LINERs, and filled stars are Starburst/HII galaxies. Limits
  are indicated by appropriately directed arrows.  }\label{figIRS_S_PAHr}
\end{figure}

We calculate the 6.2/11.3 PAH ratio using the fluxes of the features. 
The distribution of the 6.2/11.3 PAH ratio is consistent with being the 
same in the different subgroups.
In Figure~\ref{figpahratio} we plot the ratio of the 6.2 and 11.3 \m\ PAH
strength vs. the [NeV] luminosity which traces the strength of the AGN
ionizing continuum (\S~\ref{agn}).  We see no correlation of the PAH ratio with the
[NeV] luminosity. In Figure~\ref{figIRS_S_PAHr} we plot the 6.2/11.3 PAH ratio 
against the Sil 10 \m\ Strength (see \S \ref{sec:sil}).
We do not see a correlation between the PAH ratio and the Sil 10 \m\ Strength 
(Table~\ref{tab:rank}). This suggests that the PAH emission that we see in these Seyferts 
is not orientation dependent. Any effect of the AGN on the distribution
of PAH grains must be confined to a small volume so that the
integrated measurements studied here are not affected.

 
\subsubsection{PAH  Associated with  Emission from Cool Dust}\label{subseccool}

\begin{figure}
\plotone{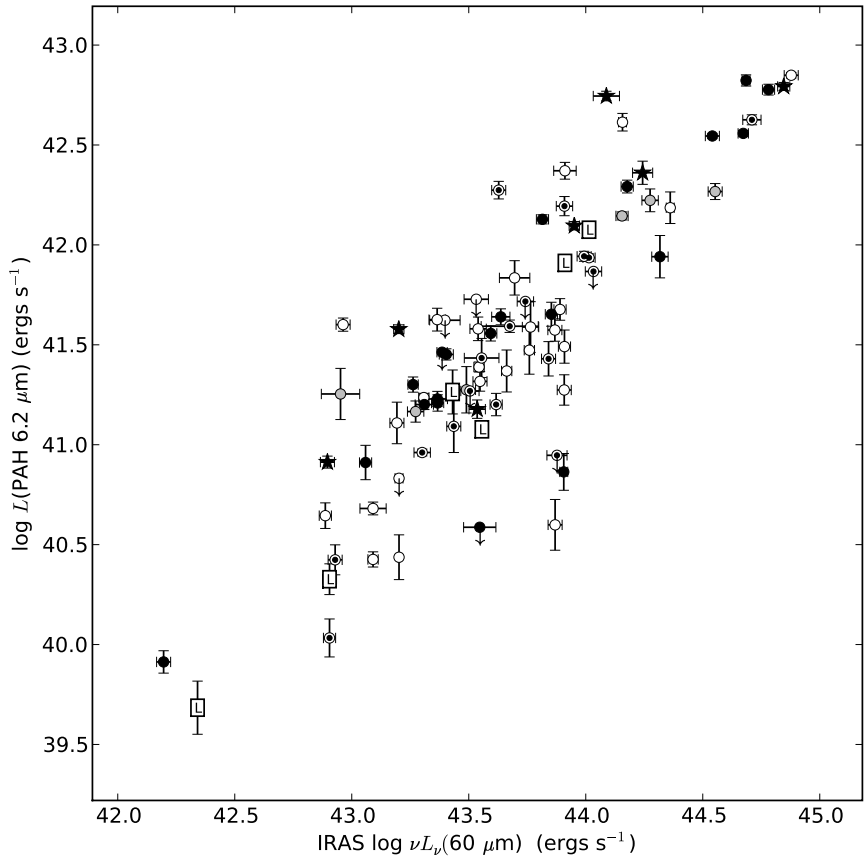}
\caption{6.2 \m\ PAH luminosity vs. IRAS \nln(60 \m).
 Open circles are S1, S1.5, and S1n, gray-filled
  circles are S1.8 and S1.9, filled circles are Sy 2,
  partially filled circles are ``hidden Seyferts'' (S1h and S1i), squares labeled
  L are LINERs, and filled stars are Starburst/HII galaxies. 
}\label{figIRS_60mic_PAH}
\end{figure}

\citet{schweitzer06}  show that QSOs and ULIRGs lie on the same 
correlation between 7.7 \m\ PAH flux and 60 \m\ continuum flux,
suggesting that the FIR is dominated by dust heated by star formation.
In Figure~\ref{figIRS_60mic_PAH} we show a strong rank correlation between
6.2 \m\ PAH luminosity and the integrated luminosity (from IRAS) at 60
\m\ for our sample. We also find a rank correlation between the {\it flux\/}
of the 6.2 \m\ PAH feature and the 60 \m\ continuum (Table ~\ref{tab:rank}).
 
\begin{figure}
\plotone{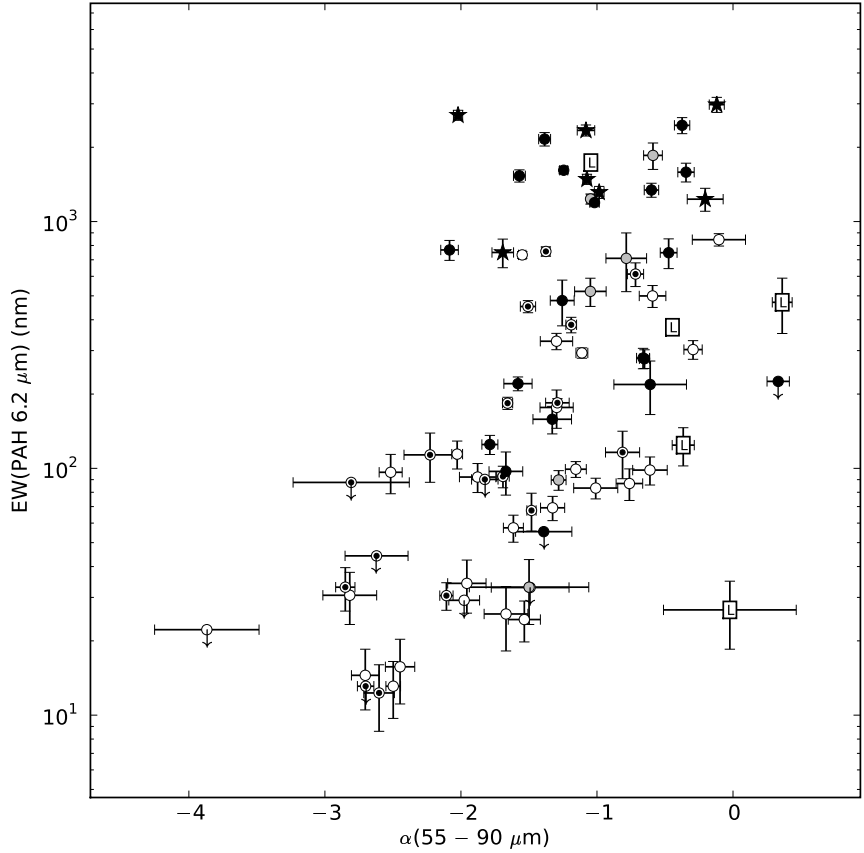}
\caption{6.2 \m\ PAH EW vs. slope of the MIPS SED.
The slope of the MIPS SED indicates the relative importance of cold vs. warm dust
to the FIR SED. Positive slope is a red MFIR SED.
 Open circles are S1, S1.5, and S1n, gray-filled
  circles are S1.8 and S1.9, filled circles are Sy 2,
  partially filled circles are ``hidden Seyferts'' (S1h and S1i), squares labeled
  L are LINERs, and filled stars are Starburst/HII galaxies. 
}\label{figIRS_MIPS_PAH}
\end{figure}

We also show the 6.2 \m\ PAH EW vs. the spectral index of the MIPS SED
55-90 \m\ in Figure \ref{figIRS_MIPS_PAH}.  The slope of the MIPS SED
indicates the relative importance of cold vs. warm dust to the FIR
SED. Positive slope is a red MFIR SED.
The starburst galaxies all lie at large PAH EW.  In the combined
sample, we see a rank correlation between PAH EW and MIPS slope, with
higher PAH strength (more star formation) proportional to more
positive (redder) MIPS slope (more emission from cool dust).  
These two correlations (Figures \ref{figIRS_60mic_PAH} and
\ref{figIRS_MIPS_PAH}) are consistent with a significant contribution from stars
to the heating of the cool dust which produces the FIR emission in our sample.  

\subsubsection{PAH EW vs. 20-30 \m\ spectral index: a measure of 
the relative strength of the starburst and AGN 
components}\label{sec:relative}

\begin{figure}
\plotone{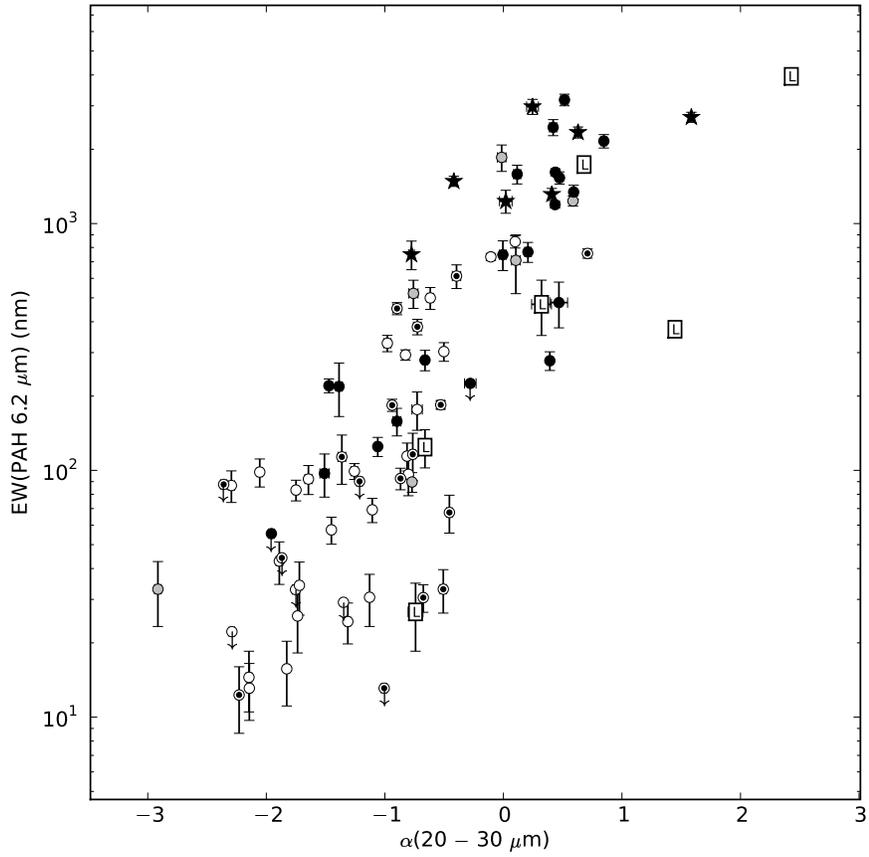}
\caption{6.2 \m\ PAH EW vs. spectral index of the continuum between 20 and 30 \m.
 Positive slope is a red MFIR SED.
 Open circles are S1, S1.5, and S1n, gray-filled
  circles are S1.8 and S1.9, filled circles are Sy 2,
  partially filled circles are ``hidden Seyferts'' (S1h and S1i), squares labeled
  L are LINERs, and filled stars are Starburst/HII galaxies.
}\label{figIRS_SLOPE_PAH}
\end{figure}

\begin{figure} 
\plotone{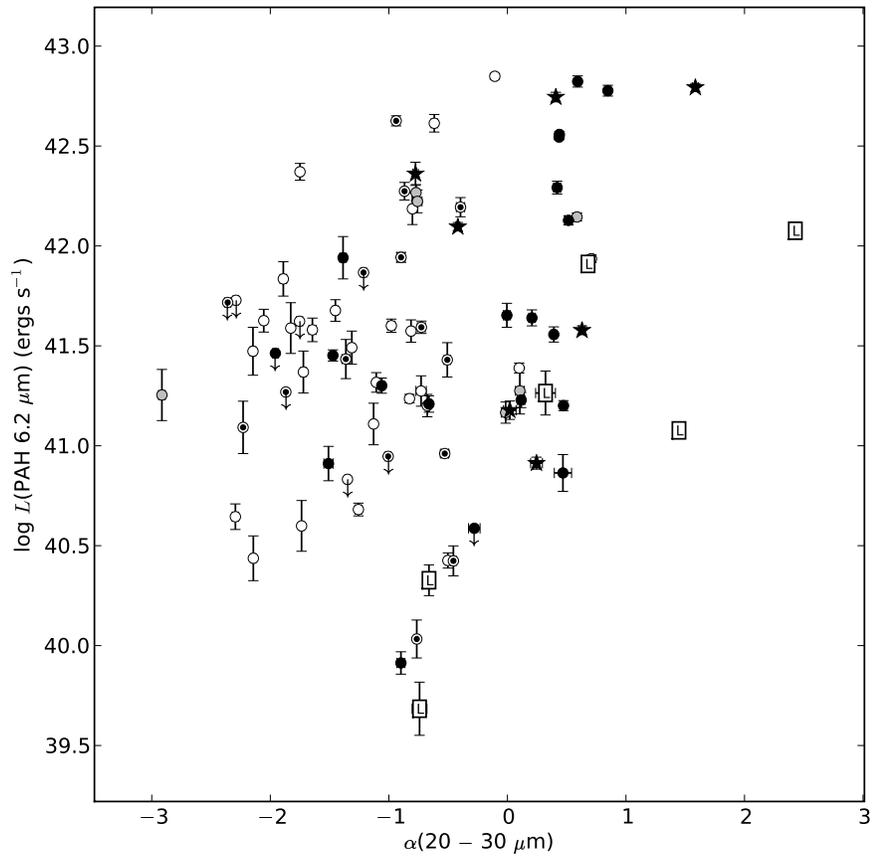}
\caption{Luminosity of the PAH 6.2 \m\ feature vs. spectral index of the 
continuum between 20 and 30 \m.
 Open circles are S1, S1.5, and S1n, gray-filled
  circles are S1.8 and S1.9, filled circles are Sy 2,
  partially filled circles are ``hidden Seyferts'' (S1h and S1i), squares labeled
  L are LINERs, and filled stars are Starburst/HII galaxies. 
}\label{IRS_SLOPE_PAH_LUM} 
\end{figure}

20 microns is just redward of the ``knee bend'' we see in the SED of
many Sy 1s and 2s \citep{buchanan06}.  The 20-30 micron spectral
index $\alpha_{20-30}$, was determined using a power law fit to the
line-subtracted continuum between 20 and 30 microns rest frame.  This
spectral index characterizes the relative contribution of warm and cool
dust to the mid-IR.  Note that the values of $\alpha_{20-30}$ used in
\citet{buchanan06}  are actually $\alpha_{20-30}$ + 2.0.  The correct
values of $\alpha_{20-30}$ are used here.  The similar 15-30
\m\ spectral index has also been found to be an effective AGN
discriminator \citep{brandl06, cleary07}.  Using a
subset of our data, \citet{deo07} show that there is a correlation
between the 6.2 \m\ PAH EW and $\alpha_{20-30}$. \citet{wu09} also find
a correlation between the the 11.3 \m\ PAH EW and the 15-30\m\ spectral 
index. We plot the 6.2 \m\ PAH EW vs.  $\alpha_{20-30}$ in Figure 
\ref{figIRS_SLOPE_PAH}.
We see the different subsamples tend to lie in different parts of the diagram.
The \syones\ and \sytwos\ with an HBLR generally have $\alpha_{20-30} < 
-0.5$ and cover a large range in 6.2 \m\ PAH EW. On the other hand 
the \sytwos\ and Liners tend to have
high values of 6.2 \m\ PAH EW and cover a large range in $\alpha_{20-30}$ with
the \sytwos\ showing a strong rank correlation ($p < 0.1\%$) between the two 
properties. We also see a rank correlation between
6.2 \m\ PAH {\it Luminosity\/} and $\alpha_{20-30}$
(Figure~\ref{IRS_SLOPE_PAH_LUM}). Thus we see rank correlations between 
the PAH EW and $\alpha_{20-30}$ which are both measures of the {\it relative\/} 
strength of the starburst and AGN components, and also  PAH luminosity 
(a measure of the absolute strength of the starburst component) and 
$\alpha_{20-30}$.  

In Figure \ref{figIRS_SLOPE_PAH} the sources in
the lower left are those with a low starburst to AGN ratio. Thus, the
value of $\alpha_{20-30}$ between -1 and -2 is likely to represent the
spectrum of the ``torus'', consistent with values calculated by
\citet{nenkova08}.

\begin{figure}
\plotone{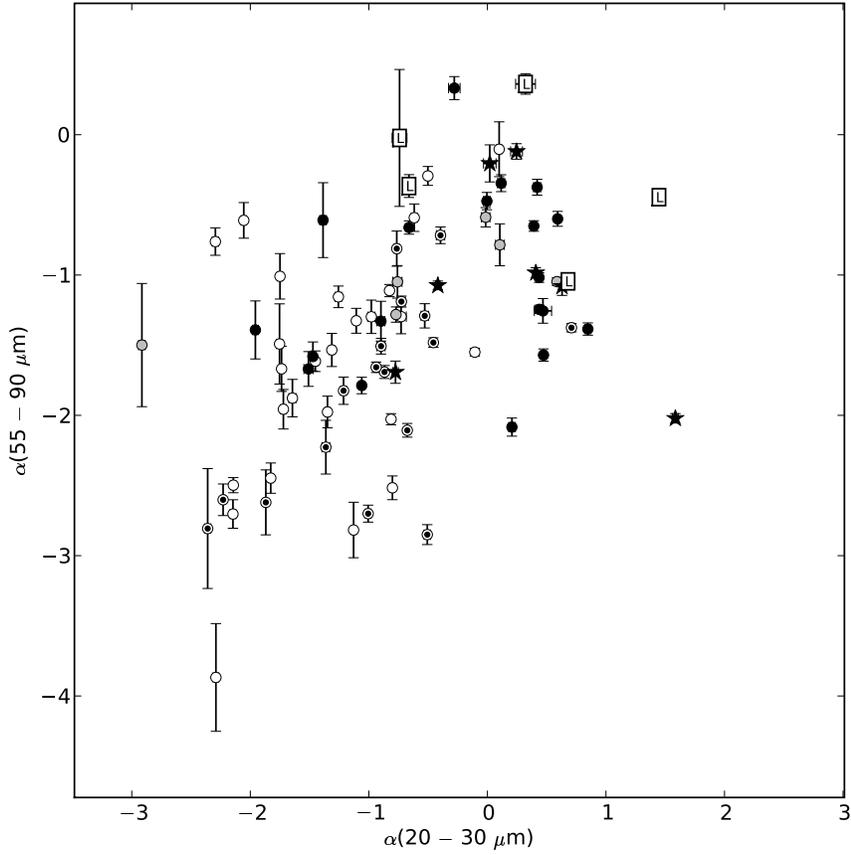}
\caption{MIPS SED slope vs. IRS 20-30 \m\ spectral index.
Sources with blue or flat MIPS SEDs tend to have flat 20-30 SEDs. Sources with red 
MIPS SEDs tend to have red 20-30 SEDs with a broad distribution.
 Open circles are S1, S1.5, and S1n, gray-filled
  circles are S1.8 and S1.9, filled circles are Sy 2,
  partially filled circles are ``hidden Seyferts'' (S1h and S1i), squares labeled
  L are LINERs, and filled stars are Starburst/HII galaxies. 
}\label{figIRS_SLOPE_MIPS}
\end{figure}

In Figure~\ref{figIRS_SLOPE_MIPS} we plot the 55-90 \m\ spectral index
vs. 20-30 \m\ spectral index.  We see a correlation in the sense that 
objects which are redder in 20-30 tend
to be redder in 55-90. This is consistent with the hypothesis that
star formation contributes to both the 20-30 and 55-90 spectral
regions. 
Thus, the 55-90 vs. 20-30 spectral index plot is also a diagnostic
of the relative contribution of Starburst to AGN. 
We see there is a large range in spectral index across the sample:  
$\Delta \alpha \sim 3$ for both indexes. 
Thus, the contribution to the IR spectrum from the AGN and starburst components
can be comparable in magnitude but the relative contribution also varies widely 
across the sample (see also \citealt{buchanan06,deo07,wu09}).

\subsubsection{The Relationship between Molecular Hydrogen and Star Formation} 

\begin{figure}
\plotone{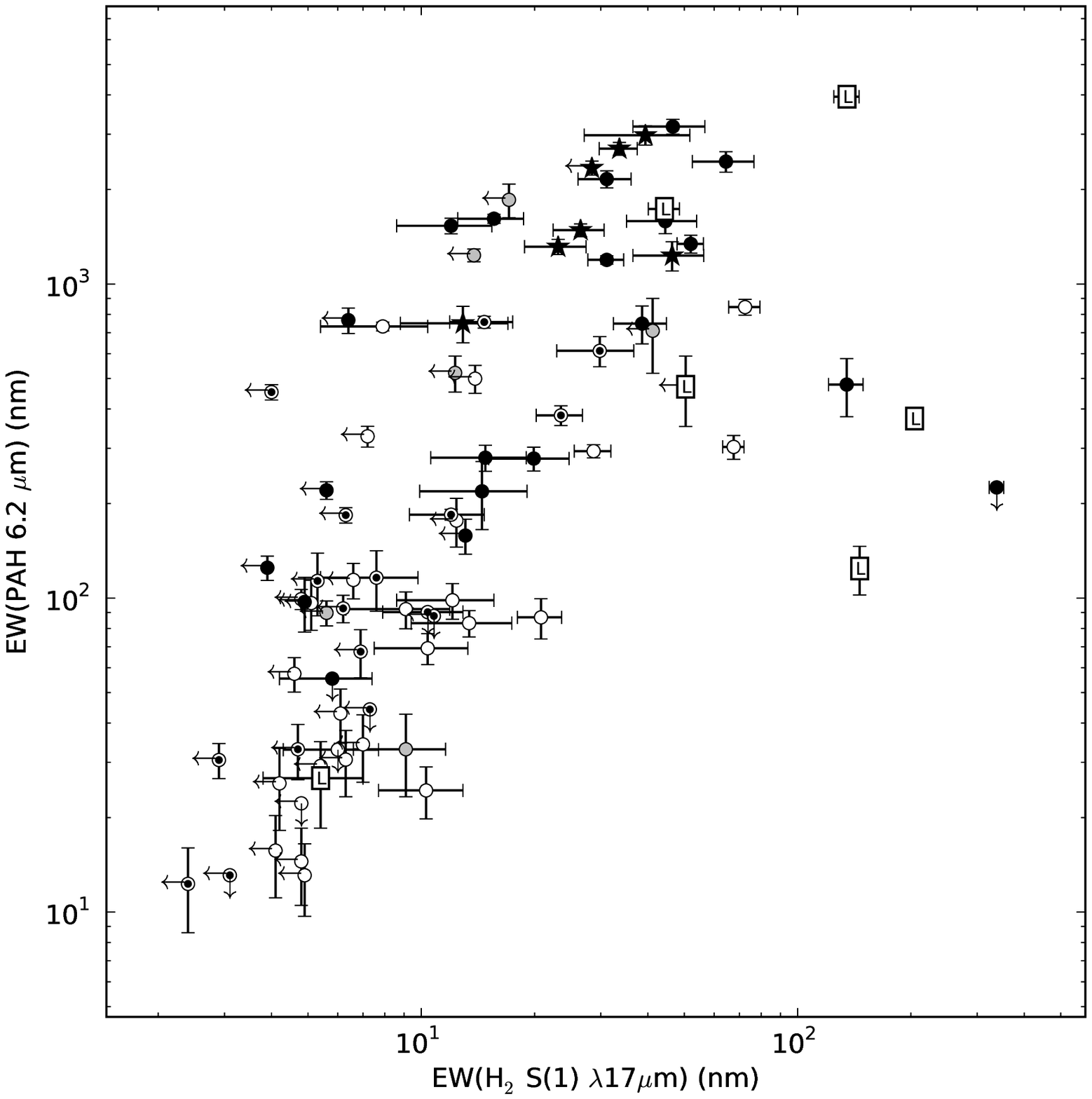}
\caption{6.2 \m\ PAH EW vs. H$_2$ S(1) 17\m\ EW.
 Open circles are S1, S1.5, and S1n, gray-filled
  circles are S1.8 and S1.9, filled circles are Sy 2,
  partially filled circles are ``hidden Seyferts'' (S1h and S1i), squares labeled
  L are LINERs, and filled stars are Starburst/HII galaxies. Limits
  are indicated by appropriately directed arrows.
}\label{figIRS_H2_PAH}
\end{figure}

Figure \ref{figIRS_H2_PAH} shows the 6.2 \m\ PAH EW vs. the \molhyd\ 
17 \m\ EW.  The \molhyd\ line has higher EW in the Starbursts and Liners than the
Seyferts.
The distribution of the  fluxes and luminosities of 
the \molhyd\    17 \m\ line are consistent with being the same in the  
subclasses of Seyferts (Tables \ref{tab:ks1}-\ref{tab:ks7}).  

We see that the \molhyd\  EW is proportional to the PAH
EW.   \citet{tommasin08} show a correlation between the
luminosities of the 11.3 \m\ PAH and \molhyd\  17 \m\ lines in a
subsample of these sources.  We also find a rank correlation between 
the luminosities of the 6.2 \m\ PAH and the \molhyd\  17 \m\ line 
(a rank correlation is also found in the fluxes; Table \ref{tab:rank}).  
We note that \citet{schweitzer06}
have also reported a correlation between PAH and H$_2$ in the higher
luminosity QUEST QSOs.  Since the PAH is a tracer of star formation
this suggests that the H$_2$ line is also associated with star
formation.

\begin{figure}
\plotone{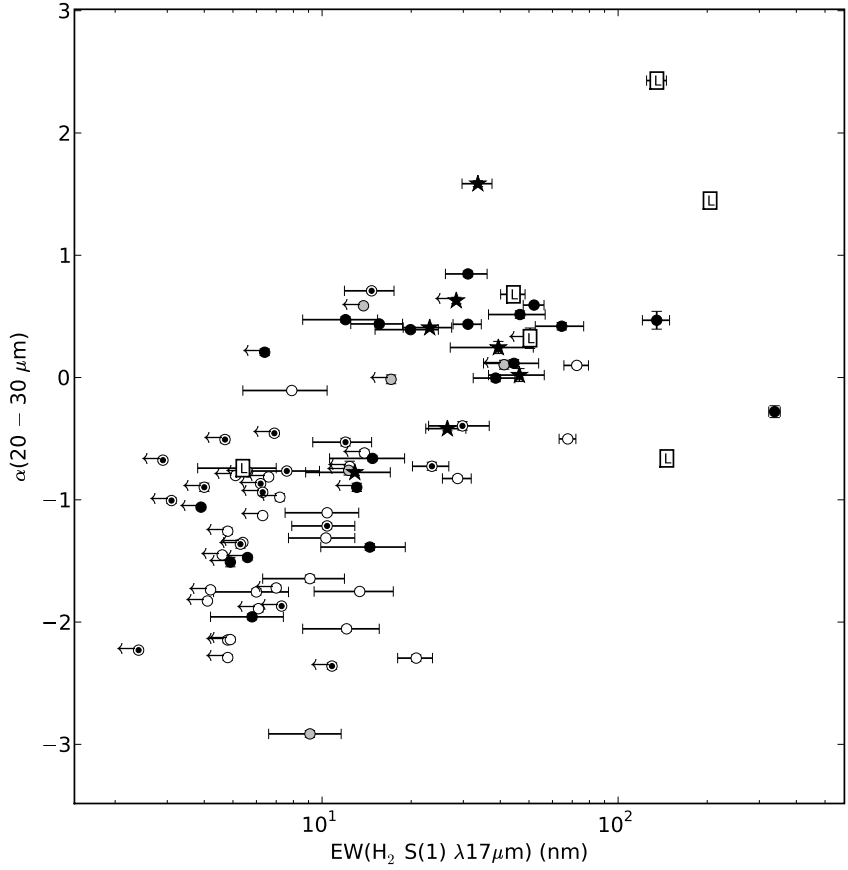}
\caption{IRS 20-30 \m\ spectral index vs. H$_2$ S(1) 17\m\ EW.
We see a trend for sources with redder $\alpha_{20-30}$ to have larger molecular line EW.
This is consistent with molecular hydrogen emission and a red SED both being signatures 
of star formation.
 Open circles are S1, S1.5, and S1n, gray-filled
  circles are S1.8 and S1.9, filled circles are Sy 2,
  partially filled circles are ``hidden Seyferts'' (S1h and S1i), squares labeled
  L are LINERs, and filled stars are Starburst/HII galaxies. Limits
  are indicated by appropriately directed arrows.
}\label{figIRS_H2_SLOPE}
\end{figure}

In Figure~\ref{figIRS_H2_SLOPE} we plot $\alpha_{20-30}$ vs. the \molhyd\
17 \m\ EW.  We see a correlation in the sense that objects with
redder continuum slopes have larger  \molhyd\  EW. 
Since red continua are associated with cold dust from stars, this correlation is
also consistent with the molecular hydrogen line being associated with
star formation.

\subsubsection{AGN Tracers}\label{agn}

Following \citet{voit92b} and \citet{spinoglio92}, we adopt the
high ionization lines [NeV] (97.1 eV) and [OIV] (54.9 eV) as tracers
of the AGN in these objects.  The vast majority of stars do not
produce photons sufficiently energetic to excite [NeV] and [OIV], with
the exception of young Wolf-Rayet stars \citep{schaerer99}.
However, \citet{abel08} show that even in AGN with a
strong starburst component, the observed ionization parameters
constrain the [NeV] emission to be dominated by the AGN.

It is also possible for [NeV] and [OIV] to be excited by shocks with
velocities $> 200$ \kms\ \citep{allen08}. However, as Allen
et al. note, the predicted line ratios are not in agreement with
observations. Detailed studies of the IR spectrum of Seyferts are
instead consistent with AGN photoionization of the NLR \citep{groves06}.
In addition, the line ratios [NeV]/[NeII] and [OIV]/[NeII] have proved 
empirically to provide good separation between AGN and Starbursts 
\citep{genzel98, sturm02, sturm06}. And the [OIV]26\m\ line has also shown
to be correlated with other AGN tracers - the [OIII]$\lambda$5007 line,
radio emission, and the 2-10 keV X-ray flux in other samples 
\citep{melendez08, stanic09}.  
 
\begin{figure} 
\plotone{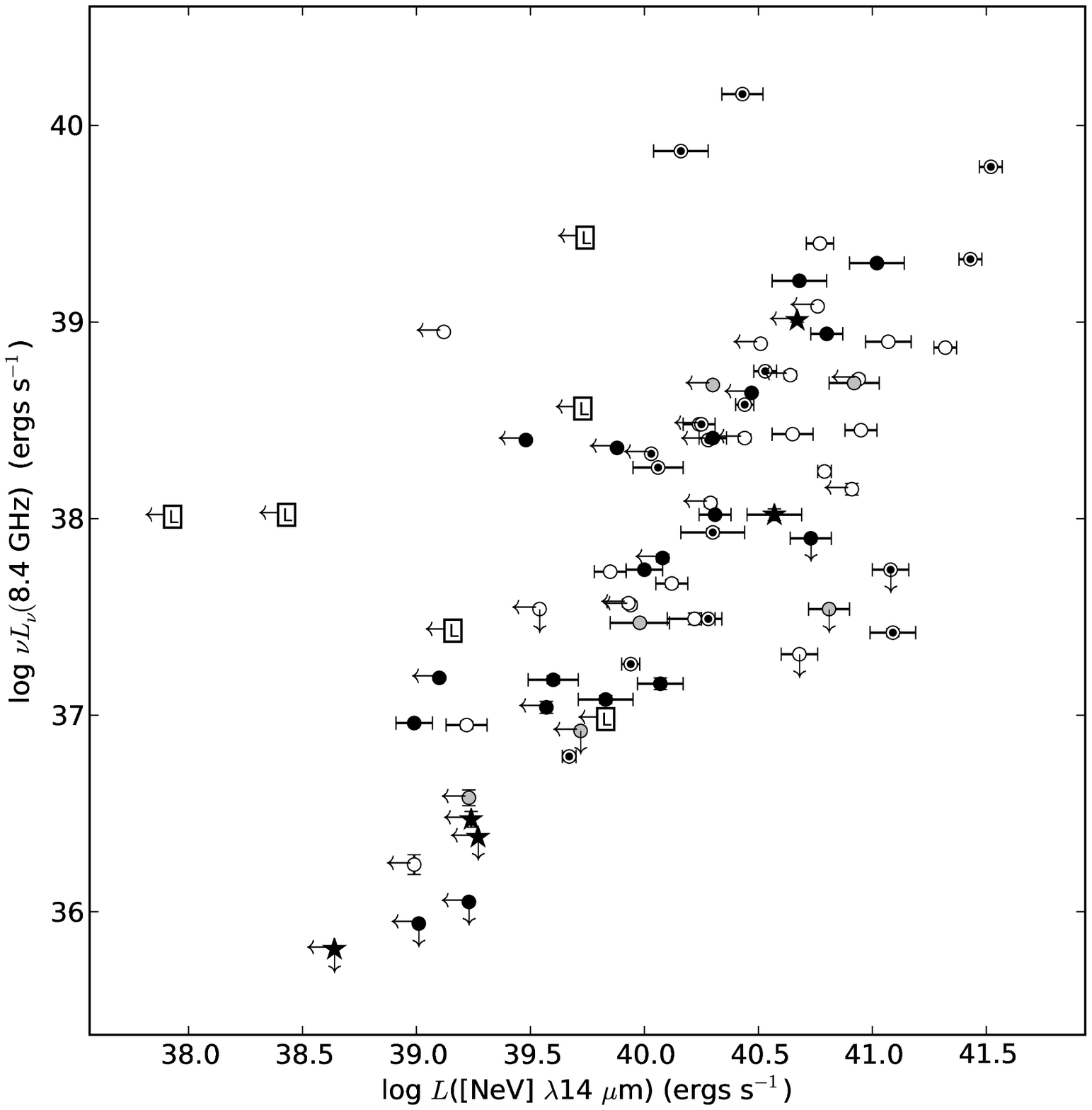}
\caption{Luminosity of the [NeV] 14\m\ line vs. the luminosity of the 
8 GHz radio continuum. 
 Open circles are S1, S1.5, and S1n, gray-filled
  circles are S1.8 and S1.9, filled circles are Sy 2,
  partially filled circles are ``hidden Seyferts'' (S1h and S1i), squares labeled
  L are LINERs, and filled stars are Starburst/HII galaxies. Limits
  are indicated by appropriately directed arrows.
 }\label{IRS_NEVl_8GHZ_L} \end{figure}

In Figure \ref{IRS_NEVl_8GHZ_L} we show a rank correlation between the
luminosity of the [NeV] 14\m\ line and the 8.4 GHz radio
continuum. A rank correlation is also present between the {\it fluxes\/} of the 
[NeV] 14\m\ line and the 8.4 GHz radio
continuum  (Table~\ref{tab:rank}).
The 8.4 GHz radio emission is not orientation dependent
\citep{thean01} and is an unbiased measure of the isotropic AGN
luminosity. The rank correlation of the [NeV] 14 \m\ line with the 8 GHz
radio emission suggests that the [NeV] line emission is also 
a useful measure of AGN luminosity (see also \S~\ref{sec:diag}).

In models of the obscuring torus, the flux in the near-mid IR (i.e., 
$\sim 3 - 10$\m) is due to hot dust heated by the AGN 
\citep{pier92,granato94,efstathiou95,vanbemmel03,nenkova02,nenkova08}.
Thus, we adopt the 6.7 \m\ continuum flux as an AGN tracer. 
In \S~\ref{sec:extdust} we show a rank correlation between the 6.7 \m\
continuum and the [NeV] 14 \m\ line emission.

\begin{figure}
\plottwo{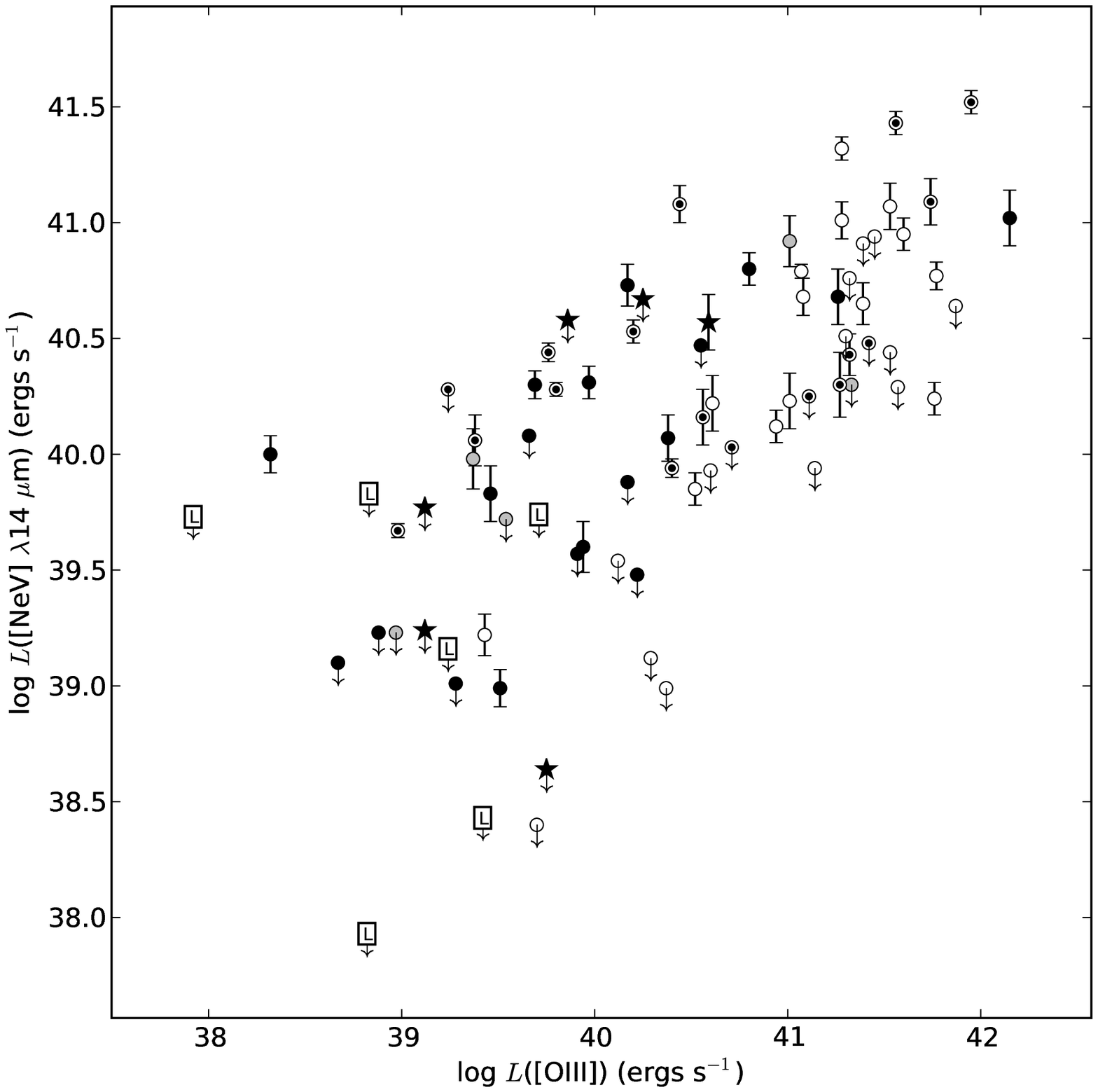}{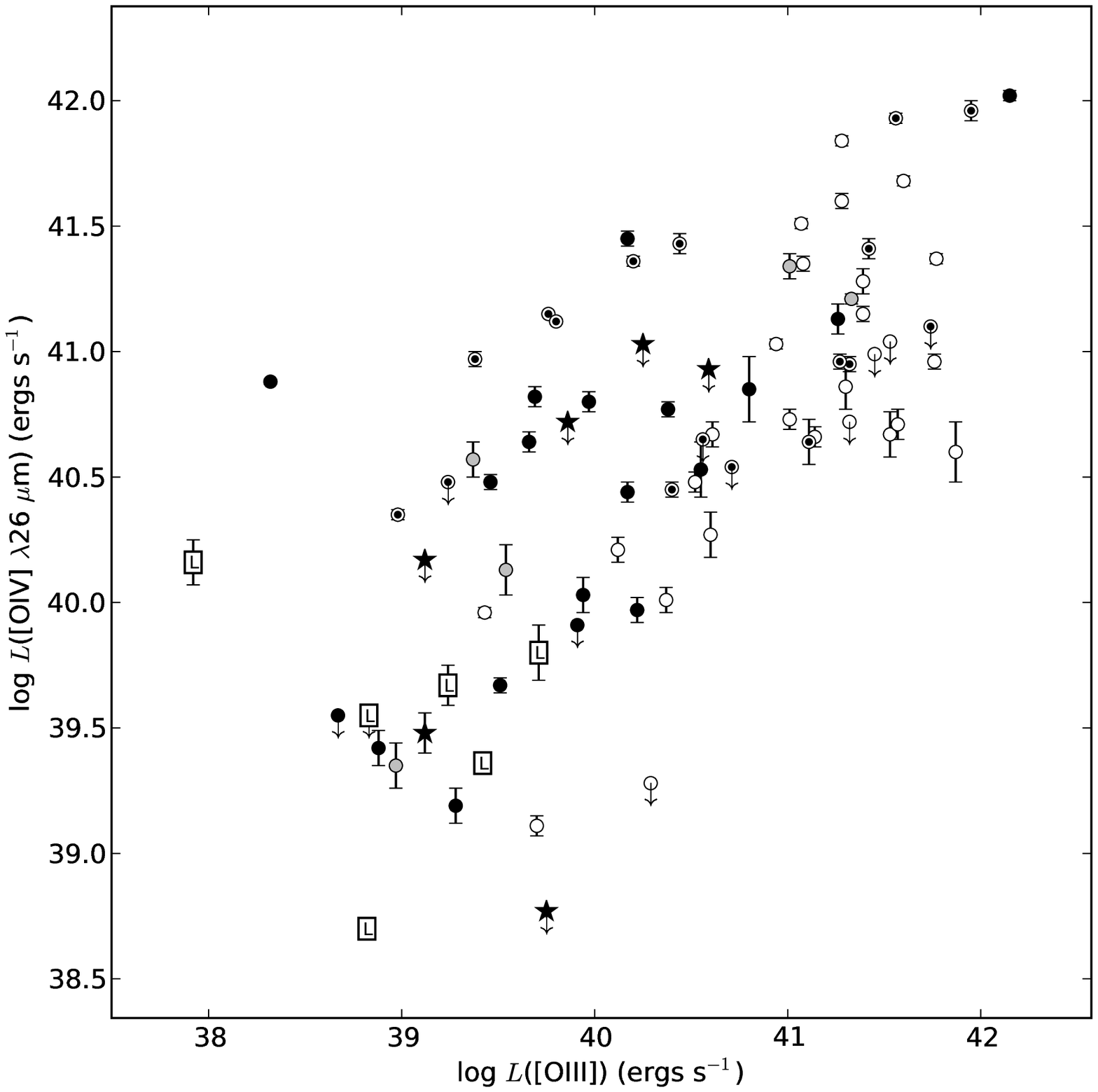}
\caption{(Left). Luminosity of the [NeV] 14\m\ line vs. the luminosity of the 
[OIII]$\lambda$5007 line.
(Right).  Luminosity of the [OIV] 26\m\ line vs. the luminosity of the
[OIII]$\lambda$5007 line.
 Open circles are S1, S1.5, and S1n, gray-filled
  circles are S1.8 and S1.9, filled circles are Sy 2,
  partially filled circles are ``hidden Seyferts'' (S1h and S1i), squares labeled
  L are LINERs, and filled stars are Starburst/HII galaxies. Limits
  are indicated by appropriately directed arrows.
 }\label{IRS_NEVl_OIII_L} \end{figure}

In Figure \ref{IRS_NEVl_OIII_L} we see  rank correlations between the luminosities
of the two IR AGN tracers the [NeV] 14\m\ and [OIV] 26\m\ lines with the
optical [OIII]$\lambda$5007 line ([OIII] measurements taken from \citep{axon09,deo07,
degrijp92,gu06,kewley01,kim95,kirhakos90,melendez08,panessa02,smith02,tran03,veron01,
winkler92,whittle92,young96}).  The rank correlations are also present but  weaker between
the fluxes of the lines. This confirms that the [NeV] 14\m\ and [OIV] 26\m\ 
are useful AGN tracers.  

\subsubsection{Comparison of Starburst and AGN Tracers}

\begin{figure}
\plotone{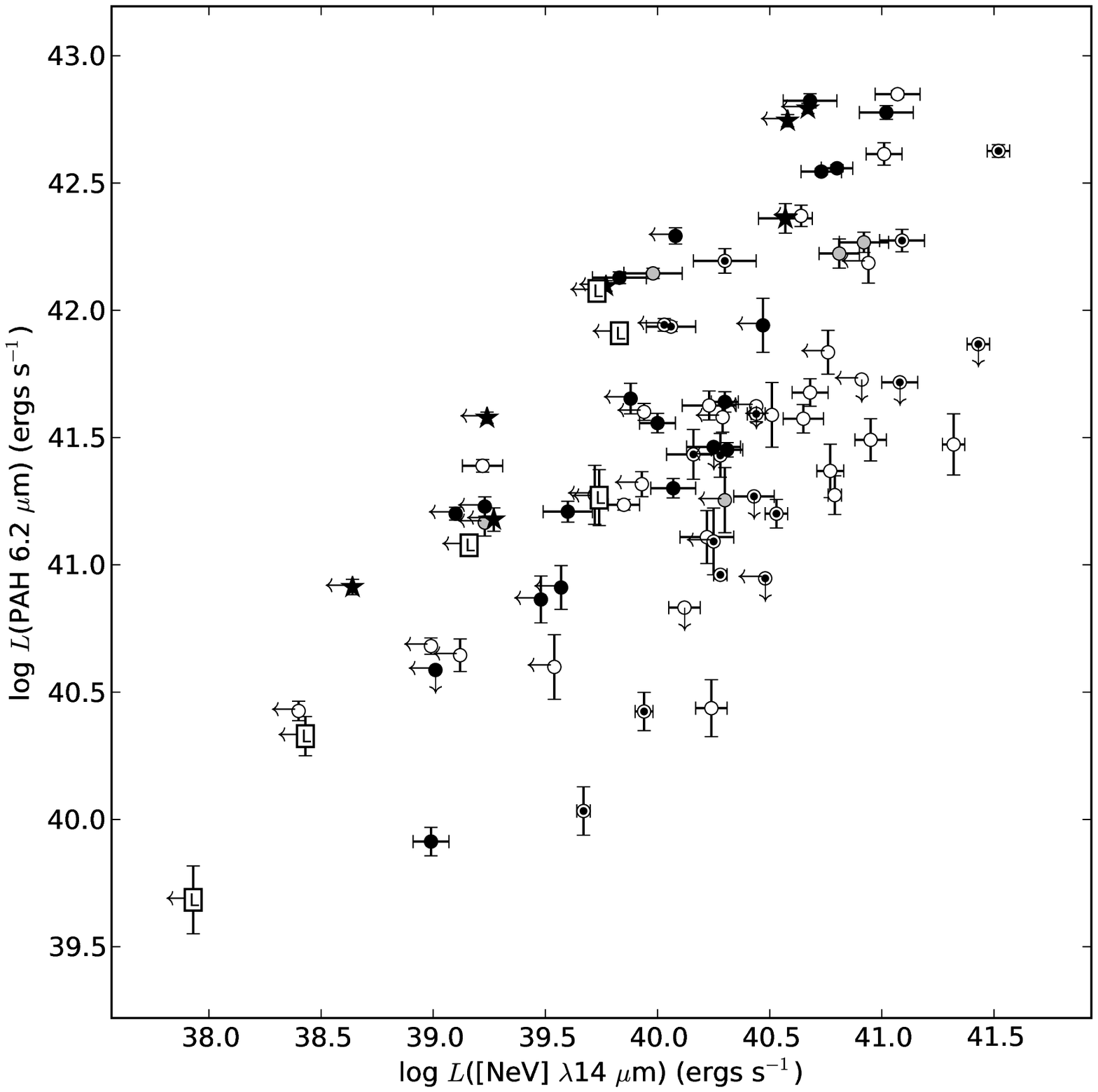} 
\caption{6.2 \m\ PAH luminosity  vs. [NeV] 14 \m\ luminosity.
 Open circles are S1, S1.5, and S1n, gray-filled
  circles are S1.8 and S1.9, filled circles are Sy 2,
  partially filled circles are ``hidden Seyferts'' (S1h and S1i), squares labeled
  L are LINERs, and filled stars are Starburst/HII galaxies. Limits
  are indicated by appropriately directed arrows.
}\label{figIRS_NEV_PAH_LUM}
\end{figure}

\begin{figure}
\plotone{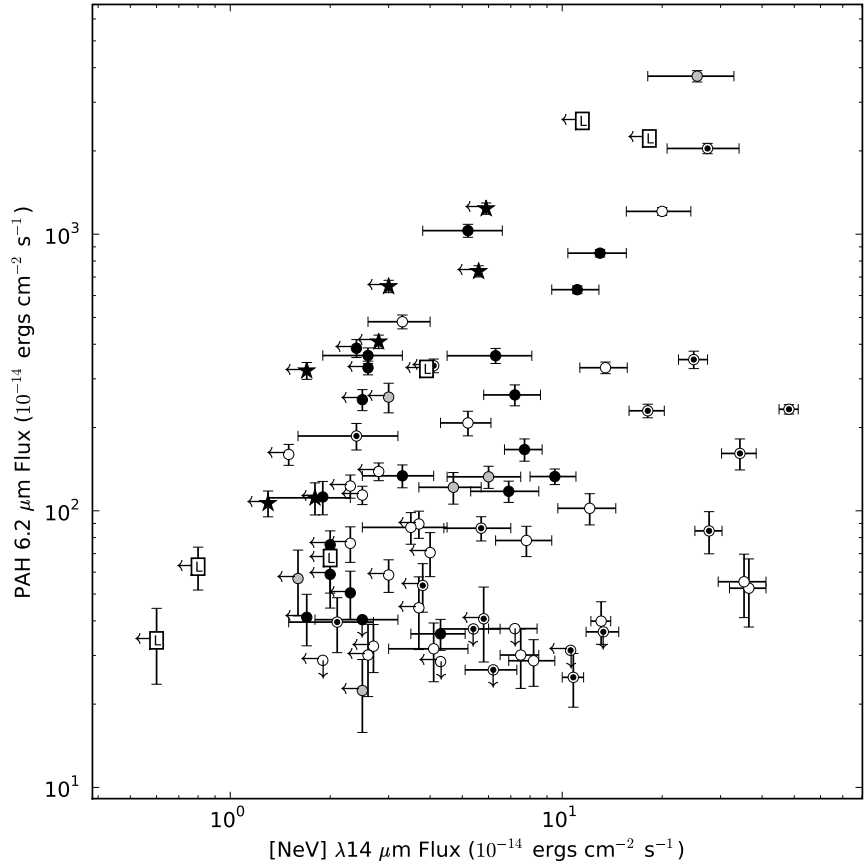}
\caption{6.2 \m\ PAH feature flux vs. [NeV] 14 \m\ line flux.  
This is an AGN vs Starburst relation.
 Open circles are S1, S1.5, and S1n, gray-filled
  circles are S1.8 and S1.9, filled circles are Sy 2,
  partially filled circles are ``hidden Seyferts'' (S1h and S1i), squares labeled
  L are LINERs, and filled stars are Starburst/HII galaxies. Limits
  are indicated by appropriately directed arrows.
}\label{figIRS_NEV_PAH_F}
\end{figure}

Correlations between AGN and starburst tracers have been 
found for QSOs \citep{haas03, schweitzer06, netzer07, maiolino07}  
and Seyferts \citep{maiolino07}.  Figure \ref{figIRS_NEV_PAH_LUM} 
shows a rank correlation between the luminosities of
the 6.2 \m\ PAH and the [NeV] 14 \m\ line in the combined sample. 
We also find  rank correlations between the luminosities of (1) the 6.2 \m\ PAH 
and the [OIV] 26 \m\ line 
(2) the [NeV] 14 \m\ line and the 60 \m\ continuum, and
(3) the 6.2 \m\ PAH feature and the 6.7\m\ continuum (not plotted, 
see Table~\ref{tab:rank}).
These correlations seem to indicate that the starburst
and AGN luminosities are correlated in these Seyferts. 
However, our statistical tests (Table \ref{tab:rank}) show that there are no
significant correlations  in the {\it flux\/} of these AGN and Starburst tracers.
Figure \ref{figIRS_NEV_PAH_F} shows that the fluxes of the 6.2 \m\ PAH and
[NeV] 14 \m\ line scatter over an order of magnitude and are uncorrelated.
So although AGN tracers are internally correlated in both flux and luminosity, 
and Starburst tracers are are internally correlated in both flux and luminosity,
the fluxes of the AGN and Starburst tracers are not correlated. 
We conclude that the correlations in Starburst and AGN
luminosity are due to the dependence on distance and are not real. 
This is consistent with our results that the relative strength of the AGN 
and Starburst components in these Seyferts vary widely across the 
sample (\S~\ref{sec:relative}). 
The lack of a correlation between the
AGN and star burst suggests that in these AGN there is no direct link between
AGN fueling, black hole growth, and star formation.  

\subsubsection{The [NeV]/[NeII] ratio vs. PAH EW Diagram}

\begin{figure}
\plotone{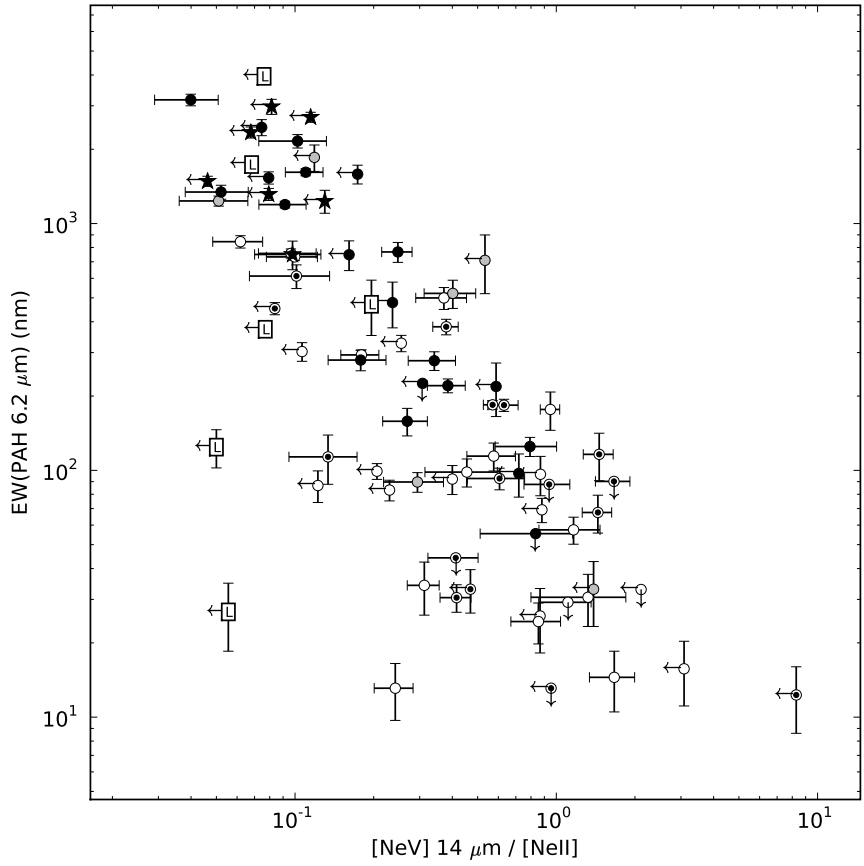}
\caption{The ratio [NeV]14\m/[NeII]12\m\ vs. 6.2\m\ PAH EW. 
Open circles are S1, S1.5, and S1n, gray-filled
  circles are S1.8 and S1.9, filled circles are Sy 2,
  partially filled circles are ``hidden Seyferts'' (S1h and S1i), squares labeled
  L are LINERs, and filled stars are Starburst/HII galaxies.
}\label{fig:neratio_v_pah}
\end{figure}

\citet{genzel98} showed that a diagram which plots the ratio of a high
to low ionization line vs. a PAH EW is able to separate
Starbursts from AGN from ULIRGs. \citet{genzel98} used [OIV]/NeII] vs.
the 7.7\m\ PAH EW. In Figure~\ref{fig:neratio_v_pah} we plot
6.2\m\ PAH EW vs. [NeV]14\m/[NeII]12\m.  We see that there are
significant differences across our sample with objects with higher
AGN to starburst ratio in the lower right and the objects with lower
AGN to starburst ratio in the upper left. Note that the Liners have
low [NeV]14\m/[NeII]12\m\ and tend to fall to the left of the 
other objects.

\subsection{Orientation, Obscuration, and Constraints on the Torus}


\subsubsection{ 6.2 \m\ PAH vs. Sil 10 \m\ Strength}\label{sec:sil}

The strength of the Sil 10 \m\ feature is potentially a powerful probe of the absorbing
material \citep{pier92, laor93, levenson07, imanishi07}. 
We adopt the definition of Sil strength 
given by \citet{spoon07},
\begin{equation}
S_{\rm Sil} = ln { f_{\rm obs} (10)  \over f_{\rm cont} (10) }
\end{equation}
where $f_{\rm obs}$ is the observed flux in the feature and $f_{\rm cont}$ is the
interpolated continuum flux at 10 \m.
Recent Spitzer results suggest that  different types of objects show systematically
different  Sil 10 \m\ strength \citep{shi06, hao07} with QSOs showing
strong Sil emission, Sy 1s showing mostly weak Sil emission,  Sy 2s showing mostly weak to
moderate Sil absorption, and ULIRGS showing strong Sil absorption.
\citet{spoon07} have shown that objects fall mainly in two branches on the
6.2 \m\ PAH vs. Sil 10 \m\ Strength diagram.

\begin{figure} 
\plotone{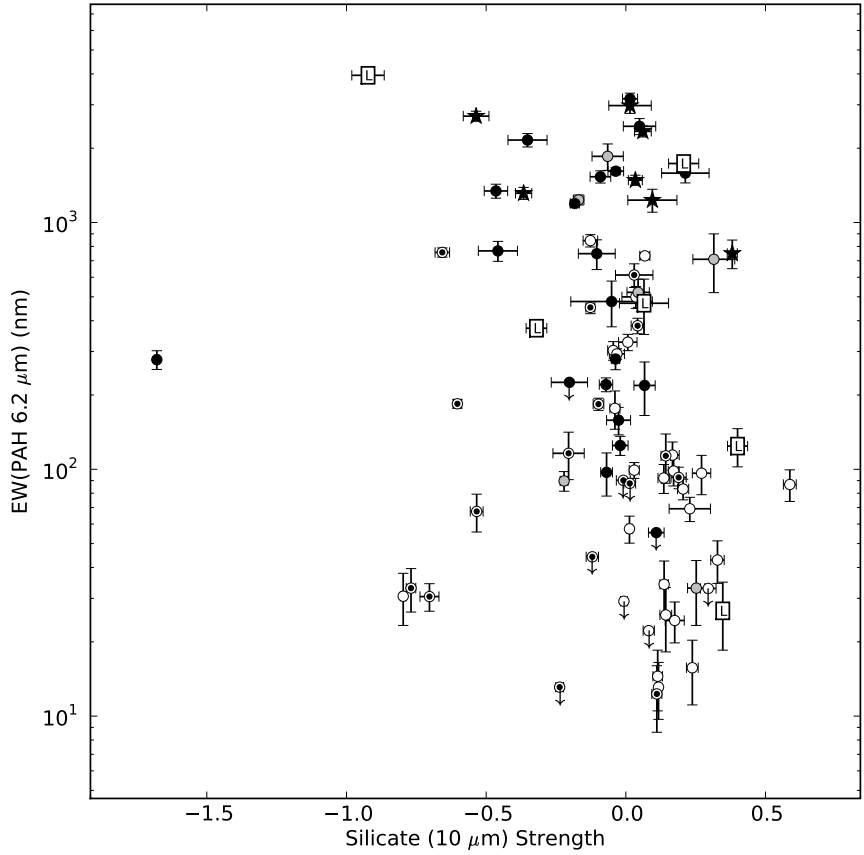}
\caption{6.2 \m\ PAH EQW vs. Sil 10 \m\ strength. 
For the Sil 10 strength, positive means emission. 
 Open circles are S1, S1.5, and S1n, gray-filled
  circles are S1.8 and S1.9, filled circles are Sy 2,
  partially filled circles are ``hidden Seyferts'' (S1h and S1i), squares labeled
  L are LINERs, and filled stars are Starburst/HII galaxies. Limits
  are indicated by appropriately directed arrows.
}\label{figIRS_S_PAH}     \end{figure}

Figure \ref{figIRS_S_PAH} shows	 the distribution on the 6.2 \m\ PAH vs. Sil 10
\m\ Strength diagram.  Our results are consistent with those of \citet{spoon07}.
As expected, the \syones\ tend to show weak Sil emission,
and the \sytwos\ tend to show Sil absorption. The KS test (Table~\ref{tab:ks1})
confirms that the distribution of Sil strength is very different in the
\syones\ and \sytwos. These results are in qualitative agreement with \citet{wu09},
though as discussed by \citet{gallimore09a} there are systematic differences
in the measured Sil strengths. 
The only significant difference between the \syones\ and
the \sytwos\ with an HBLR is the Sil strength (Table~\ref{tab:ks4}).

Two of the HBLR \sytwos\ show Sil emission - F01475-0740 and M-3-58-7. 
In these objects, the obscuring material (torus) may be oriented at an 
intermediate inclination. 
In addition, four objects classified as S1.8-1.9 show Sil absorption  - 
NGC1194 (S1.9),  M-2-40-4 (Sy1.9), NGC1365 (S1.8), and NGC5033 (S1.8).
These are also likely to be fairly edge-on (\S~\ref{sec:comp}).
About half of the objects identified as Liners and Starbursts show weak Sil 
features, with strong absorption seen in MRK938 (HII), NGC3079 (LINER), 
NGC5005 (LINER), and strong emission seen in NGC4579 (LINER), NGC5494 (LINER),
and CGCG381-051 (HII). 

\subsubsection{Sil 10 \m\ Strength vs. Sil 18 \m\ Strength}

\begin{figure} 
\plotone{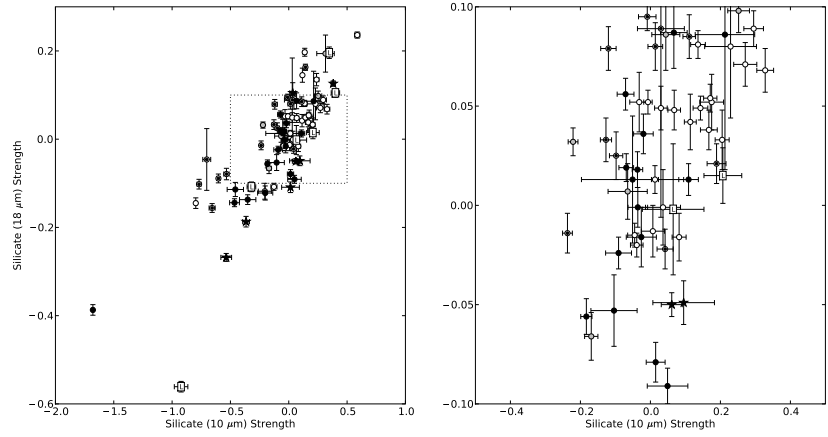}
\caption{Sil 18 \m\ Strength vs. Sil 10 \m\ strength.  For the Sil
  strength, positive means emission.  (Left) All the points. Blow up
  of inset region is on the right.  (Right) Blow up of region near Sil
  Strength of zero.  Open circles are S1, S1.5, and S1n, gray-filled
  circles are S1.8 and S1.9, filled circles are Sy 2, partially filled
  circles are ``hidden Seyferts'' (S1h and S1i), squares labeled L are
  LINERs, and filled stars are Starburst/HII galaxies. Limits are
  indicated by appropriately directed arrows.  }\label{figsilsilplot3}
\end{figure}

\begin{figure} 
\plotone{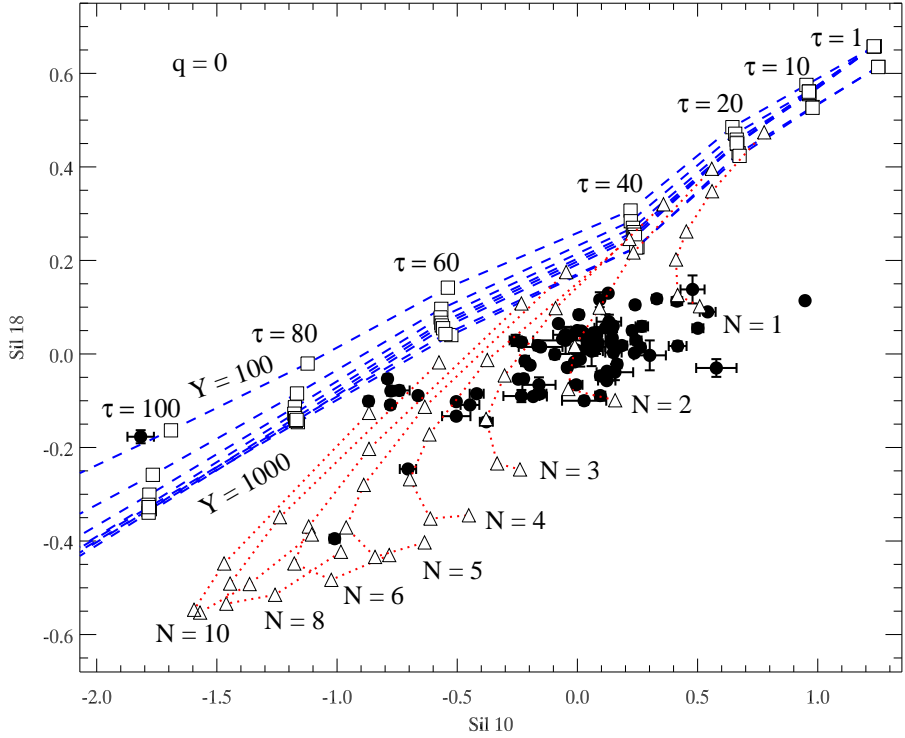}
\caption{Sil 18 \m\ Strength vs. Sil 10 \m\ strength.  For the Sil
  strength, positive means emission.  Theoretical curves for spherical
  obscuration models \citep{sirocky08} are shown.  Dotted lines are
  clumpy models and dashed lines are smooth models. Model parameters
  are annotated: $q$ is the power law index of the radial density
  distribution (here, flat); $Y$ is the ratio of the outer radius to
  the inner radius of the obscuring region (spherical models, the
  inner radius is set by dust sublimation); $\tau$ refers to
  $\tau_{\rm V}$, the total opacity through the spherical shell
  (spherical models); and $N$ refers to the average number of clumps
  along the sight-line (clumpy models). Spherical models range over $Y
  = $ 100, 200, 300, 400, 500, 600, 800, \& 1000; $Y=30$ for the
  clumpy models. The clumpy models are further
  distinguished by the (unlabeled) opacity through individual clumps: $\tau_{\rm
    V} = $ 10, 20, 30, 40, 60, \& 80, with lower opacities oriented to
  the upper right on this diagram. }\label{figsilsilplot1} \end{figure}

\citet{sirocky08} showed that the observed relationship between
the two Sil features 18 \m\ and 10 \m\ can be a powerful probe of the
properties of the obscuring material.  Figure~\ref{figsilsilplot3} shows
the distribution of the Sil strengths for the different subsamples of
galaxies. We see the \syones\ mainly to the upper right with emission
in both Sil features, and the \sytwos\ mainly to the lower left with
absorption in both Sil features.  In Figure~\ref{figsilsilplot1} we
compare our results for the two Sil Strength features with simple
spherical obscuration models \citep{sirocky08}. The spherical models
have the advantage that they reduce the number of free parameters
and still give good agreement with the more detailed clumpy torus models 
\citep{nenkova02, nenkova08}. 
We see that the data lie below the smooth torus models and within the 
parameter space of the clumpy torus models. Our data are inconsistent
with the smooth torus models but are consistent with the clumpy torus
models. Within the context of the spherical clumpy models, there
is a preference for a small number of clumps along the line of sight 
($N \sim 1-3$).

\subsubsection{Density and Obscuration Diagnostics}\label{sec:diag}

The [NeV] 14/24 and  [SIII] 18/33  line ratios are density diagnostics 
\citep{voit92b,spinoglio92}, and in principle, in the 
low density limit can be used as extinction
indicators \citep{voit92b, alexander99, dudik07}. 
The [NeV] lines will probe high excitation gas, most likely
in the narrow line region, while the [SIII] can be excited by stars and may
be produced on a larger spatial scale. 
Sturm et al (2002) examined the [NeV] 14/24 ratio in a sample of Seyferts using ISO data
and also concluded that the [NeV] emitting gas in both Seyfert types was in the 
low density limit. 
On the other hand \citet{dudik07} find that type 2 AGNs tend to have lower
ratios than type 1 AGN, and some of the type 2 AGNs are below the low density limit
possibly due to differential absorption. 
Tommasin \etal\ (2008) using high resolution IRS spectra on a subsample of the 12 \m\ sample,
report [NeV] line ratios slightly above the low density limit, giving densities of $n_e \sim  
10^3$ cm\mthree. 

\begin{figure}
\plotone{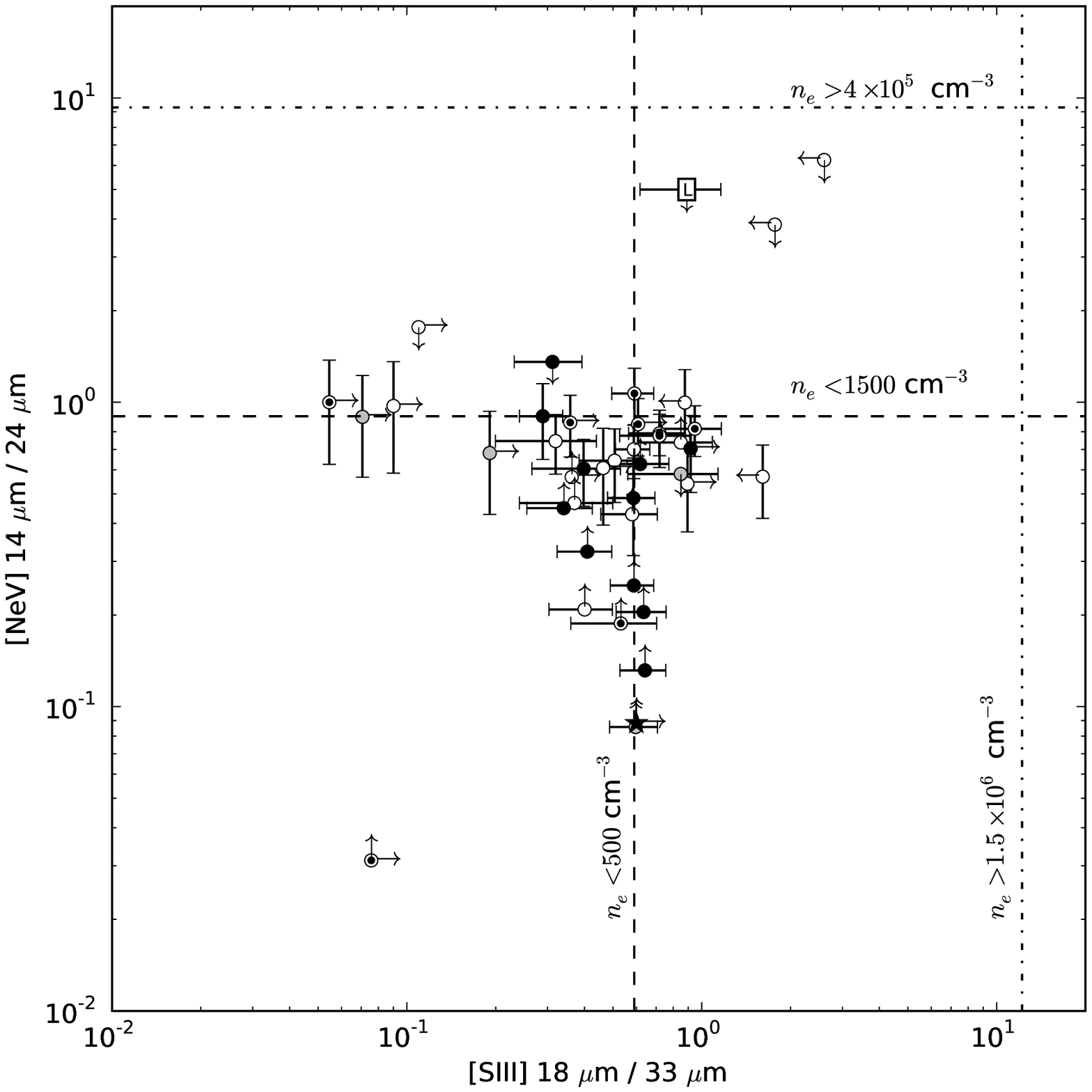} 
\caption{[NeV] 14\m/24\m\ vs. [SIII] 18\m/33\m.
 Open circles are S1, S1.5, and S1n, gray-filled
  circles are S1.8 and S1.9, filled circles are Sy 2,
  partially filled circles are ``hidden Seyferts'' (S1h and S1i), squares labeled
  L are LINERs, and filled stars are Starburst/HII galaxies. Limits
  are indicated by appropriately directed arrows.
}\label{figIRS_SIII_NEV} 
\end{figure}

Our data (Figure \ref{figIRS_SIII_NEV}) and Tables \ref{tab:median} - \ref{tab:ks4}) 
are consistent with \syones, \sytwos\ and hidden \syones\ having 
the same distribution of these diagnostic ratios. 
Both line ratios seem to scatter around their low density values 
giving for [NeV] $n_e \sim 10^3$ cm\mthree\ and for [SIII], $n_e \sim $ few hundred cm\mthree.
We also see no evidence 
for a correlation between the values of these ratios (Table \ref{tab:rank}).  

\begin{figure} 
\plotone{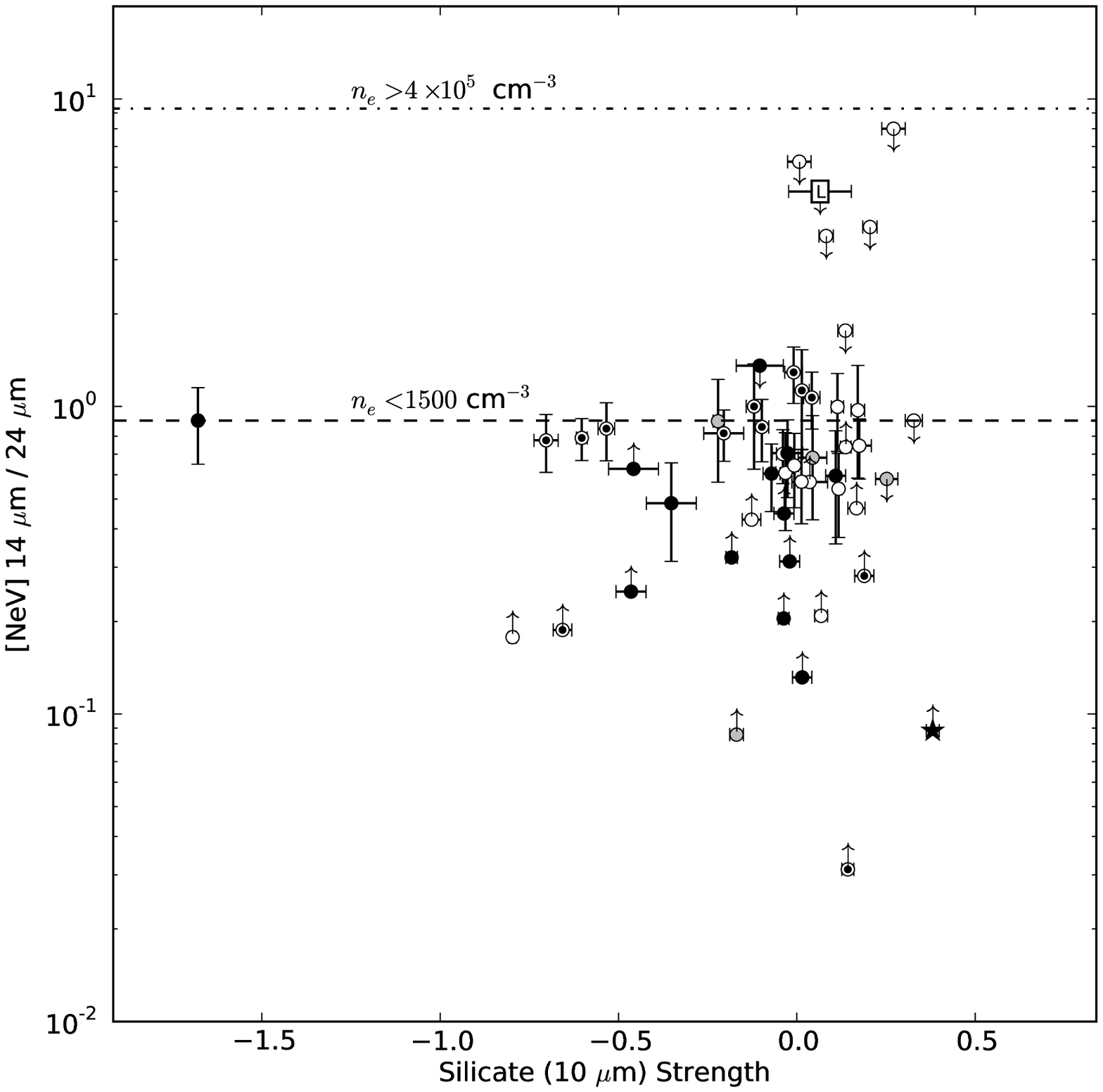}
\caption{[NeV] 14\m/24\m\ vs. Sil 10 \m\ strength.
 Open circles are S1, S1.5, and S1n, gray-filled
  circles are S1.8 and S1.9, filled circles are Sy 2,
  partially filled circles are ``hidden Seyferts'' (S1h and S1i), squares labeled
  L are LINERs, and filled stars are Starburst/HII galaxies. Limits
  are indicated by appropriately directed arrows.
}\label{figIRS_S_NEVr} \end{figure}

In Figure \ref{figIRS_S_NEVr} we plot the [NeV] 14/24 line ratio against the Sil 
10 \m\ feature which is an indicator of absorption along the line of sight 
(\S \ref{sec:sil}).  We see that there is no  correlation between these two 
quantities (Table~\ref{tab:rank}). This suggests that extinction does not  have a 
significant effect on the observed [NeV] 14/24 line ratio. 

\begin{figure} 
\plotone{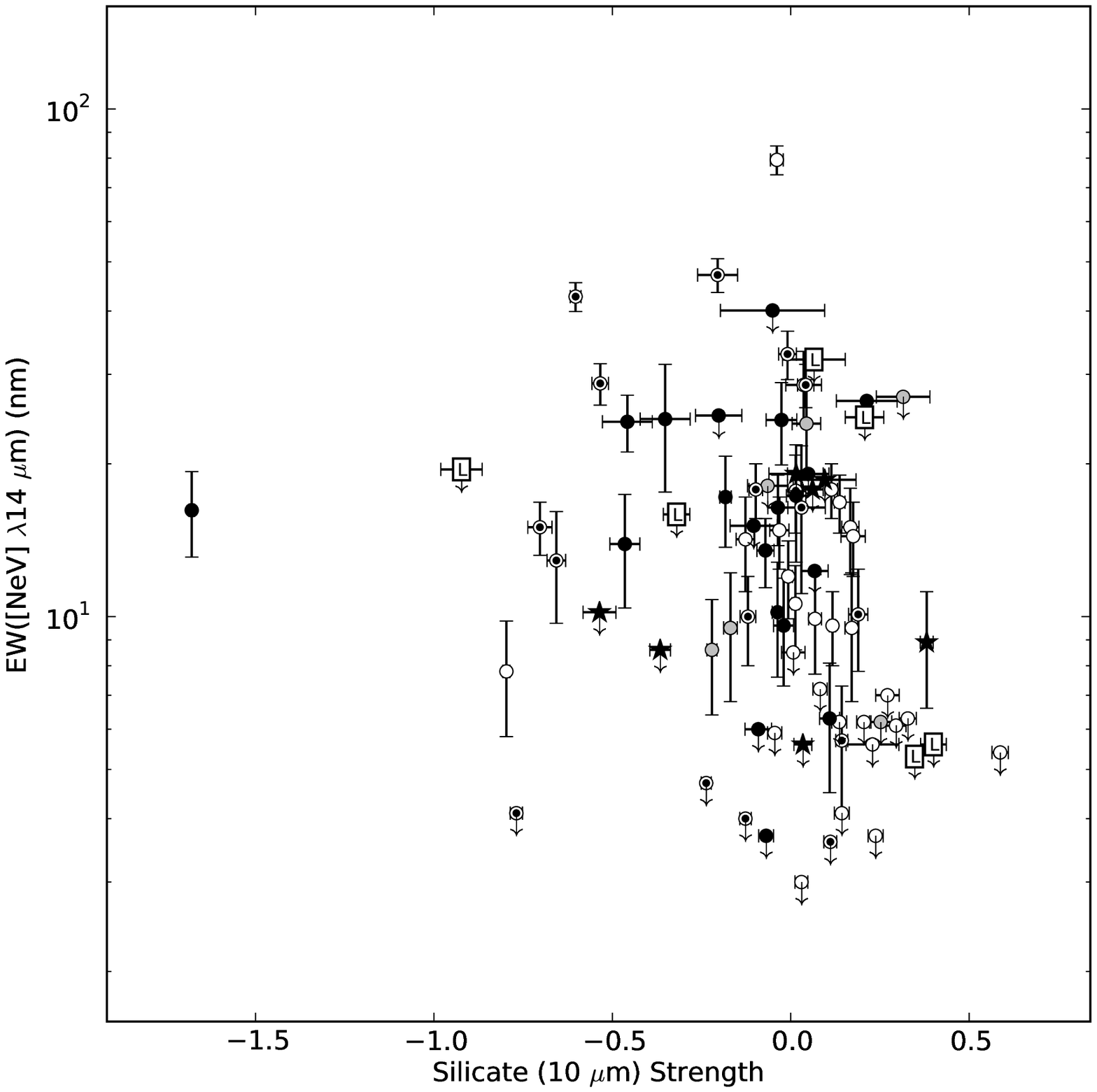}
\caption{[NeV] 14\m\ EW vs. Sil 10 \m\ strength.
 Open circles are S1, S1.5, and S1n, gray-filled
  circles are S1.8 and S1.9, filled circles are Sy 2,
  partially filled circles are ``hidden Seyferts'' (S1h and S1i), squares labeled
  L are LINERs, and filled stars are Starburst/HII galaxies. Limits
  are indicated by appropriately directed arrows.
 }\label{figIRS_S_NEVe}  \end{figure}

In Figure \ref{figIRS_S_NEVe}  
we show the [NeV] 14 \m\ EW vs. Sil 10 \m\ Strength. We see that the distribution of 
[NeV] EW is similar in \syones, \sytwos\ and hidden \syones\  (Table~\ref{tab:ks1},
\ref{tab:ks2},\ref{tab:ks3},\ref{tab:ks4}). 
There is no correlation of [NeV] EW or luminosity with Sil strength (Table~\ref{tab:rank}). 
This is consistent with our results above that (1) the [NeV] luminosity is
proportional to the radio power, and (2) the effect of extinction on the [NeV] 14/24
ratio is small. This suggests that  either the torus is not very opaque at these 
wavelengths (14-24 \m), or [NeV] is  produced primarily outside the torus, e.g., in 
the NLR. However, we expect the high ionization lines 
[NeV] and [OIII] to be produced in similar locations around the AGN. 

\begin{figure} 
\plottwo{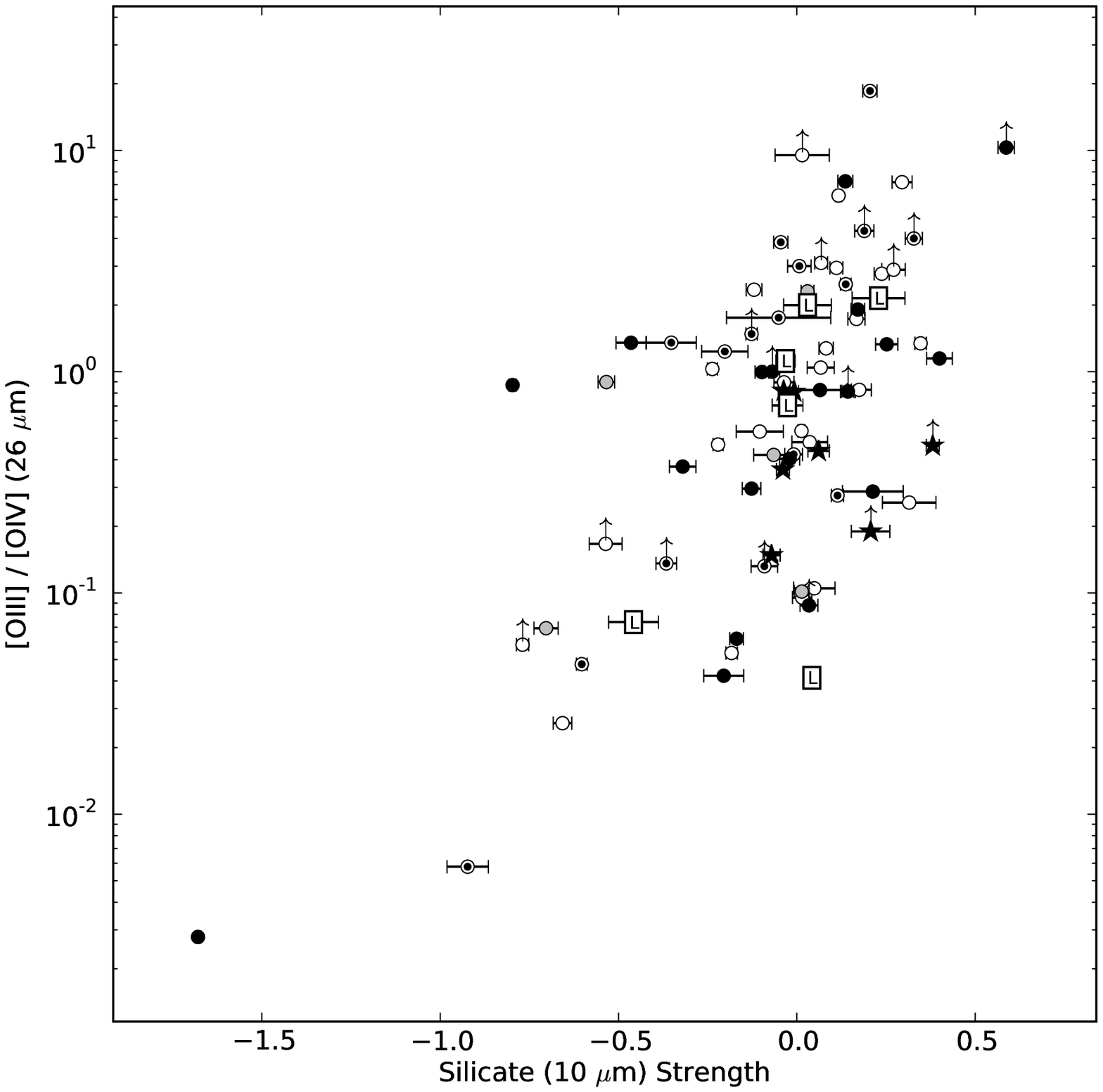}{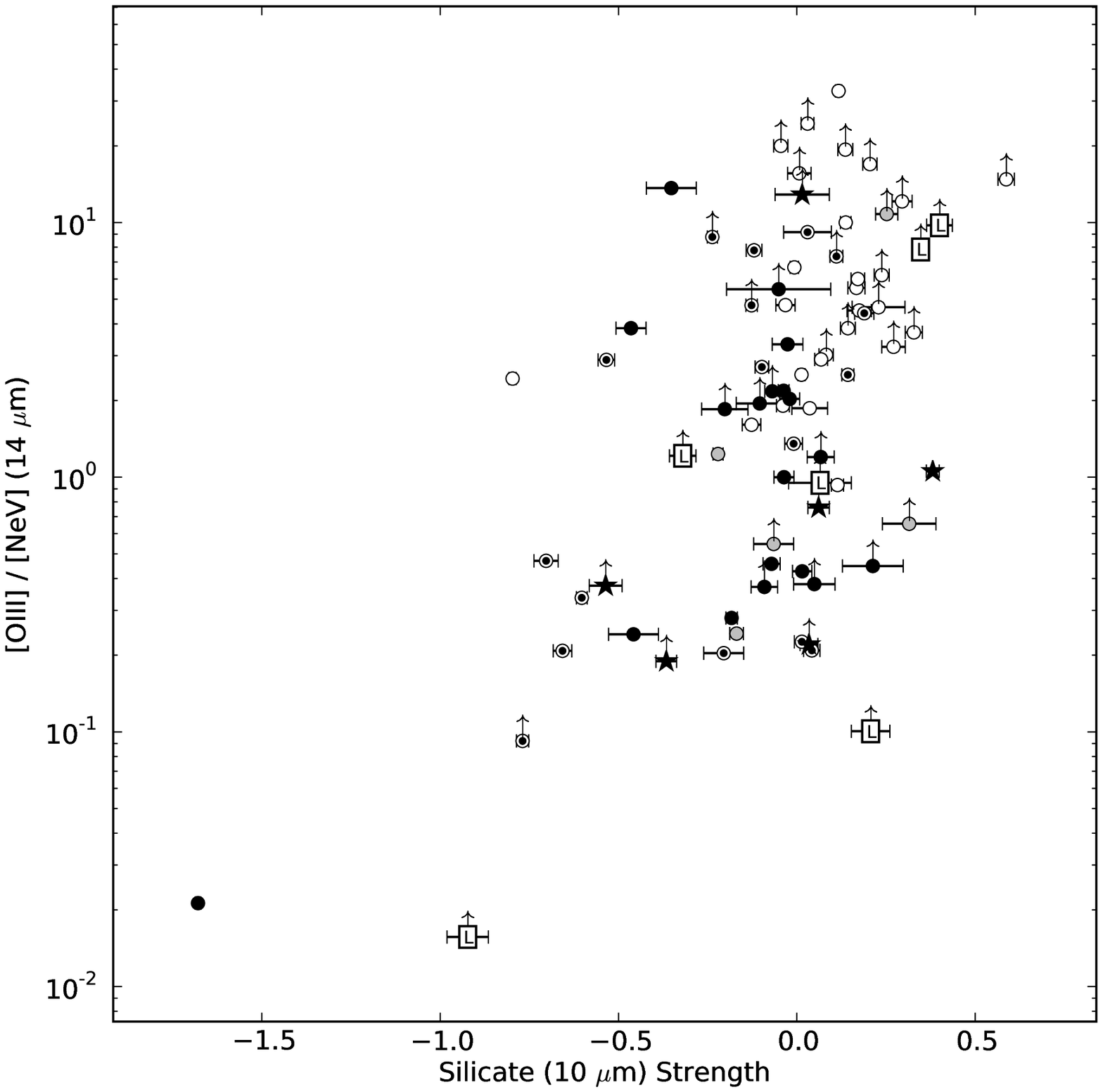}
\caption{(Left) The ratio [OIII]5007/[OIV]26\m\ vs.  Sil 10 \m\ strength.
(Right) The ratio [OIII]5007/[NeV]14\m\ vs.  Sil 10 \m\ strength.
 Open circles are S1, S1.5, and S1n, gray-filled
  circles are S1.8 and S1.9, filled circles are Sy 2,
  partially filled circles are ``hidden Seyferts'' (S1h and S1i), squares labeled
  L are LINERs, and filled stars are Starburst/HII galaxies. Limits
  are indicated by appropriately directed arrows.
 }\label{figIRS_S_ratio}	 \end{figure}

\citet{haas05}  have shown that [OIII] is
very orientation dependent in quasars and radio galaxies, suggesting that much of the
[OIII] is produced in regions obscured by the torus. In Figure \ref{figIRS_S_ratio}
we show the ratios [OIII]$\lambda$5007/[OIV]26\m\ and [OIII]$\lambda$5007/[NeV]14\m\ 
vs. the Sil Strength. We see a strong rank correlation for [OIII]$\lambda$5007/[OIV]26\m\
and a weaker one for [OIII]$\lambda$5007/[NeV]14\m\ in the sense that the 
ratio is higher in objects with Sil emission (roughly face on) and lower in 
objects with Sil absorption (roughly edge on). We find that [OIII] is orientation 
dependent in our sample of Seyferts suggesting that much of the [OIII] emission 
is produced in a region which is subject to orientation dependent obscuration, 
e.g., the torus.  Thus,  if [NeV] is also primarily produced in regions obscured 
by the torus, then our result suggests that the opacity of the torus at  
14 \m\ is not large.  This is consistent with the results of \citet{buchanan06}.


\subsubsection{An Extended Hot Dust Component in Seyferts?}\label{sec:extdust}

\begin{figure}
\plotone{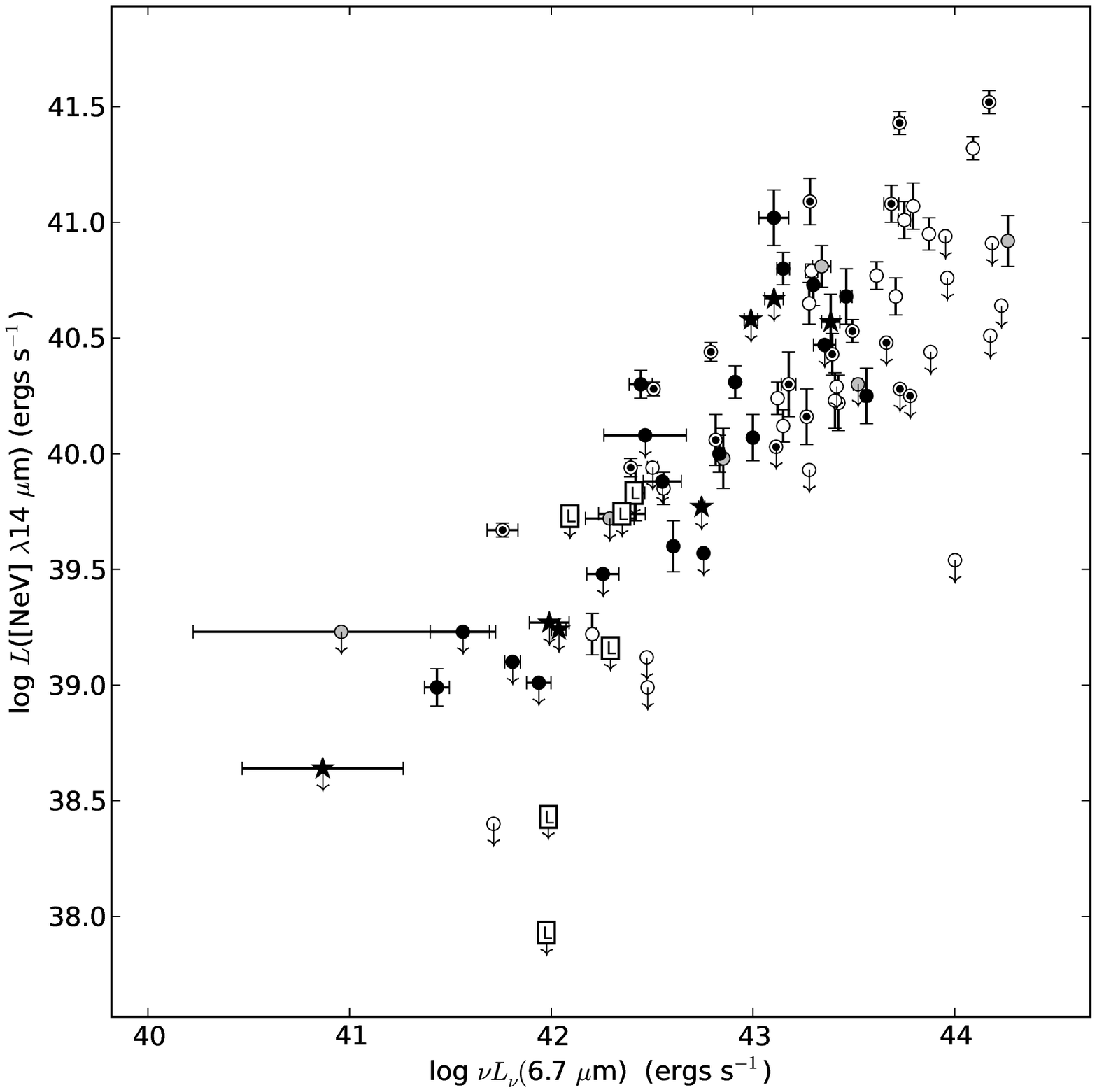}
\caption{Luminosity of the [NeV] 14\m\ line vs. the luminosity of the continuum at 6.7 \m.
 Open circles are S1, S1.5, and S1n, gray-filled
  circles are S1.8 and S1.9, filled circles are Sy 2,
  partially filled circles are ``hidden Seyferts'' (S1h and S1i), squares labeled
  L are LINERs, and filled stars are Starburst/HII galaxies. Limits
  are indicated by appropriately directed arrows.
}\label{IRS_DUST_NEV_L}
\end{figure}

High resolution mid-IR observations show extended hot dust associated
with the NLR in NGC1068 (Galliano et al. 2005; Mason et al. 2006) and
the Circinus galaxy (Packham et al. 2005).
Figure~\ref{IRS_DUST_NEV_L} shows a rank correlation between the luminosities in the
[NeV] 14\m\ line and the continuum determined at 6.7 \m\ in our Seyferts. 
The rank correlation is also present between the fluxes (Table \ref{tab:rank}).
In \S~\ref{sec:diag} we show that the [NeV] emission is not strongly orientation dependent. 
However, in models
where the 6.7 \m\ continuum originates in the hot dust in the inner torus,
we would expect the 6.7 \m\ continuum to be orientation dependent. 
In the context of standard uniform torus models we would expect that
the 6.7 \m\ emission should be high in the \syones,  but
low in the \sytwos\ where it is obscured by the torus. Instead we see
a strong rank correlation followed by both \syones\ and \sytwos. 
This suggests that in addition to any hot dust in the torus,
we are seeing an extended component of hot dust
whose luminosity scales with the AGN luminosity (as measured by [NeV]).
This might be due to dust in the NLR or a wind. \citet{mor09} also find evidence
for hot dust emission associated with the NLR in luminous QSOs.


\subsection{Comparison of Seyfert Subsamples}\label{sec:comp}

Here we discuss the properties of subsamples of the Seyfert population. 
Table~\ref{tab:median} gives the median values of a range of properties for the
continuum emission, Sil feature, star formation tracers, AGN tracers, and diagnostic
line ratios. Tables~\ref{tab:ks1}-\ref{tab:ks8} give the results of statistical 
tests for the significance of differences in the distribution of those properties. 

The comparison of the \syones\ with the \sytwos\ (both with and without a detected
HBLR) (Table~\ref{tab:ks1}) shows significant differences ($p < 1\%$) between 
the Sil feature, the 20-30 \m\ spectral index, 6.7 \m\ continuum flux density, 
and the EW of the 6.2 \m\ PAH feature.  As discussed in \S \ref{sec:sil},
the \syones\ exhibit weak Sil emission and the \sytwos\ show Sil absorption. 
The other differences are consistent with being due to a higher
 contribution from hot dust in the spectra of the \syones\ which causes 
higher 6.7 \m\ continuum,
bluer 20-30 \m\ spectral index, and lower EW of the 6.2 \m\ PAH feature.
Note that although the EWs are different, the luminosities are similar.
When we compare the \syones\  to just those \sytwos\ without an HBLR (Table~\ref{tab:ks2})
we see the same differences as between the \syones\ and all the \sytwos, with the addition
of a difference in the luminosity of the 6.7 \m\ continuum.

The \sytwos\ with and without HBLRs show differences
in 20-30 \m\ spectral index, 6.7 \m\ continuum, and EW of the 6.2 \m\ PAH feature 
and the \molhyd\ line  (Table \ref{tab:ks3}). These differences are consistent
with a higher contribution from a hot dust component in the HBLR \sytwos.
In addition, the \sytwos\ with an HBLR are very similar to the
the \syones; the only significant difference is the Sil strength (Table \ref{tab:ks4}).
Thus, the \sytwos\ with an HBLR are more similar to the \syones\
than they are to the \sytwos\ without an HBLR. 
The differences are in the sense that
the \sytwos\ with an HBLR have a higher ratio of AGN to Starburst contribution
to the SED. This may contribute to the detection of the HBLR in polarized light.
This is consistent with \citet{alexander01} who noted that there was a weaker contribution
from the galaxy in the HBLR \sytwos\ than the non-HBLR \sytwos.
We cannot rule out the possibility that the differences between the HBLR and
non-HBLR \sytwos\ reflect intrinsic differences in the AGN.

\citet{deo07} find that the mid-IR spectra of Sy 1.8-1.9 and are similar to those
of \sytwos\ with significant star formation. 
The Sy 1.8--1.9 are the smallest subgroup with only six objects in this study. 
The statistics are of course severely affected by small numbers, but we include 
the analyses for completeness.
The Sy 1.8--1.9 show no significant differences with the \sytwos\ 
without an HBLR (Table \ref{tab:ks5}) or  with an HBLR (Table \ref{tab:ks6}).
or the \syones\ (Table \ref{tab:ks7}).
This lack of difference between the Sy 1.8--1.9 and the other
Seyfert types suggests that the Sy 1.8-1.9 may have intermediate properties
between the \syones\ and \sytwos\ (as suggested by \citet{deo07}). 
However, it may also be due to the small number of Sy 1.8--1.9 in this study.

As noted above Sy 1.8-1.9 have mid-IR spectra which are similar to those of
\sytwos\ \citep{deo07} and HBLR \sytwos\ have mid-IR spectra which are closer 
to those of \syones\ than non-HBLR \sytwos. 
Thus, we also created a modified class of AGN1 consisting of the \syones\ plus 
the \sytwos\ with an HBLR, and a modified class of AGN2 consisting of the
\sytwos\ without an HBLR plus the Sy 1.8-1.9. The test for differences in the
properties of these two modified classes is shown in Table \ref{tab:ks8}.
We compare these results (Table \ref{tab:ks8}) to the comparison between 
the \syones\ and \sytwos\ (Table \ref{tab:ks1}). In the modified AGN1 and AGN2 
the differences in 20-30 \m\ spectral index, 6.7 \m\ continuum flux and luminosity,
and EW of the 6.2 \m\ PAH feature are stronger; the differences in Sil strength 
are weaker; and the modified samples show a new difference in the EW of the \molhyd\ line.
These findings are consistent with our results that the \sytwos\ with
an HBLR have a larger contribution from a hot dust component than the the \sytwos\ without
an  HBLR.


\section{SUMMARY}

We present an analysis of Spitzer IRS spectroscopy of 83 active galaxies, 
including 73 Seyferts, 6 LINERS, and 4 HII galaxies from the
extended 12\m\ sample of \citet{rush93}.  We fit the spectra using
a modified version of the \citet{smith07a} PAHFIT program and
examine selected diagnostic spectral features.

\syones\ and \sytwos\ and the \citet{brandl06} sample of starbursts have
similar distributions of the luminosity of the 6.2\m\ PAH feature. 
The EW of the 6.2\m\ PAH feature is slightly larger in the \sytwos\
and much larger in the \citet{brandl06} sample of starbursts than in
the \syones\ due to the differences in the underlying continuum. 

We see no difference in the 6.2/11.3 PAH ratio between the different sub samples
or any dependence of the ratio on [NeV] luminosity or Sil 10\m\ strength.
Thus, any effect of the AGN on the distribution of PAH grains must be confined to 
a small volume so that the integrated measurements studied here are not affected. 

The PAH EW is larger in sources with redder mid and far IR spectra.
In addition, the luminosity of the 6.2\m\ PAH feature is rank correlated
with the luminosity of the continuum at 60\m. The correlation is also
present in the fluxes. Both of these results
suggest that (1) The PAH feature is a reliable tracer of star formation
and (2) the there is a significant contribution to the heating
of the cool dust by  stars.  
The 55-90 vs. 20-30 spectral index plot is also a diagnostic
of the relative contribution of Starburst to AGN. 
We see there is a large change in spectral index across the sample:  
$\Delta \alpha \sim 3$ for both indexes. 
Thus, the contribution to the IR spectrum from the AGN and starburst 
components can be comparable in magnitude but the relative contribution 
also varies widely across the sample. 

We see a rank correlation between the \molhyd\  17 \m\ EW and two measures 
of the relative strength of the Starburst to AGN component - the 6.2\m\ PAH EW
and the 20-30\m\ spectral index. We also see are rank correlation
between the luminosities of the 6.2\m\ PAH feature and the \molhyd\  17 \m\ line.
The rank correlation is also present in the fluxes. These results suggest that the H$_2$ 
emission is also primarily excited by star formation.

We find rank correlations between several AGN tracers - 8 GHz  vs. [NeV]14\m\,
[NeV]14\m\ vs. [OIII]$\lambda$5007, [OIV]26\m\ vs. [OIII]$\lambda$5007,
and [NeV]14\m\ vs. 6.7\m\ continuum. The correlations are present both in
the fluxes and luminosities of these AGN tracers.  
The rank correlation between [NeV]14\m\ vs. 6.7\m\ continuum suggests that there
is a component to the 6.7\m\ continuum from hot dust which is extended, 
e.g., in the NLR.
We find rank correlations of the ratios [OIII]$\lambda$5007/[OIV]26\m\ and
[OIII]$\lambda$5007/[NeV]14\m\ with the Sil strength orientation indicator.
This suggests that some of the [OIII]$\lambda$5007 emission in these
Seyferts is subject to orientation dependent obscuration as found
by \citet{haas05} for radio galaxies and quasars. 
There is no correlation of [NeV] EW with the orientation indicator Sil 
10\m\ strength indicating that the [NeV] emission is not strongly 
orientation dependent.  This suggests with the obscuring material 
(e.g., torus) is not very  optically thick at 14\m\ consistent with
the results of \citet{buchanan06}.

We searched for correlations between AGN and Starburst tracers
(i.e., 6.2 \m\ PAH vs. [NeV] 14\m\ or [OIV] 26 \m\ or 6.7\m\ continuum, 
and [NeV] 14\m\ vs. 60 \m\ continuum). 
Although there are apparent correlations in luminosity (likely due to the
dependence on distance), there are no significant
correlations in flux, and we conclude that the AGN and Starburst tracers are 
not correlated. This is consistent with our conclusion that the relative strength
of the AGN and Starburst components varies widely across the sample. 
The lack of a correlation between the
AGN and star burst suggests that in these AGN there is no direct link between
AGN fueling, black hole growth, and star formation.  

The density diagnostic [NeV] 14/24 \m\ and [SIII] 18/33 \m\ line
ratios are consistent with the gas being near the low density limit,
i.e., $\sim 10^3$ cm\mthree\ for [NeV] and $n_e \sim $ few hundred
cm\mthree\ for [SIII].

The distribution of Sil 10 \m\ and 18 \m\ strengths is consistent
with the clumpy torus models of \citet{sirocky08}. 

We find a rank correlation between the [NeV] 14\m\ line and the continuum 
determined at 6.7 \m\ in our Seyferts.
This suggests that in addition to any hot dust in the torus,
we are seeing an extended component of hot dust
whose luminosity scales with the AGN luminosity (as measured by [NeV]).
This might be due to dust in the NLR or a wind. 

We have compared the distribution of several starburst and AGN tracers
between the different subsamples of Seyferts.  
Our comparison of the \syones\ with the \sytwos\ without an HBLR
shows that in addition to the strong difference in the Sil strength, 
the \sytwos\ without an HBLR have $2-3\sigma$ differences in the 
20-30 \m\ spectral index, 6.7 \m\ continuum, and the EW of the 
6.2 \m\ PAH feature and [NeV] 14\m\ line.  
The \sytwos\ with an HBLR have similar distributions of properties to
the \syones; the only significant difference is the Sil strength.
Thus, the \sytwos\ with an HBLR are more similar to the \syones\
than they are to the \sytwos\ without an HBLR.
The differences are in the sense that
the \sytwos\ with an HBLR have a higher ratio of AGN to Starburst contribution
to the SED. This may contribute to the detection of the HBLR in polarized light.
This is consistent with \citet{alexander01} who noted that there was a weaker contribution
from the galaxy in the HBLR \sytwos\ than the non-HBLR \sytwos.

\acknowledgments This work is based in part on observations made with
the Spitzer Space Telescope, which is operated by the Jet Propulsion
Laboratory, California Institute of Technology under a contract with
NASA. Support for this work at Bucknell University and Rochester
Institute of Technology was provided by NASA through an award issued
by JPL/Caltech.  We are grateful to Henrik Spoon for helpful
discussions and to an anonymous referee for constructive comments.
This research made use of (1) the NASA/IPAC Extragalactic Database (NED), 
which is operated by the Jet Propulsion Laboratory, California Institute 
of Technology, under contract with the National Aeronautics and Space 
Administration, and (2) NASA's Astrophysics Data System Abstract Service.

{}
\clearpage


\begin{deluxetable}{lrrrrr}
\tablewidth{0pt}
\rotate
\tablecaption{Median Values of Seyfert Properties \label{tab:median}}
\tablehead{
\colhead{Property} &
\colhead{Entire Sample} &
\colhead{Sy 1.0--1.5, 1n } &
\colhead{Sy 1.8--1.9 } &
\colhead{Sy 1h, 1i} &
\colhead{Sy 2}
\\
\colhead{ (1)} &
\colhead{ (2)} &
\colhead{ (3)} &
\colhead{ (4)} &
\colhead{ (5)} &
\colhead{ (6)}
}
\startdata
Count                                    &     83 &     27 &      6 &     17 &     20 \\
Distance                                 &$  50.4^{\,\,  83.2}_{\,\,  22.0}$&$  69.9^{\,\, 107.6}_{\,\,  34.6}$&$  67.2^{\,\, 107.1}_{\,\,  21.8}$&$  50.4^{\,\,  77.7}_{\,\,  26.3}$&$  40.4^{\,\,  67.1}_{\,\,  32.4}$\\
$\alpha$(20 $-$ 30 $\mu$m)               &$ -0.76^{\,\,  0.11}_{\,\, -1.38}$&$ -1.35^{\,\, -0.82}_{\,\, -1.81}$&$ -0.39^{\,\,  0.10}_{\,\, -0.77}$&$ -0.87^{\,\, -0.52}_{\,\, -1.25}$&$  0.16^{\,\,  0.45}_{\,\, -0.98}$\\
$\alpha$(55 $-$ 90 $\mu$m)               &$ -1.31^{\,\, -0.74}_{\,\, -1.81}$&$ -1.54^{\,\, -1.11}_{\,\, -2.03}$&$ -1.05^{\,\, -0.78}_{\,\, -1.28}$&$ -1.69^{\,\, -1.35}_{\,\, -2.61}$&$ -1.24^{\,\, -0.60}_{\,\, -1.53}$\\
$S_{\nu}$(6.7 $\mu$m)                    &$ 0.097^{\,\, 0.163}_{\,\, 0.038}$&$ 0.146^{\,\, 0.186}_{\,\, 0.094}$&$ 0.064^{\,\, 0.209}_{\,\, 0.013}$&$ 0.124^{\,\, 0.265}_{\,\, 0.060}$&$ 0.037^{\,\, 0.082}_{\,\, 0.023}$\\
log $\nu L_{\nu}($6.7 $\mu$m)            &$ 43.11^{\,\, 43.55}_{\,\, 42.42}$&$ 43.42^{\,\, 43.94}_{\,\, 43.13}$&$ 43.10^{\,\, 43.52}_{\,\, 42.29}$&$ 43.28^{\,\, 43.70}_{\,\, 42.81}$&$ 42.68^{\,\, 43.13}_{\,\, 42.34}$\\
log $\nu L_{\nu}$(8.4 GHz)               &$ 38.04^{\,\, 38.70}_{\,\, 37.19}$&$ 38.40^{\,\, 38.86}_{\,\, 37.57}$&$ 37.87^{\,\, 38.69}_{\,\, 36.69}$&$ 38.37^{\,\, 39.11}_{\,\, 37.47}$&$ 37.75^{\,\, 38.41}_{\,\, 37.07}$\\
IRAS log $\nu L_{\nu}$(60 $\mu$m)        &$ 43.63^{\,\, 43.94}_{\,\, 43.32}$&$ 43.55^{\,\, 43.88}_{\,\, 43.23}$&$ 43.82^{\,\, 44.27}_{\,\, 43.27}$&$ 43.67^{\,\, 43.93}_{\,\, 43.49}$&$ 43.62^{\,\, 44.25}_{\,\, 43.37}$\\
Sil 10 $\mu$m                            &$ 0.014^{\,\, 0.137}_{\,\,-0.143}$&$ 0.117^{\,\, 0.198}_{\,\, 0.009}$&$-0.010^{\,\, 0.252}_{\,\,-0.169}$&$-0.120^{\,\, 0.033}_{\,\,-0.551}$&$-0.060^{\,\,-0.003}_{\,\,-0.193}$\\
EW(PAH 6.2 $\mu$m)                       &$ 184.2^{\,\, 762.9}_{\,\,  56.9}$&$  84.5^{\,\, 191.1}_{\,\,  25.5}$&$ 615.4^{\,\,1234.7}_{\,\,  89.7}$&$ 113.8^{\,\, 411.3}_{\,\,  31.9}$&$ 557.0^{\,\,1547.9}_{\,\, 190.6}$\\
PAH 6.2 $\mu$m Flux                      &$ 112.8^{\,\, 328.8}_{\,\,  44.2}$&$  69.7^{\,\, 120.0}_{\,\,  31.9}$&$ 127.1^{\,\, 257.6}_{\,\,  56.9}$&$ 161.2^{\,\, 339.0}_{\,\,  40.4}$&$ 183.6^{\,\, 368.8}_{\,\,  69.6}$\\
log $L$(PAH 6.2 $\mu$m)                  &$ 41.44^{\,\, 42.08}_{\,\, 41.09}$&$ 41.46^{\,\, 41.63}_{\,\, 40.98}$&$ 41.71^{\,\, 42.22}_{\,\, 41.25}$&$ 41.42^{\,\, 41.94}_{\,\, 40.85}$&$ 41.57^{\,\, 42.32}_{\,\, 41.20}$\\
EW(H$_2$ S(1) $\lambda$17$\mu$m)         &$  12.0^{\,\,  42.5}_{\,\,   1.8}$&$   7.3^{\,\,  20.5}_{\,\,   1.7}$&$   6.5^{\,\,   8.2}_{\,\,   2.8}$&$   4.2^{\,\,  18.7}_{\,\,   0.7}$&$  20.4^{\,\,  47.1}_{\,\,   7.2}$\\
log $L$(H$_2$ S(1) $\lambda$17$\mu$m)    &$ 39.95^{\,\, 40.50}_{\,\, 39.39}$&$ 39.79^{\,\, 40.32}_{\,\, 39.29}$&$ 32.67^{\,\, 40.12}_{\,\,  6.53}$&$ 39.88^{\,\, 40.81}_{\,\, 38.90}$&$ 40.01^{\,\, 40.57}_{\,\, 39.44}$\\
log $L$(\siii\  $\lambda$18 $\mu$m)      &$ 40.15^{\,\, 40.69}_{\,\, 39.70}$&$ 40.15^{\,\, 40.59}_{\,\, 39.39}$&$ 40.80^{\,\, 40.94}_{\,\, 40.27}$&$ 40.57^{\,\, 40.67}_{\,\, 40.00}$&$ 40.11^{\,\, 40.93}_{\,\, 39.88}$\\
EW(\nev\  $\lambda$14 $\mu$m)            &$  10.4^{\,\,  17.7}_{\,\,   2.7}$&$   9.8^{\,\,  15.9}_{\,\,   1.6}$&$   8.7^{\,\,  10.7}_{\,\,   6.7}$&$  15.7^{\,\,  28.8}_{\,\,   6.8}$&$  13.7^{\,\,  17.3}_{\,\,   7.7}$\\
log $L$(\nev\  $\lambda$14 $\mu$m)       &$ 40.07^{\,\, 40.71}_{\,\, 38.31}$&$ 40.22^{\,\, 40.77}_{\,\, 38.71}$&$ 39.79^{\,\, 40.83}_{\,\, 24.52}$&$ 40.30^{\,\, 41.01}_{\,\, 40.01}$&$ 40.25^{\,\, 40.79}_{\,\, 39.49}$\\
log $L$(\oiv\  $\lambda$26 $\mu$m)       &$ 40.65^{\,\, 41.14}_{\,\, 39.97}$&$ 40.71^{\,\, 41.31}_{\,\, 40.39}$&$ 40.73^{\,\, 41.21}_{\,\, 40.13}$&$ 40.97^{\,\, 41.41}_{\,\, 40.49}$&$ 40.63^{\,\, 40.88}_{\,\, 39.95}$\\
\nev\  14 $\mu$m / 24 $\mu$m             &$  0.78^{\,\,  0.93}_{\,\,  0.62}$&$  0.69^{\,\,  0.86}_{\,\,  0.60}$&$  0.66^{\,\,  0.79}_{\,\,  0.58}$&$  0.86^{\,\,  1.08}_{\,\,  0.81}$&$  0.64^{\,\,  0.77}_{\,\,  0.60}$\\
\siii\  18 $\mu$m / 33 $\mu$m            &$  0.60^{\,\,  0.72}_{\,\,  0.47}$&$  0.58^{\,\,  0.68}_{\,\,  0.40}$&$  0.60^{\,\,  0.79}_{\,\,  0.57}$&$  0.69^{\,\,  0.73}_{\,\,  0.55}$&$  0.60^{\,\,  0.69}_{\,\,  0.40}$\\
PAH 6.2 $\mu$m / 11.3 $\mu$m             &$  1.02^{\,\,  1.34}_{\,\,  0.72}$&$  0.83^{\,\,  1.00}_{\,\,  0.64}$&$  1.25^{\,\,  1.48}_{\,\,  0.75}$&$  0.78^{\,\,  1.07}_{\,\,  0.52}$&$  1.26^{\,\,  1.43}_{\,\,  0.90}$\\

\enddata
\tablecomments{Median values of the properties of the 4 subgroups of
  Seyferts.  Col 1. The property.  Col 2. Median values for the whole
  sample. Col 3. Median value for S1 -- S1.5, and S1n. Col 4. Median
  value for S1.8--1.9. Col 5. Median value for S1h and S1i (hidden
  BLR). Col 6. Median value for S2. Luminosities are reported in
  \lumunits, fluxes in \fluxunits, and EW in nm. Medians were
  calculated using the Kaplan-Meier estimator where data include lower
  and upper limits (interval-censored data). Declination limits reduce
  the sample slightly for the VLA measurements. Several
  planned MIPS observations never executed, reducing slightly the
  number of $\alpha$(55-90~\m) measurements. Non-detections of both
  lines also affect the sample size for the line ratio measurements.
}
\end{deluxetable}

\begin{deluxetable}{lrrrrr}
\tablewidth{0pt}
\tablecaption{Comparison of \syone\ and \sytwo\ (non-HBLR + HBLR) 
Properties \label{tab:ks1}}
\tablehead{
\colhead{Property} &
\colhead{$n_1$} &
\colhead{$n_2$} &
\colhead{Test Statistic } &
\colhead{Value } &
\colhead{Two-sided significance (\%) }\\
\colhead{ (1)} &
\colhead{ (2)} &
\colhead{ (3)} &
\colhead{ (4)} &
\colhead{ (5)} &
\colhead{ (6)}
}
\startdata 
Distance&  27 &  37 & KS      &     0.29 &    15.8 \\
$\alpha$(20 $-$ 30 $\mu$m)&  27 &  37 & KS      &     0.42 &     0.5 \\
$\alpha$(55 $-$ 90 $\mu$m)&  26 &  36 & KS      &     0.16 &    75.3 \\
$S_{\nu}$(6.7 $\mu$m) &  27 &  37 & KS      &     0.41 &     0.7 \\
log $\nu L_{\nu}($6.7 $\mu$m) &  27 &  37 & KS      &     0.35 &     3.0 \\
IRAS log $\nu L_{\nu}$(60 $\mu$m) &  27 &  37 & KS      &     0.16 &    82.4 \\
Sil 10 $\mu$m&  27 &  37 & KS      &     0.52 &  $<0.1$ \\
log $\nu L_{\nu}$(8.4 GHz) &  24 &  35 & logrank & MC &   88.8 \\
EW(PAH 6.2 $\mu$m)&  27 &  37 & logrank & MC &    0.6 \\
log $L$(PAH 6.2 $\mu$m)&  27 &  37 & logrank & MC &   75.0 \\
EW(H$_2$ S(1) $\lambda$17$\mu$m)&  27 &  37 & logrank & MC &   32.0 \\
log $L$(H$_2$ S(1) $\lambda$17$\mu$m)&  27 &  37 & logrank & MC &   45.2 \\
log $L$(\siii\  $\lambda$18 $\mu$m)&  27 &  37 & logrank & MC &   14.6 \\
EW(\nev\  $\lambda$14 $\mu$m)&  27 &  37 & logrank & MC &    3.8 \\
log $L$(\nev\  $\lambda$14 $\mu$m)&  27 &  37 & logrank & MC &   75.6 \\
log $L$(\oiv\  $\lambda$26 $\mu$m)&  27 &  37 & logrank & MC &   76.4 \\
\nev\  14 $\mu$m / 24 $\mu$m&  20 &  26 & logrank & MC &   16.2 \\
\siii\  18 $\mu$m / 33 $\mu$m&  19 &  28 & logrank & MC &   64.2 \\
PAH 6.2 $\mu$m / 11.3 $\mu$m&  26 &  36 & logrank & MC &   37.4 \\

\enddata
\tablecomments{The results of statistical tests to determine if the
  measured parameters of the \syones\ and \sytwos\ are consistent with
  being drawn from the same distribution. The samples are defined
  based on optical classifications: \syone\ includes S1 -- 1.5 and S1n;
  \sytwo\ includes S2, S1h, S1i.  Col 1. Property being tested.	 Col
  2. Number of objects in first sample (\syone).  Col 3. Number of
  objects in the second sample (\sytwo).  Col 4. Type of statistical
  test: KS is the Kolmogorov-Smirnov test, and logrank is that test
  from survival analysis, adapted to accommodate lower and upper limits
  (interval-censored data).	 Col 5. The value of the test statistic; MC
  indicates that the probability in Col. 6 was determined by 1000
  Monte Carlo trials.  Col
  6. The probability that the two samples are drawn from the same
  parent distribution of object properties; low values indicate a
  significant difference between the samples. Lower limits were
  removed from the sample of line ratios, but upper limits were
  included in the analysis. }
\end{deluxetable}

\begin{deluxetable}{lrrrrr}
\tablewidth{0pt}
\tablecaption{Comparison of \syone\ and \sytwo\ (non-HBLR) Properties \label{tab:ks2}}
\tablehead{
\colhead{Property} &
\colhead{$n_1$} &
\colhead{$n_2$} &
\colhead{Test Statistic } &
\colhead{Value } &
\colhead{Two-sided significance (\%)}\\
\colhead{ (1)} &
\colhead{ (2)} &
\colhead{ (3)} &
\colhead{ (4)} &
\colhead{ (5)} &
\colhead{ (6)}
}
\startdata 
Distance&  27 &  20 & KS      &     0.32 &    15.3 \\
$\alpha$(20 $-$ 30 $\mu$m)&  27 &  20 & KS      &     0.58 &  $<0.1$ \\
$\alpha$(55 $-$ 90 $\mu$m)&  26 &  19 & KS      &     0.33 &    13.8 \\
$S_{\nu}$(6.7 $\mu$m) &  27 &  20 & KS      &     0.65 &  $<0.1$ \\
log $\nu L_{\nu}($6.7 $\mu$m) &  27 &  20 & KS      &     0.54 &     0.1 \\
IRAS log $\nu L_{\nu}$(60 $\mu$m) &  27 &  20 & KS      &     0.23 &    60.1 \\
Sil 10 $\mu$m&  27 &  20 & KS      &     0.56 &  $<0.1$ \\
log $\nu L_{\nu}$(8.4 GHz) &  24 &  19 & logrank & MC &   34.4 \\
EW(PAH 6.2 $\mu$m)&  27 &  20 & logrank & MC &    0.2 \\
log $L$(PAH 6.2 $\mu$m)&  27 &  20 & logrank & MC &   48.8 \\
EW(H$_2$ S(1) $\lambda$17$\mu$m)&  27 &  20 & logrank & MC &    2.2 \\
log $L$(H$_2$ S(1) $\lambda$17$\mu$m)&  27 &  20 & logrank & MC &   18.8 \\
log $L$(\siii\  $\lambda$18 $\mu$m)&  27 &  20 & logrank & MC &    8.4 \\
EW(\nev\  $\lambda$14 $\mu$m)&  27 &  20 & logrank & MC &    6.0 \\
log $L$(\nev\  $\lambda$14 $\mu$m)&  27 &  20 & logrank & MC &   37.8 \\
log $L$(\oiv\  $\lambda$26 $\mu$m)&  27 &  20 & logrank & MC &   70.8 \\
\nev\  14 $\mu$m / 24 $\mu$m&  20 &  13 & logrank & MC &    4.8 \\
\siii\  18 $\mu$m / 33 $\mu$m&  19 &  16 & logrank & MC &   30.4 \\
PAH 6.2 $\mu$m / 11.3 $\mu$m&  26 &  20 & logrank & MC &   19.8 \\

\enddata
\tablecomments{The results of statistical tests to determine if the measured
parameters of the \syones\ and \sytwos\ are consistent with being drawn from the same
distribution. The samples are defined based on optical classifications:
\syone\ includes S1--1.5 and S1n; \sytwo\ includes S2 only.
Col 1. Property being tested.
Col 2. Number of objects in first sample (\syone).
Col 3. Number of objects in the second sample (\sytwo).
Col 4. Type of statistical test: KS is the Kolmogorov-Smirnov test,
and logrank is that test from survival analysis, adapted to
accommodate lower and aupper limits (interval-censored data).
Col 5. The value of the test statistic. MC
  indicates that the probability in Col. 6 was determined by 1000
  Monte Carlo trials.  
Col 6. The probability that the two samples are drawn from the same
parent distribution of object properties; low values indicate a
significant difference between the samples. Lower limits were removed
from the sample of line ratios, but upper limits were included in the
analysis. }
\end{deluxetable}

\begin{deluxetable}{lrrrrr}
\tablewidth{0pt}
\tablecaption{Comparison of HBLR and non-HBLR \sytwo\ Properties \label{tab:ks3}}
\tablehead{
\colhead{Property} &
\colhead{$n_1$} &
\colhead{$n_2$} &
\colhead{Test Statistic } &
\colhead{Value } &
\colhead{Two-sided significance (\%)}\\
\colhead{ (1)} &
\colhead{ (2)} &
\colhead{ (3)} &
\colhead{ (4)} &
\colhead{ (5)} &
\colhead{ (6)}
}
\startdata 
Distance&  17 &  20 & KS      &     0.20 &    86.8 \\
$\alpha$(20 $-$ 30 $\mu$m)&  17 &  20 & KS      &     0.59 &     0.2 \\
$\alpha$(55 $-$ 90 $\mu$m)&  17 &  19 & KS      &     0.44 &     4.1 \\
$S_{\nu}$(6.7 $\mu$m) &  17 &  20 & KS      &     0.61 &     0.1 \\
log $\nu L_{\nu}($6.7 $\mu$m) &  17 &  20 & KS      &     0.46 &     2.9 \\
IRAS log $\nu L_{\nu}$(60 $\mu$m) &  17 &  20 & KS      &     0.24 &    58.2 \\
Sil 10 $\mu$m&  17 &  20 & KS      &     0.24 &    55.4 \\
log $\nu L_{\nu}$(8.4 GHz) &  16 &  19 & logrank & MC &    8.6 \\
EW(PAH 6.2 $\mu$m)&  17 &  20 & logrank & MC &    0.4 \\
log $L$(PAH 6.2 $\mu$m)&  17 &  20 & logrank & MC &   28.4 \\
EW(H$_2$ S(1) $\lambda$17$\mu$m)&  17 &  20 & logrank & MC &    0.2 \\
log $L$(H$_2$ S(1) $\lambda$17$\mu$m)&  17 &  20 & logrank & MC &   16.8 \\
log $L$(\siii\  $\lambda$18 $\mu$m)&  17 &  20 & logrank & MC &   24.8 \\
EW(\nev\  $\lambda$14 $\mu$m)&  17 &  20 & logrank & MC &   76.6 \\
log $L$(\nev\  $\lambda$14 $\mu$m)&  17 &  20 & logrank & MC &    6.2 \\
log $L$(\oiv\  $\lambda$26 $\mu$m)&  17 &  20 & logrank & MC &   23.4 \\
\nev\  14 $\mu$m / 24 $\mu$m&  13 &  13 & logrank & MC &   34.6 \\
\siii\  18 $\mu$m / 33 $\mu$m&  12 &  16 & logrank & MC &   18.8 \\
PAH 6.2 $\mu$m / 11.3 $\mu$m&  16 &  20 & logrank & MC &    6.6 \\

\enddata
\tablecomments{The results of statistical tests to determine if the measured
parameters of hidden \syones\ and standard \sytwos\ are consistent with being drawn from the same
distribution. The samples are defined based on optical
classifications, with HBLR comprising S1i and S1h, and non-HBLR
comprising only S2.
Col 1. Property being tested.
Col 2. Number of objects in first sample (\syone).
Col 3. Number of objects in the second sample (\sytwo).
Col 4. Type of statistical test: KS is the Kolmogorov-Smirnov test,
and logrank is that test from survival analysis, adapted to
accommodate lower and upper limits (interval-censored data).
Col 5. The value of the test statistic. MC
  indicates that the probability in Col. 6 was determined by 1000
  Monte Carlo trials.  
Col 6. The probability that the two samples are drawn from the same
parent distribution of object properties; low values indicate a
significant difference between the samples. Lower limits were removed
from the sample of line ratios, but upper limits were included in the
analysis. }
\end{deluxetable}

\begin{deluxetable}{lrrrrr}
\tablewidth{0pt}
\tablecaption{Comparison of \syone\ and HBLR \sytwo\ Properties \label{tab:ks4}}
\tablehead{
\colhead{Property} &
\colhead{$n_1$} &
\colhead{$n_2$} &
\colhead{Test Statistic } &
\colhead{Value } &
\colhead{Two-sided significance (\%)}\\
\colhead{ (1)} &
\colhead{ (2)} &
\colhead{ (3)} &
\colhead{ (4)} &
\colhead{ (5)} &
\colhead{ (6)}
}
\startdata 
Distance&  27 &  17 & KS      &     0.26 &    47.4 \\
$\alpha$(20 $-$ 30 $\mu$m)&  27 &  17 & KS      &     0.37 &     8.2 \\
$\alpha$(55 $-$ 90 $\mu$m)&  26 &  17 & KS      &     0.19 &    76.6 \\
$S_{\nu}$(6.7 $\mu$m) &  27 &  17 & KS      &     0.22 &    61.7 \\
log $\nu L_{\nu}($6.7 $\mu$m) &  27 &  17 & KS      &     0.31 &    20.9 \\
IRAS log $\nu L_{\nu}$(60 $\mu$m) &  27 &  17 & KS      &     0.22 &    67.0 \\
Sil 10 $\mu$m&  27 &  17 & KS      &     0.51 &     0.8 \\
log $\nu L_{\nu}$(8.4 GHz) &  24 &  16 & logrank & MC &   24.8 \\
EW(PAH 6.2 $\mu$m)&  27 &  17 & logrank & MC &   70.8 \\
log $L$(PAH 6.2 $\mu$m)&  27 &  17 & logrank & MC &   95.4 \\
EW(H$_2$ S(1) $\lambda$17$\mu$m)&  27 &  17 & logrank & MC &   32.6 \\
log $L$(H$_2$ S(1) $\lambda$17$\mu$m)&  27 &  17 & logrank & MC &   61.2 \\
log $L$(\siii\  $\lambda$18 $\mu$m)&  27 &  17 & logrank & MC &   60.6 \\
EW(\nev\  $\lambda$14 $\mu$m)&  27 &  17 & logrank & MC &   20.0 \\
log $L$(\nev\  $\lambda$14 $\mu$m)&  27 &  17 & logrank & MC &   18.4 \\
log $L$(\oiv\  $\lambda$26 $\mu$m)&  27 &  17 & logrank & MC &   26.8 \\
\nev\  14 $\mu$m / 24 $\mu$m&  20 &  13 & logrank & MC &   72.8 \\
\siii\  18 $\mu$m / 33 $\mu$m&  19 &  12 & logrank & MC &   66.8 \\
PAH 6.2 $\mu$m / 11.3 $\mu$m&  26 &  16 & logrank & MC &   90.2 \\

\enddata
\tablecomments{The results of statistical tests to determine if the measured
parameters of \syones\ and \sytwos\ with HBLR are consistent with being drawn from the same
distribution. The samples are defined based on optical
classifications.
Col 1. Property being tested.
Col 2. Number of objects in first sample (\syone).
Col 3. Number of objects in the second sample (\sytwo).
Col 4. Type of statistical test: KS is the Kolmogorov-Smirnov test,
and logrank is that test from survival analysis, adapted to
accommodate lower and upper limits (interval-censored data).
Col 5. The value of the test statistic.  MC
  indicates that the probability in Col. 6 was determined by 1000
  Monte Carlo trials.  
Col 6. The probability that the two samples are drawn from the same
parent distribution of object properties; low values indicate a
significant difference between the samples. Lower limits were removed
from the sample of line ratios, but upper limits were included in the
analysis. }
\end{deluxetable}

\begin{deluxetable}{lrrrrr}
\tablewidth{0pt}
\tablecaption{Comparison of Sy 1.8--1.9 and non-HBLR \sytwo\ Properties \label{tab:ks5}}
\tablehead{
\colhead{Property} &
\colhead{$n_1$} &
\colhead{$n_2$} &
\colhead{Test Statistic } &
\colhead{Value } &
\colhead{Two-sided significance (\%)}\\
\colhead{ (1)} &
\colhead{ (2)} &
\colhead{ (3)} &
\colhead{ (4)} &
\colhead{ (5)} &
\colhead{ (6)}
}
\startdata 
Distance&   6 &  20 & KS      &     0.35 &    52.8 \\
$\alpha$(20 $-$ 30 $\mu$m)&   6 &  20 & KS      &     0.38 &    40.9 \\
$\alpha$(55 $-$ 90 $\mu$m)&   6 &  19 & KS      &     0.26 &    82.2 \\
$S_{\nu}$(6.7 $\mu$m) &   6 &  20 & KS      &     0.33 &    59.3 \\
log $\nu L_{\nu}($6.7 $\mu$m) &   6 &  20 & KS      &     0.35 &    52.8 \\
IRAS log $\nu L_{\nu}$(60 $\mu$m) &   6 &  20 & KS      &     0.20 &    97.2 \\
Sil 10 $\mu$m&   6 &  20 & KS      &     0.33 &    59.3 \\
log $\nu L_{\nu}$(8.4 GHz) &   6 &  19 & logrank & MC &   68.6 \\
EW(PAH 6.2 $\mu$m)&   6 &  20 & logrank & MC &   59.4 \\
log $L$(PAH 6.2 $\mu$m)&   6 &  20 & logrank & MC &   91.2 \\
EW(H$_2$ S(1) $\lambda$17$\mu$m)&   6 &  20 & logrank & MC &   31.6 \\
log $L$(H$_2$ S(1) $\lambda$17$\mu$m)&   6 &  20 & logrank & MC &   25.2 \\
log $L$(\siii\  $\lambda$18 $\mu$m)&   6 &  20 & logrank & MC &   32.2 \\
EW(\nev\  $\lambda$14 $\mu$m)&   6 &  20 & logrank & MC &   38.2 \\
log $L$(\nev\  $\lambda$14 $\mu$m)&   6 &  20 & logrank & MC &   49.2 \\
log $L$(\oiv\  $\lambda$26 $\mu$m)&   6 &  20 & logrank & MC &   72.6 \\
\nev\  14 $\mu$m / 24 $\mu$m&   4 &  13 & KS      &     0.29 &    91.9 \\
\siii\  18 $\mu$m / 33 $\mu$m&   5 &  16 & logrank & MC &   13.2 \\
PAH 6.2 $\mu$m / 11.3 $\mu$m&   6 &  20 & logrank & MC &   37.6 \\

\enddata
\tablecomments{The results of statistical tests to determine if the measured
parameters of Sy 1.8--1.9 and \sytwos\ without HBLR are consistent with being drawn from the same
distribution. The samples are defined based on optical
classifications.
Col 1. Property being tested.
Col 2. Number of objects in first sample (\syone).
Col 3. Number of objects in the second sample (\sytwo).
Col 4. Type of statistical test: KS is the Kolmogorov-Smirnov test,
and logrank is that test from survival analysis, adapted to
accommodate lower and upper limits (interval-censored data).
Col 5. The value of the test statistic.  MC
  indicates that the probability in Col. 6 was determined by 1000
  Monte Carlo trials.  
Col 6. The probability that the two samples are drawn from the same
parent distribution of object properties; low values indicate a
significant difference between the samples. Lower limits were removed
from the sample of line ratios, but upper limits were included in the
analysis.
}
\end{deluxetable}

\begin{deluxetable}{lrrrrr}
\tablewidth{0pt}
\tablecaption{Comparison of Sy 1.8--1.9 and HBLR \sytwo\ Properties \label{tab:ks6}}
\tablehead{
\colhead{Property} &
\colhead{$n_1$} &
\colhead{$n_2$} &
\colhead{Test Statistic } &
\colhead{Value } &
\colhead{Two-sided significance (\%)}\\
\colhead{ (1)} &
\colhead{ (2)} &
\colhead{ (3)} &
\colhead{ (4)} &
\colhead{ (5)} &
\colhead{ (6)}
}
\startdata 
Distance&   6 &  17 & KS      &     0.26 &    84.7 \\
$\alpha$(20 $-$ 30 $\mu$m)&   6 &  17 & KS      &     0.44 &    25.0 \\
$\alpha$(55 $-$ 90 $\mu$m)&   6 &  17 & KS      &     0.66 &     2.4 \\
$S_{\nu}$(6.7 $\mu$m) &   6 &  17 & KS      &     0.44 &    25.0 \\
log $\nu L_{\nu}($6.7 $\mu$m) &   6 &  17 & KS      &     0.27 &    79.3 \\
IRAS log $\nu L_{\nu}$(60 $\mu$m) &   6 &  17 & KS      &     0.44 &    25.0 \\
Sil 10 $\mu$m&   6 &  17 & KS      &     0.35 &    54.1 \\
log $\nu L_{\nu}$(8.4 GHz) &   6 &  16 & logrank & MC &   17.6 \\
EW(PAH 6.2 $\mu$m)&   6 &  17 & logrank & MC &    3.2 \\
log $L$(PAH 6.2 $\mu$m)&   6 &  17 & logrank & MC &   54.4 \\
EW(H$_2$ S(1) $\lambda$17$\mu$m)&   6 &  17 & logrank & MC &    7.4 \\
log $L$(H$_2$ S(1) $\lambda$17$\mu$m)&   6 &  17 & logrank & MC &   81.2 \\
log $L$(\siii\  $\lambda$18 $\mu$m)&   6 &  17 & logrank & MC &   67.4 \\
EW(\nev\  $\lambda$14 $\mu$m)&   6 &  17 & logrank & MC &   72.0 \\
log $L$(\nev\  $\lambda$14 $\mu$m)&   6 &  17 & logrank & MC &   38.2 \\
log $L$(\oiv\  $\lambda$26 $\mu$m)&   6 &  17 & logrank & MC &   44.8 \\
\nev\  14 $\mu$m / 24 $\mu$m&   4 &  13 & logrank & MC &   38.8 \\
\siii\  18 $\mu$m / 33 $\mu$m&   5 &  12 & logrank & MC &   39.0 \\
PAH 6.2 $\mu$m / 11.3 $\mu$m&   6 &  16 & logrank & MC &    3.4 \\

\enddata
\tablecomments{The results of statistical tests to determine if the measured
parameters of Sy 1.8-1.9 and \sytwos\ with HBLR are consistent with being drawn from the same
distribution. The samples are defined based on optical
classifications.
Col 1. Property being tested.
Col 2. Number of objects in first sample (\syone).
Col 3. Number of objects in the second sample (\sytwo).
Col 4. Type of statistical test: KS is the Kolmogorov-Smirnov test,
and logrank is that test from survival analysis, adapted to
accommodate lower and upper limits (interval-censored data).
Col 5. The value of the test statistic.  MC
  indicates that the probability in Col. 6 was determined by 1000
  Monte Carlo trials.  
Col 6. The probability that the two samples are drawn from the same
parent distribution of object properties; low values indicate a
significant difference between the samples. Lower limits were removed
from the sample of line ratios, but upper limits were included in the
analysis. }
\end{deluxetable}

\begin{deluxetable}{lrrrrr}
\tablewidth{0pt}
\tablecaption{Comparison of Sy 1.8--1.9 and \syone\ Properties \label{tab:ks7}}
\tablehead{
\colhead{Property} &
\colhead{$n_1$} &
\colhead{$n_2$} &
\colhead{Test Statistic } &
\colhead{Value } &
\colhead{Two-sided significance (\%)}\\
\colhead{ (1)} &
\colhead{ (2)} &
\colhead{ (3)} &
\colhead{ (4)} &
\colhead{ (5)} &
\colhead{ (6)}
}
\startdata 
Distance&   6 &  27 & KS      &     0.17 &    99.6 \\
$\alpha$(20 $-$ 30 $\mu$m)&   6 &  27 & KS      &     0.65 &     1.8 \\
$\alpha$(55 $-$ 90 $\mu$m)&   6 &  26 & KS      &     0.54 &     7.9 \\
$S_{\nu}$(6.7 $\mu$m) &   6 &  27 & KS      &     0.46 &    18.5 \\
log $\nu L_{\nu}($6.7 $\mu$m) &   6 &  27 & KS      &     0.31 &    62.2 \\
IRAS log $\nu L_{\nu}$(60 $\mu$m) &   6 &  27 & KS      &     0.39 &    44.8 \\
Sil 10 $\mu$m&   6 &  27 & KS      &     0.43 &    26.6 \\
log $\nu L_{\nu}$(8.4 GHz) &   6 &  24 & logrank & MC &   16.8 \\
EW(PAH 6.2 $\mu$m)&   6 &  27 & logrank & MC &    1.2 \\
log $L$(PAH 6.2 $\mu$m)&   6 &  27 & logrank & MC &   59.8 \\
EW(H$_2$ S(1) $\lambda$17$\mu$m)&   6 &  27 & logrank & MC &   30.4 \\
log $L$(H$_2$ S(1) $\lambda$17$\mu$m)&   6 &  27 & logrank & MC &   57.0 \\
log $L$(\siii\  $\lambda$18 $\mu$m)&   6 &  27 & logrank & MC &   95.0 \\
EW(\nev\  $\lambda$14 $\mu$m)&   6 &  27 & logrank & MC &   39.6 \\
log $L$(\nev\  $\lambda$14 $\mu$m)&   6 &  27 & logrank & MC &   84.4 \\
log $L$(\oiv\  $\lambda$26 $\mu$m)&   6 &  27 & logrank & MC &   65.6 \\
\nev\  14 $\mu$m / 24 $\mu$m&   4 &  20 & logrank & MC &   48.8 \\
\siii\  18 $\mu$m / 33 $\mu$m&   5 &  19 & logrank & MC &   35.2 \\
PAH 6.2 $\mu$m / 11.3 $\mu$m&   6 &  26 & logrank & MC &   20.4 \\

\enddata
\tablecomments{The results of statistical tests to determine if the measured
parameters of Sy 1.8--1.9 and \syones\  are consistent with 
being drawn from the same distribution. The samples are defined based on optical
classifications.
Col 1. Property being tested.
Col 2. Number of objects in first sample (\syone).
Col 3. Number of objects in the second sample (\sytwo).
Col 4. Type of statistical test: KS is the Kolmogorov-Smirnov test,
and logrank is that test from survival analysis, adapted to
accommodate lower and upper limits (interval-censored data).
Col 5. The value of the test statistic.  MC
  indicates that the probability in Col. 6 was determined by 1000
  Monte Carlo trials.  
Col 6. The probability that the two samples are drawn from the same
parent distribution of object properties; low values indicate a
significant difference between the samples. Lower limits were removed
from the sample of line ratios, but upper limits were included in the
analysis. }
\end{deluxetable}

\begin{deluxetable}{lrrrrr}
\tablewidth{0pt}
\tablecaption{Comparison of (Sy 1 + HBLR Sy 2) and (Sy 1.8--1.9 +
  non-HBLR Sy 2) Properties \label{tab:ks8}}
\tablehead{
\colhead{Property} &
\colhead{$n_1$} &
\colhead{$n_2$} &
\colhead{Test Statistic } &
\colhead{Value } &
\colhead{Two-sided significance (\%)}\\
\colhead{ (1)} &
\colhead{ (2)} &
\colhead{ (3)} &
\colhead{ (4)} &
\colhead{ (5)} &
\colhead{ (6)}
}
\startdata 
Distance&  44 &  26 & KS      &     0.18 &    65.2 \\
$\alpha$(20 $-$ 30 $\mu$m)&  44 &  26 & KS      &     0.55 &  $<0.1$ \\
$\alpha$(55 $-$ 90 $\mu$m)&  43 &  25 & KS      &     0.39 &     1.1 \\
$S_{\nu}$(6.7 $\mu$m) &  44 &  26 & KS      &     0.55 &  $<0.1$ \\
log $\nu L_{\nu}($6.7 $\mu$m) &  44 &  26 & KS      &     0.44 &     0.2 \\
IRAS log $\nu L_{\nu}$(60 $\mu$m) &  44 &  26 & KS      &     0.26 &    23.7 \\
Sil 10 $\mu$m&  44 &  26 & KS      &     0.35 &     3.5 \\
log $\nu L_{\nu}$(8.4 GHz) &  40 &  25 & logrank & MC &    5.6 \\
EW(PAH 6.2 $\mu$m)&  44 &  26 & logrank & MC &    0.2 \\
log $L$(PAH 6.2 $\mu$m)&  44 &  26 & logrank & MC &   28.2 \\
EW(H$_2$ S(1) $\lambda$17$\mu$m)&  44 &  26 & logrank & MC &    0.6 \\
log $L$(H$_2$ S(1) $\lambda$17$\mu$m)&  44 &  26 & logrank & MC &   19.6 \\
log $L$(\siii\  $\lambda$18 $\mu$m)&  44 &  26 & logrank & MC &   14.4 \\
EW(\nev\  $\lambda$14 $\mu$m)&  44 &  26 & logrank & MC &   28.4 \\
log $L$(\nev\  $\lambda$14 $\mu$m)&  44 &  26 & logrank & MC &   13.2 \\
log $L$(\oiv\  $\lambda$26 $\mu$m)&  44 &  26 & logrank & MC &   34.0 \\
\nev\  14 $\mu$m / 24 $\mu$m&  33 &  17 & logrank & MC &    8.2 \\
\siii\  18 $\mu$m / 33 $\mu$m&  31 &  21 & logrank & MC &   42.0 \\
PAH 6.2 $\mu$m / 11.3 $\mu$m&  42 &  26 & logrank & MC &    3.8 \\

\enddata
\tablecomments{The results of statistical tests to determine if the measured
parameters of the combined groups (Sy 1 + HBLR Sy 2) and (Sy 1.8--1.9
and non-HBLR Sy 2)  are consistent with being drawn from the same
distribution. The samples are defined based on optical classifications.
Col 1. Property being tested.
Col 2. Number of objects in first sample (\syone).
Col 3. Number of objects in the second sample (\sytwo).
Col 4. Type of statistical test: KS is the Kolmogorov-Smirnov test,
and logrank is that test from survival analysis, adapted to
accommodate lower and upper limits (interval-censored data).
Col 5. The value of the test statistic.	 MC
  indicates that the probability in Col. 6 was determined by 1000
  Monte Carlo trials.
Col 6. The probability that the two samples are drawn from the same
parent distribution of object properties; low values indicate a
significant difference between the samples. Lower limits were removed
from the sample of line ratios, but upper limits were included in the
analysis. }
\end{deluxetable}

\begin{deluxetable}{llrr}
\tablewidth{0pt}
\tablecaption{Tests for Correlations in Seyfert Properties  \label{tab:rank}}
\tablehead{
\colhead{First Property} &
\colhead{Second Property} &
\colhead{$z$ } &
\colhead{$p$(\%)} \\
\colhead{ (1)} &
\colhead{ (2)} &
\colhead{ (3)} &
\colhead{ (4)}
}
\startdata 
EW(PAH 6.2 $\mu$m)                       & Distance                                 & $ -2.058 $ & $   4.0 $ \\ 
EW(PAH 6.2 $\mu$m)                       & Sil 10 $\mu$m                            & $ -2.311 $ & $   2.1 $ \\ 
EW(PAH 6.2 $\mu$m)                       & $\alpha$(20 $-$ 30 $\mu$m)               & $  7.691 $ & $ < 0.1 $ \\ 
EW(PAH 6.2 $\mu$m)                       & $\alpha$(55 $-$ 90 $\mu$m)               & $  5.663 $ & $ < 0.1 $ \\ 
EW(PAH 6.2 $\mu$m)                       & EW(H$_2$ S(1) $\lambda$17$\mu$m)         & $  4.888 $ & $ < 0.1 $ \\ 
EW(PAH 6.2 $\mu$m)                       & EW(\nev\  $\lambda$14 $\mu$m)            & $  2.220 $ & $   2.6 $ \\ 
EW(PAH 6.2 $\mu$m)                       & log $L$(\nev\  $\lambda$14 $\mu$m)       & $ -0.668 $ & $  50.4 $ \\ 
PAH 6.2 $\mu$m Flux                      & \nev\  $\lambda$14 $\mu$m Flux           & $  1.933 $ & $   5.3 $ \\ 
log $L$(PAH 6.2 $\mu$m)                  & log $L$(\nev\  $\lambda$14 $\mu$m)       & $  4.205 $ & $ < 0.1 $ \\ 
PAH 6.2 $\mu$m Flux                      & H$_2$ S(1) $\lambda$17$\mu$m Flux        & $  4.969 $ & $ < 0.1 $ \\ 
log $L$(PAH 6.2 $\mu$m)                  & log $L$(H$_2$ S(1) $\lambda$17$\mu$m)    & $  5.333 $ & $ < 0.1 $ \\ 
log $L$(PAH 6.2 $\mu$m)                  & $\alpha$(20 $-$ 30 $\mu$m)               & $  2.671 $ & $   0.8 $ \\ 
log $L$(PAH 6.2 $\mu$m)                  & $\alpha$(55 $-$ 90 $\mu$m)               & $  1.370 $ & $  17.1 $ \\ 
PAH 6.2 $\mu$m Flux                      & IRAS $S_{\nu}$(60 $\mu$m)                & $  7.260 $ & $ < 0.1 $ \\ 
log $L$(PAH 6.2 $\mu$m)                  & IRAS log $\nu L_{\nu}$(60 $\mu$m)        & $  7.491 $ & $ < 0.1 $ \\ 
PAH 6.2 $\mu$m Flux                      & \oiv\  $\lambda$26 $\mu$m Flux           & $  1.029 $ & $  30.3 $ \\ 
log $L$(PAH 6.2 $\mu$m)                  & log $L$(\oiv\  $\lambda$26 $\mu$m)       & $  2.744 $ & $   0.6 $ \\ 
$\alpha$(20 $-$ 30 $\mu$m)               & $\alpha$(55 $-$ 90 $\mu$m)               & $  4.288 $ & $ < 0.1 $ \\ 
$\alpha$(20 $-$ 30 $\mu$m)               & EW(H$_2$ S(1) $\lambda$17$\mu$m)         & $  4.292 $ & $ < 0.1 $ \\ 
$\alpha$(20 $-$ 30 $\mu$m)               & IRAS log $\nu L_{\nu}$(60 $\mu$m)        & $  2.788 $ & $   0.5 $ \\ 
$\alpha$(20 $-$ 30 $\mu$m)               & EW(\nev\  $\lambda$14 $\mu$m)            & $  2.933 $ & $   0.3 $ \\ 
$\alpha$(20 $-$ 30 $\mu$m)               & log $L$(\nev\  $\lambda$14 $\mu$m)       & $ -0.317 $ & $  75.2 $ \\ 
$\alpha$(20 $-$ 30 $\mu$m)               & log $L$(\siii\  $\lambda$18 $\mu$m)      & $  2.861 $ & $   0.4 $ \\ 
$\alpha$(20 $-$ 30 $\mu$m)               & Sil 10 $\mu$m                            & $ -4.674 $ & $ < 0.1 $ \\ 
$S_{\nu}$(6.7 $\mu$m)                    & \nev\  $\lambda$14 $\mu$m Flux           & $  3.228 $ & $   0.1 $ \\ 
log $\nu L_{\nu}($6.7 $\mu$m)            & log $L$(\nev\  $\lambda$14 $\mu$m)       & $  5.136 $ & $ < 0.1 $ \\ 
$S_{\nu}$(6.7 $\mu$m)                    & PAH 6.2 $\mu$m Flux                      & $ -0.690 $ & $  49.0 $ \\ 
log $\nu L_{\nu}($6.7 $\mu$m)            & log $L$(PAH 6.2 $\mu$m)                  & $  2.884 $ & $   0.4 $ \\ 
EW(\nev\  $\lambda$14 $\mu$m)            & Sil 10 $\mu$m                            & $ -2.997 $ & $   0.3 $ \\ 
\nev\  $\lambda$14 $\mu$m Flux           & $S_{\nu}$(8.4 GHz)                       & $  3.118 $ & $   0.2 $ \\ 
log $L$(\nev\  $\lambda$14 $\mu$m)       & log $\nu L_{\nu}$(8.4 GHz)               & $  4.371 $ & $ < 0.1 $ \\ 
\nev\  $\lambda$14 $\mu$m Flux           & IRAS $S_{\nu}$(60 $\mu$m)                & $  2.724 $ & $   0.6 $ \\ 
log $L$(\nev\  $\lambda$14 $\mu$m)       & IRAS log $\nu L_{\nu}$(60 $\mu$m)        & $  4.808 $ & $ < 0.1 $ \\ 
log $L$(\nev\  $\lambda$14 $\mu$m)       & Sil 10 $\mu$m                            & $ -0.410 $ & $  68.1 $ \\ 
\nev\  14 $\mu$m / 24 $\mu$m             & \siii\  18 $\mu$m / 33 $\mu$m            & $ -0.171 $ & $  86.4 $ \\ 
\nev\  14 $\mu$m / 24 $\mu$m             & Sil 10 $\mu$m                            & $ -0.914 $ & $  36.1 $ \\ 
PAH 6.2 $\mu$m / 11.3 $\mu$m             & log $L$(\nev\  $\lambda$14 $\mu$m)       & $  0.501 $ & $  61.7 $ \\ 
PAH 6.2 $\mu$m / 11.3 $\mu$m             & Sil 10 $\mu$m                            & $ -0.006 $ & $  99.5 $ \\ 
X-ray Flux (2$-$10 keV)                  & \nev\  $\lambda$14 $\mu$m Flux           & $  1.685 $ & $   9.2 $ \\ 
\oiii\  Flux                             & \nev\  $\lambda$14 $\mu$m Flux           & $  1.811 $ & $   7.0 $ \\ 
\nev\  14 $\mu$m / [NeII]                & EW(PAH 6.2 $\mu$m)                       & $ -5.261 $ & $ < 0.1 $ \\ 
\oiii\  / \nev\  (14 $\mu$m)             & Sil 10 $\mu$m                            & $  4.197 $ & $ < 0.1 $ \\ 
\oiii\  / \oiv\  (26 $\mu$m)             & Sil 10 $\mu$m                            & $  4.121 $ & $ < 0.1 $ \\ 
log $L$(\oiv\  $\lambda$26 $\mu$m)       & log $L$(\oiii\ )                         & $  4.847 $ & $ < 0.1 $ \\ 
\oiv\  $\lambda$26 $\mu$m Flux           & \oiii\  Flux                             & $  2.564 $ & $   1.0 $ \\ 
log $L$(\nev\  $\lambda$14 $\mu$m)       & log $L$(\oiii\ )                         & $  4.655 $ & $ < 0.1 $ \\ 
\nev\  $\lambda$14 $\mu$m Flux           & \oiii\  Flux                             & $  1.811 $ & $   7.0 $ \\ 
log $L$(\nev\  $\lambda$14 $\mu$m)       & Sil 10 $\mu$m                            & $ -0.410 $ & $  68.1 $ \\ 
$N_{\rm H}$ (cm$^{-2}$)                  & Sil 10 $\mu$m                            & $ -2.372 $ & $   1.8 $ \\ 

\enddata
\tablecomments{The results of non-parametric rank tests to look for correlations among
the properties of the combined	sample.
Col 1 \& 2. Properties tested for correlation.
Col 3. Kendall's generalized rank correlation coefficient,
adjusted to account for upper and lower limits. Reported is the
$z$-score, following the method outlined by \citet{isobe1986}.
Col 4. Two-sided significance, giving the probability of no
correlation; small values ($p < 1\%$) indicate evidence for correlation.}
\end{deluxetable}


\clearpage

\clearpage


\begin{thebibliography}{}

\bibitem[Abel \& Satyapal (2008)]{abel08}
Abel, N. P., \& Satyapal, S., 2008, ApJ, 678, 686

\bibitem[Alexander (2001)]{alexander01}
Alexander, D. M., 2001, MNRAS, 320, 320, L15

\bibitem[Alexander et al.\ (1999)]{alexander99}
Alexander, T.,  Sturm, E.,  Lutz, D., Sternberg, A.,  Netzer, H.,  \& Genzel, R.,
1999, ApJ, 512, 204

\bibitem[Alen et al. (2008)]{allen08}
Allen, M. G., Groves, B. A., Dopita, M. A., Sutherland, R. S., Kewley, L. J.,
2008, ApJS, 178, 20 


\bibitem[Anscombe (1973)]{anscombe73}
Anscombe, F. J. 1973, American Statistician, 27, 17

\bibitem[Armus et al. (2007)]{armus07}
Armus, L., et al., 2007, 656, 148

\bibitem[Axon 2009, private communication]{axon09}
Axon, D., 2009, private communication 

\bibitem[Brandl et al. (2006)]{brandl06}
Brandl, B. R., et al., 2006, ApJ, 653, 1129, erratum ApJ, 665, 884

\bibitem[Chiar \& Tielens (2006)]{chiar06}
Chiar, J. E., \& Tielens, A. G. G. M., 2006, ApJ, 637, 774

\bibitem[Buchanan et al. (2006)]{buchanan06}
Buchanan, C. L., Gallimore, J. F., O'Dea, C. P., Baum, S. A., Axon, D. J., Robinson, A.,
Elitzur, M., \& Elvis, M., 2006, AJ, 132, 401

\bibitem[Buchanan et al. (2008)]{buchanan08}
Buchanan, C. L., Gallimore, J. F., O'Dea, C. P., Baum, S. A., Axon, D. J., Robinson, A.,
Noel-Storr, J., Yzaguirre, A., Elitzur, M., Elvis, M., \& Tadhunter, C.,
2008, Mem. S. A. It., 75, 282

\bibitem[Buchanan et al. (2009)]{buchanan09}
Buchanan, C. L., et al., (2009) in preparation.

\bibitem[Calzetti et al.(2007)]{calzetti07}
Calzetti, D., et al. 2007, ApJ, 666, 870

\bibitem[Clavel et al.(2000)]{clavel00}
Clavel, J., Schulz, B., Altieri, B., Barr, P., Claes, P., Heras, A., Leech, K., Metcalfe, L.,
\& Salama, A., 2000, A\&A, 357, 839

\bibitem[Cleary et al.(2007)]{cleary07}
Cleary, K., Lawrence, C. R., Marshall, J. A., Hao, L., \& Meier, D., 2007, ApJ, 660, 117

\bibitem[Dale et al.(2006)]{dale06}
Dale, D. A., et al., 2006, ApJ, 646, 161

\bibitem[de Grijp et al.(1992)]{degrijp92}
de Grijp, M. H. K., Keel, W. C., Miley, G. K., Goudfrooij, P., \& Lub, J.,
1992, \aaps, 96, 389

\bibitem[Deo et al.(2007)]{deo07}
Deo, R. P., Crenshaw, D. M., Kraemer, S. B., Dietrich, M., Elitzur, M., Teplitz, H., 
\& Turner, T. J., 2007, ApJ, 671, 124.

\bibitem[Diamond-Stanic et al. (2009)]{stanic09}
Diamond-Stanic, A. M., Rieke, G. H., \& Rigby, J. R., 2009, \apj, 698, 623

\bibitem[Draine \& Li (2007)]{draine07a}
Draine, B. T., \& Li, A., 2007, ApJ, 657, 810

\bibitem[Draine et al.(2007)]{draine07b}
Draine, B. T., et al., 2007, ApJ, 663, 866

\bibitem[Dudik et al.(2007)]{dudik07}
Dudik, R. P., Weingartner, J. C., Satyapal, S., Fischer, Jacqueline, Dudley, C. C., \& 
O'Halloran, B., 2007, ApJ, 664, 71

\bibitem[Edelson et al. (1987)]{edelson87}
Edelson, R. A., Malkan, M. A., \& Rieke, G. H., 1987, ApJ, 321, 233

\bibitem[Efstathiou \& Rowan-Robinson (1995)]{efstathiou95}
Efstathiou, A. \& Rowan-Robinson, M., 1995, \mnras, 273, 649

\bibitem[Efstathiou (2006)]{efstathiou06}
Efstathiou, A., 2006, MNRAS, 371, L70

\bibitem[Farrah et al.(2007)]{farrah07}
Farrah, D., et al., 2007, ApJ, 667, 149

\bibitem[Feigelson \& Nelson (1985)]{feigelson1985}
Feigelson, E. D., \& Nelson, P. I. 1985, \apj, 293, 192

\bibitem[Finkelstein (1986)]{finkelstein1986}
Finkelstein, D. M. 1986, Biometrics, 42, 845

\bibitem[Galliano (2006)]{galliano06}
Galliano, F., 2006, to appear in the proceedings of ''Studying 
Galaxy Evolution with Spitzer and Herschel'', astro-ph/0610852

\bibitem[Galliano et al.(2005)]{galliano05}
Galliano, E., Pantin, E., Alloin, D., \& Lagage, P. O., 2005, \mnras, 363, L1

\bibitem[Gallimore et al.\ (2009a)]{gallimore09a}
Gallimore, J. F., Yzaguirre, A., Jakoboski, J., Stevenosky, M. J., 
Axon, D. J., Baum, S. A., Buchannan, C. L., Elitzur, M., Elvis, M.,
O'Dea, C. P., Robinson, A., 2009, \apjs, submitted

\bibitem[Gallimore et al.\ (2009b)]{gallimore09b}
Gallimore, J. F., Buchannan, C. L., Axon, D. J., Baum, S. A., Elitzur, M., Elvis, M.,
O'Dea, C. P., Robinson, A. X., 2009, in preparation.


\bibitem[Gebhardt et al.\ (2000)]{gebhardt00} Gebhardt, K., et al.\ 2000, \apjl, 539,
L13

\bibitem[Genzel et al.(1998)]{genzel98}
Genzel, R., et al., 1998, ApJ, 498, 579

\bibitem[Granato \& Danese (1994)]{granato94}
Granato, G. L. \& Danese, L., 1994, \mnras, 268, 235

\bibitem[Granato et al.(2004)]{granato04}
Granato, G. L., De Zotti, G., Silva, L., Bressan, A., Danese, L., 2004, ApJ, 600, 580

\bibitem[Groves, Dopita \& Sutherland (2006)]{groves06}
Groves, B., Dopita, M., \& Sutherland, R., 2006, \aap, 458, 405

\bibitem[Gu et al.(2006)]{gu06}
Gu, Q., Melnick, J., Cid Fernandes, R., Kunth, D., Terlevich, E.; Terlevich, R.,
2006, \mnras, 366, 480

\bibitem[Haas et al.(2003)]{haas03}
Haas, M., et al., 2003, A\&A, 402, 87

\bibitem[Haas et al.(2005)]{haas05}
Haas, M., Siebenmorgen, R., Schulz, B., Kr\"ugel, E., \& Chini, R., 2005, A\&A, 442, L39

\bibitem[Hao et al.(2007)]{hao07}
Hao, L.,  Weedman, D. W., Spoon, H. W. W., Marshall, J. A., Levenson, N. A., Elitzur, M.,
\& Houck, J. R., 2007, ApJ, 655, L77

\bibitem[Hernquist \& Mihos  (1995)]{hernquist95}
Hernquist, L. \&  Mihos, J. C.,  1995, ApJ, 448, 41

\bibitem[Ho et al.(1997)]{ho97} Ho, L.~C., Filippenko, A.~V., \& Sargent,
W.~L.~W.\ 1997, \apjs, 112, 315

\bibitem[Houck et al.(2004)]{houck04} Houck, J.~R., et al.\ 2004, \apjs, 154, 18

\bibitem[Imanishi et al.(2007)]{imanishi07}
Imanishi, M., Dudley, C. C., Maiolino, R., Maloney, P. R., Nakagawa, T., \& Risaliti, G.,
2007, ApJS, 171, 72

\bibitem[Isobe, Feigelson, \& Nelson (1986)]{isobe1986}
Isobe, T., Feigelson, E. D., \& Nelson, P. I. 1986, \apj, 306, 490

\bibitem[Kauffmann et al.(2003)]{kauffman03} Kauffmann, G., et al.\ 2003, \mnras,
346, 1055

\bibitem[Kendall (1938)]{kendall38}
Kendall, M. 1938, Biometrika, 30, 81

\bibitem[Kewley et al.(2001)]{kewley01}
Kewley, L. J.; Heisler, C. A.; Dopita, M. A.; Lumsden, S.,
2001, \apjs, 132, 37

\bibitem[Kewley et al. (2006)]{kewley06}
Kewley, L. J., Groves, B.,  Kauffmann, G., \& Heckman, T., 2006, \mnras, 372, 961

\bibitem[Kim et al.(1995)]{kim95}
Kim, D.-C., Sanders, D. B., Veilleux, S., Mazzarella, J. M., Soifer, B. T.,
1995, \apjs, 98, 129

\bibitem[Kirhakos \& Steiner (1990)]{kirhakos90}
Kirhakos, S. D., \& Steiner, J. E., 1990, \aj, 99, 1722

\bibitem[Laor \& Draine (1993)]{laor93} Laor, A., \&  Draine, B. T., 1993, ApJ, 402, 441

\bibitem[Laurent et al.(2000)]{laurent00}
Laurent, O., Mirabel, I. F., Charmandaris, V., Gallais, P., Madden, S. C., 
Sauvage, M., Vigroux, L., \& Cesarsky, C., 2000, A\&A, 359. 887

\bibitem[Levenson et al.(2007)]{levenson07}
Levenson, N. A., Sirocky, M. M., Hao, L., Spoon, H. W. W., Marshall, J. A., 
Elitzur, M., Houck, J. R., 2007, ApJ, 654, L45

\bibitem[Lu et al. (2008)]{lu08}
Lu, N., et al., 2008, \pasp, 120, 328

\bibitem[Maiolino et al.(1995)]{maiolino95}
Maiolino, R., Ruiz, M., Rieke, G. H., Keller, L. D., 1995, ApJ, 446, 561

\bibitem[Maiolino et al.(2007)]{maiolino07}
Maiolino, R., Shemmer, O., Imanishi, M., Netzer, H., Oliva, E., Lutz, D., 
\&  Sturm, E., 2007, A\&A, 468, 979

\bibitem[Mason et al. (2006)]{mason06}
Mason, R. E., Geballe, T. R., Packham, C., Levenson, N. A., Elitzur, M., Fisher, R. S., 
Perlman, E., 2006, ApJ, 640, 612

\bibitem[Mel\'endez et al.(2008)]{melendez08}
Mel\'endez, M., et al., 2008, \apj, 682, 94

\bibitem[Merritt \& Ferrarese(2001)]{merritt01} Merritt, D., \& Ferrarese, L.\
2001, \apj, 547, 140

\bibitem[Mihos  \& Hernquist (1996)]{mihos96}
Mihos, J. C., \& Hernquist, L., 1996, ApJ, 464, 641

\bibitem[Mor et al. (2009)]{mor09}
Mor, R., Netzer, H., \&  Elitzur, M., 2009, astroph/0907.1654 

\bibitem[Nenkova et al. (2002)]{nenkova02}
Nenkova, M., Ivez\'ic, Z., \& Elitzur, M., 2002, \apj, 570, L9

\bibitem[Nenkova et al. (2008)]{nenkova08}
Nenkova, M., Sirocky, M. M., Nikutta, R., Ivez\'ic, Z., \& Elitzur, M., 2008,
\apj, 685, 160

\bibitem[Netzer et al.(2007)]{netzer07}
Netzer, H., et al., 2007, ApJ, 666, 806

\bibitem[Ogle et al.(2006)]{ogle06}
Ogle, P., Whysong, D., \&  Antonucci, R., 2006, ApJ, 647, 161

\bibitem[Packham et al. (2005)]{packham05}
Packham, C.,  Radomski, J. T., Roche, P. F., Aitken, D. K., Perlman, E., 
Alonso-Herrero, A., Colina, L., \& Telesco, C. M., 2005, ApJ, 618, L17

\bibitem[Panessa \& Bassani (2002)]{panessa02}
Panessa, F. \&  Bassani, L., 2002, \aap, 394, 435

\bibitem[Peeters et al.(2004)]{peeters04}
Peeters, E., Spoon, H. W. W., \& Tielens, A. G. G. M., 2004, ApJ, 613, 986

\bibitem[Phillips(1979)]{phillips79} Phillips, M.~M.\ 1979, \apjl, 227, L121

\bibitem[Pier \& Krolik (1992)]{pier92}
Pier, E. A., \& Krolik, J. H., 1992, ApJ, 401, 99

\bibitem[Roussel et al.(2001)]{roussel01}
Roussel, H., Sauvage, M., Vigroux, L., \& Bosma, A., 2001, A\&A, 372, 427

\bibitem[Rush et al.(1993)]{rush93}
Rush, B., Malkan, M. A., \& Spinoglio, L., 1993, ApJS, 89, 1

\bibitem[Schaerer \& Stasi\'nska (1999)]{schaerer99}
Schaerer, D., \&  Stasi\'nska, G., 1999, \aap, 345, L17

\bibitem[Schweitzer et al.(2006)]{schweitzer06}
Schweitzer, M., et al., 2006, ApJ, 649, 79 

\bibitem[Shi et al.(2006)]{shi06}
Shi, Y., Rieke, G. H., Hines, D. C., Gorjian, V., Werner, M. W., Cleary, K., Low, F. J., 
Smith, P. S., \& Bouwman, J., 2006, ApJ, 653, 127

\bibitem[Shi et al.(2007)]{shi07}
Shi, Y., Ogle, P., Rieke, G. H., Antonucci, R., Hines, D. C., Smith, P. S., Low, F. J., 
Bouwman, J.,  Willmer, C., 2007, ApJ, 669, 841

\bibitem[Shlosman et al.(1990)]{shlosman90}
Shlosman, I., Begelman, M. C., \&  Frank, J., 1990, Nature, 345, 679 

\bibitem[Siebenmorgen et al.(1997)]{siebenmorgen97}
Siebenmorgen, R., Moorwood, A., Freudling, W., \& K\"aeufl, H. U.,
1997, A\&A, 325, 450

\bibitem[Siebenmorgen et al.(2004)]{siebenmorgen04}
Siebenmorgen, R., Kr\"ugel, E., \& Spoon, H. W. W., 2004, A\&A, 414, 123

\bibitem[Sirocky et al.(2008)]{sirocky08}
Sirocky, M. M., Levenson, N. A., Elitzur, M., Spoon, H. W. W., \& Armus, L.,
2008, ApJ, 678, 729

\bibitem[Smith et al.(2002)]{smith02}
Smith, J. E., Young, S., Robinson, A., Corbett, E. A., Giannuzzo, M. E., 
Axon, D. J., \& Hough, J. H., 2002, \mnras, 335, 773

\bibitem[Smith  et al.(2007a)]{smith07a}
Smith, J. D., et al., 2007, ApJ, 656, 770

\bibitem[Smith	et al.(2007b)]{smith07b}
Smith, J. D., et al., 2007, PASP, 119, 1133 


\bibitem[Spinoglio \& Malkan(1989)]{spinoglio89} Spinoglio, L., \& Malkan,
  M. A. 1989, \apj, 342, 83

\bibitem[Spinoglio  \& Malkan (1992)]{spinoglio92}
Spinoglio, L., \& Malkan, M. A., ApJ, 399, 504

\bibitem[Spitzer Science Center (2006)]{spizer06}
Spitzer Science Center, 2006, SPICE Spitzer IRS Custom Extraction (Pasadena: SSC)


\bibitem[Spoon	et al.(2007)]{spoon07}
Spoon, H. W. W., Marshall, J. A., Houck, J. R., Elitzur, M., Hao, L., Armus, L., 
Brandl, B. R., \& Charmandaris, V., 2007, ApJ, 654, L49

\bibitem[Sprent \& Smeeton (2000)]{sprent00}
Sprent, P. \& Smeeton, N. C. 2000, Applied Nonparametric Statistical Methods,
(Chapman \& Hall/CRC Press: Boca Raton, FL

\bibitem[Sturm	et al.(2002)]{sturm02}
Sturm, E., Lutz, D., Verma, A., Netzer, H., Sternberg, A., Moorwood, A. F. M., Oliva, E., 
Genzel, R., 2002, A\&A, 393, 821

\bibitem[Sturm	et al.(2006)]{sturm06}
Sturm, E., et al., 2006, ApJ, 653, L13

\bibitem[Sun (1996)]{sun1996}
Sun, J. 1996, Statistics in Medicine, 15, 1387


\bibitem[Thean et al. (2001)]{thean01}
Thean, A., Pedlar, A., Kukula, M. J., Baum, S. A., O'Dea, C. P., 2001, MNRAS, 
325, 737


\bibitem[Tielens et al.(1999)]{tielens99}
Tielens, A. G. G. M., Hony, S., van Kerckhoven, C., \& Peeters, E., 1999,
The Universe as Seen by ISO. Eds. P. Cox \& M. F. Kessler. ESA-SP 427., p. 579

\bibitem[Tommasin et al. (2008)]{tommasin08}
Tommasin, S., Spinoglio, L., Malkan, M. A., Smith, H., Charmandaris, V., 2007, 
\apj, 676, 836

\bibitem[Tran(2003)]{tran03} Tran, H.~D.\ 2003, \apj, 583, 632

\bibitem[van Bemmel \& Dullemond (2003)]{vanbemmel03}
van Bemmel, I. M. \& Dullemond, C. P., 2003, \aap, 404, 1

\bibitem[Veilleux et al.(1995)]{veilleux95} Veilleux, S., Kim, D.-C., Sanders,
D.~B., Mazzarella, J.~M., \& Soifer, B.~T.\ 1995, \apjs, 98, 171

\bibitem[V{\' e}ron-Cetty \& V{\' e}ron(2001)]{veron01} V{\' e}ron-Cetty, M.-P.,
\& V{\' e}ron, P.\ 2001, \aap, 372, 730


\bibitem[V{\' e}ron-Cetty \& V{\' e}ron(2003)]{veron03} V{\' e}ron-Cetty, M.-P.,
\& V{\' e}ron, P.\ 2003, \aap, 412, 399


\bibitem[Voit (1992a)]{voit92a}
Voit, G. M., 1992a, MNRAS, 258, 841

\bibitem[Voit (1992b)]{voit92b}
Voit, G. M., 1992b, ApJ, 399, 495 

\bibitem[Weedman et al. (2005)]{weedman05}
Weedman, D. W., Hao, L., Higdon, S. J. U., Devost, D., Wu, Y.,  Charmandaris, V., Brandl, B., 
Bass, E., \& Houck, J. R., 2005, ApJ, 633, 706

\bibitem[Whittle (1992)]{whittle92}
Whittle, M., 1992, \apjs, 79, 49

\bibitem[Williams et al. (1999)]{williams99}
Williams, R. J. R., Baker, A. C., \& Perry, J. J., 1999, MNRAS, 310, 913

\bibitem[Winkler (1992)]{winkler92}
Winkler, H., 1992, \mnras, 257, 677

\bibitem[Wu et al. (2009)]{wu09}
Wu, Y., Charmandaris, V., Huang, J., Spinoglio, L., \& Tommasin, S., 2009, ApJ,
701, 658

\bibitem[Young et al.(1996)]{young96}
Young, S., Hough, J. H., Efstathiou, A., Wills, B. J., Bailey, J. A., 
Ward, M. J., \& Axon, D. J., 1996, \mnras, 281, 1206

\end{thebibliography}
\end{document}